# Character of couple and couple-stress in continuum mechanics

Ali R. Hadjesfandiari

*Department of Mechanical and Aerospace Engineering*
*University at Buffalo, State University of New York*
*Buffalo, NY 14260 USA*

ah@buffalo.edu

January 10, 2022

**Abstract**

In this paper, the concept of moment and couple in mechanics is examined from a fundamental perspective. It turns out that although representing a couple by its moment vector is very useful in rigid body mechanics and strength of materials, it has been very misleading in continuum mechanics. To specify the effect of a concentrated couple in continuum mechanics, not only the couple moment, but also the line of action of its constituent parallel opposite forces must be specified. However, in the governing equations of equilibrium or motion of a continuum, only moment of body couple, moment of couple-tractions, and moment of couple-stresses appear without specifying the line of action of any couple density forces. This results in non-uniqueness of the state of stresses and deformation in the continuum, which has shown itself in the indeterminacy of the couple-stress tensor. Nevertheless, the physical state of stress and deformation in the continuum is unique and determinate. Therefore, this imposes some restrictions on the form of body couple, couple-traction and couple-stresses. Here, the uniqueness of interactions in the continuum is used to establish that the continuum does not support a distribution of body couple, and a distribution of surface twisting couple-traction with normal moment. From this the mechanism of action of the couple-traction as a double layer of shear force-tractions, and the skew-symmetric character of the couple-stress moment tensor in continuum mechanics are also established.

## 1. Introduction

Couple-stresses $\mu_{ij}$ inevitably appear along with force-stresses $\sigma_{ij}$ in a complete continuum mechanics. As a result, the force-stress tensor $\sigma_{ij}$ is not symmetric as is the case in classical theory. Voigt (1887) was the first who postulated the existence of couple-stresses in continuum mechanics. Later, Cosserat and Cosserat (1909) developed the original mathematical model for couple stress continuum mechanics. Mindlin and Tiersten (1962) and Koiter (1964) developed an initial incomplete version of couple stress theory, which uses the four foundational mechanical quantities (i.e., force, displacement, couple, rotation) compatible with linear and angular momentum principles in continuum mechanics. Subsequently, Stokes (1966) brought this formulation into fluid mechanics to model the size dependency effect in fluids. However, this original couple stress theory suffers from some fundamental inconsistencies, which are mainly:

1. The indeterminacy in the spherical part of the couple-stress moment tensor;
2. The inconsistency in boundary conditions, since the normal component of couple-traction moment vector appears in the formulation;
3. The appearance of the body couple moment in the relation for the force-stress tensor.

The appearance of the indeterminacy of the spherical part of the couple-stress moment tensor is troublesome in most cases, especially those with torsional deformation. Surprisingly, Koiter (1964) and Stokes (1966) make the claim that without any loss of generality the indeterminate spherical part may be taken zero to make the couple-stress tensor deviatoric. However, a deviatoric couple stress theory, where $\mu_{ii} = \mu_{11} + \mu_{22} + \mu_{33} = 0$, is a non-physical theory and still suffers from ill-posed boundary conditions. Surprisingly, in this deviatoric theory, a uniaxial torsional deformation with a normal couple-stress becomes impossible. Based on physical grounds, if one can exert torsional couple-stresses $\mu_{11}$, $\mu_{22}$ and $\mu_{33}$ on some element of the body, these three components must be independent of each other. However, this contradicts the mathematical deviatoric condition, $\mu_{ii} = \mu_{11} + \mu_{22} + \mu_{33} = 0$. Enforcing this condition is as absurd as enforcing the mathematical constraint $\sigma_{ii} = \sigma_{11} + \sigma_{22} + \sigma_{33} = 0$ on the force-stress tensor in a classical continuum mechanics theory. The three components $\sigma_{11}$, $\sigma_{22}$ and $\sigma_{33}$ acting on some element of the body are generally independent of each other.

Eringen (1968) realized this indeterminacy as a major mathematical problem in Mindlin, Tiersten and Koiter couple-stress theory. As a result, he called this theory indeterminate couple stress theory. It is obvious that for having a consistent couple stress theory, it is necessary to resolve the inconsistencies, especially the criticism of Eringen (1968) about the indeterminacy of couple-stress moment tensor. However, instead of understanding the source of the indeterminacy and ill-posed boundary condition issues, and resolving them in a fundamental level, researchers tried to avoid these issues by inventing new concepts and theories in continuum mechanics. Therefore, not resolving the inconsistencies of couple stress theory has cost continuum mechanics a great deal and confused its progress by creating many non-physical concepts and theories.

Recently, Hadjesfandiari and Dargush (2011) have resolved all of these inconsistencies by developing the consistent couple stress theory. Interestingly, this progress demonstrates the subtle skew-symmetric character of the couple-stress moment tensor, and the impossibility of an independent body couple distribution in the continuum. This has been achieved by examining kinematics, well-posed boundary conditions, and the virtual work principle. Elements of establishing this character are based on Mindlin and Tiersten (1962) and Koiter (1964), which established the impossibility of normal couple-traction (surface couple-traction with normal moment) in a continuum. It is remarkable to note that the well posed form of boundary conditions imposes the skew-symmetric character of the couple-stress moment tensor (Hadjesfandiari and Dargush, 2011; Hadjesfandiari et al., 2015), which was missed by Mindlin and Tiersten (1962) and Koiter (1964). Interestingly, this new development shows that there is an interrelationship between the consistent mechanical boundary conditions, and the determinacy of the couple-stress moment tensor; resolving one, resolves the other (Hadjesfandiari and Dargush, 2011, 2015).

Although the discovery of the skew-symmetric character of the couple-stress moment tensor resolves the quest for the consistent continuum mechanics (Hadjesfandiari and Dargush, 2011; Hadjesfandiari et al., 2015), its form of establishment seems very intriguing. One might ask why it is necessary to use the concept of energy and kinematics, and specify the independent degrees of freedom as well as their conjugate generalized forces, or if there exists any other method to establish this statement. Moreover, this method of proof does not specify the mechanism of action of couple-tractions and couple-stresses. Experience shows that there are usually a few different methods to prove a lemma. Here it is demonstrated that this is the case and is established the skew-symmetric character of the couple-stress moment tensor by a different fundamental method, which does not depend on using the energy concept and kinematics. This character is systematically established by examining the concepts of moment and couple and the fundamental governing equations. Remarkably, this new fundamental method of proof is more complete, because it reveals the mechanism of action of couple-traction as a double layer of shear-force tractions.

First it is shown that representing moment of a force by a vector in mechanics, although it is very convenient, has been very misleading. This is because it usually gives impression that moment is a vector exerted to a point (moment center) similar to a concentrated force. However, it should be noted that the vector moment of a force is a pseudo-vector without any real vectorial character. Interestingly, moment of a force is in reality a skew-symmetric second-order true tensor. Since working with a skew-symmetric moment tensor is awkward in practice, its dual pseudo-vector has been standard. Remarkably, this vectorial representation of moment also simplifies the governing equations in mechanics.

It turns out that using this misperception has been more misleading for a couple system. As known, the moment of a couple remains the same for all points. However, the representation of the effect of couple by this constant moment has been the main reason for failure in the progress of couple stress continuum mechanics in the last century. It is only in rigid body mechanics that the effect of a couple is completely represented by its moment vector, because it directly appears in the governing equations. Note that replacing this couple by a different equipollent couple does not change the state of equilibrium or motion of the rigid body. Therefore, in rigid body mechanics, couple is considered equivalent to its moment, which is also treated as a free vector. However, when one studies the deformation and internal stresses in continuum mechanics, the couple moment is not a free vector and cannot completely describe its effect. This is also the case for a concentrated couple, where the line of action of its opposite parallel concentrated forces are approaching each other. Therefore, in continuum mechanics the specification of concentrated forces is necessary to represent this couple. To specify the effect of a concentrated couple in continuum mechanics, its moment and the line of action of its opposite parallel forces must be specified. This clearly shows that why the incomplete representation of a general couple by its pseudo-vector moment has been so misleading in the progress of continuum mechanics from beginning of the last century. Cosserats, Mindlin, Tiersten, Koiter, Eringen assumed that the general distributions of body couple and couple-stresses can exist and can be completely represented by their moment densities without any restriction. However, they did not notice that couple is in reality a system of two opposite parallel concentrated forces acting on the body.

Contrary to assumption of Cosserat and Cosserat (1909), Mindlin and Tiersten (1962) and Koiter (1964) general distribution of body couple and couple-stresses cannot be completely represented with their moments. However, in the governing equations of equilibrium or motion of continuum, boundary conditions, and constitutive relations for internal interactions, only moment of body couple, moment of couple-tractions and moment of couple-stresses appear without specifying the line of action of any couple density forces. This results in non-uniqueness of the state of stresses and deformation in the continuum, which has shown itself in the indeterminacy of the couple-stress tensor in Mindlin-Tiersten-Koiter couple stress theory. Nevertheless, the physical state of stress and deformation in the continuum is unique and determinate. Obviously, this character enforces some restrictions on the form of distribution of body couple, couple-traction and couple-stresses so that their effect is completely described by their moment densities without requiring the specification of the line of action of opposite parallel constituent couple forces. In this paper, the uniqueness of interactions in the continuum is used to establish:

1. An independent distribution of body couple does not exist in continuum mechanics;
2. A distributions of surface twisting couple-traction (with normal moment) does not exist on any arbitrary surface;
3. A distributions of surface bending couple-traction (with tangential moment) can exist;
4. The surface bending couple-traction is a double layer of shear force-tractions ;
5. The pseudo moment tensor of couple-stresses is skew-symmetric, and has a true vectorial character.

This development shows more fundamentally why original couple stress theory of Mindlin and Tiersten (1962) and Koiter (1964) is indeterminate and suffers from inconsistencies. The arbitrary surface and body couple distributions cannot be completely represented by their moment densities in this theory. Importantly, this development also reveals the character of tangential bending couple-traction as a double layer of shear force-tractions. Interestingly, this is the tensorial analogy of double layer in electrostatics, where a single tangential shear force-traction distribution is analogous to a single layer of electric charge.

The remainder of the paper is organized as follows. Section 2 provides an overview of some important aspects of mechanics. This includes the review of forces and their moments, governing equations of motion for system of particles, and equipollent system of forces. Section 3 briefly presents the concepts of a couple and its moment, a concentrated couple, and reduction of system of forces to an equipollent system of one force and one couple. It is seen that in rigid body mechanics and to some extent in strength of material and structural mechanics equipollent systems of forces are equivalent. Section 4 considers the state of loading in continuum mechanics by reviewing body force, body couple, surface force-traction, and couple-traction. Section 5 provides the governing equations in continuum mechanics by introducing force- and couple-stresses. In this section, the uniqueness of interactions in the continuum mechanics is discussed. Afterwards, by using this uniqueness or determinacy character, the impossibility of a distribution of body couple, and a distribution of surface twisting couple-traction is established in Section 6. From this the mechanism of action of the couple-traction as a double layer of shear force-tractions, and the skew-symmetric character of couple-stress moment tensor in continuum mechanics are established. Finally, a summary and some general conclusions is presented in Section 7.

## 2. Preliminaries

Let us consider the three dimensional orthogonal right-handed (positive) coordinate system $x_1x_2x_3$ as the reference frame with unit base vectors $\mathbf{e}_1$, $\mathbf{e}_2$ and $\mathbf{e}_3$. This is the coordinate system used to represent the components of fundamental vectors, tensors, and vector and tensor equations.

## 2.1. Forces and their moments

Point forces acting on individual particles and bodies are concentrated forces. This means a force vector $\mathbf{F}$ is characterized by its magnitude, point of application and its direction.

Consider the force $\mathbf{F}$ acting point *A* with the position vector $\mathbf{r}$ relative to fixed point *O*, as shown in Figure 1.

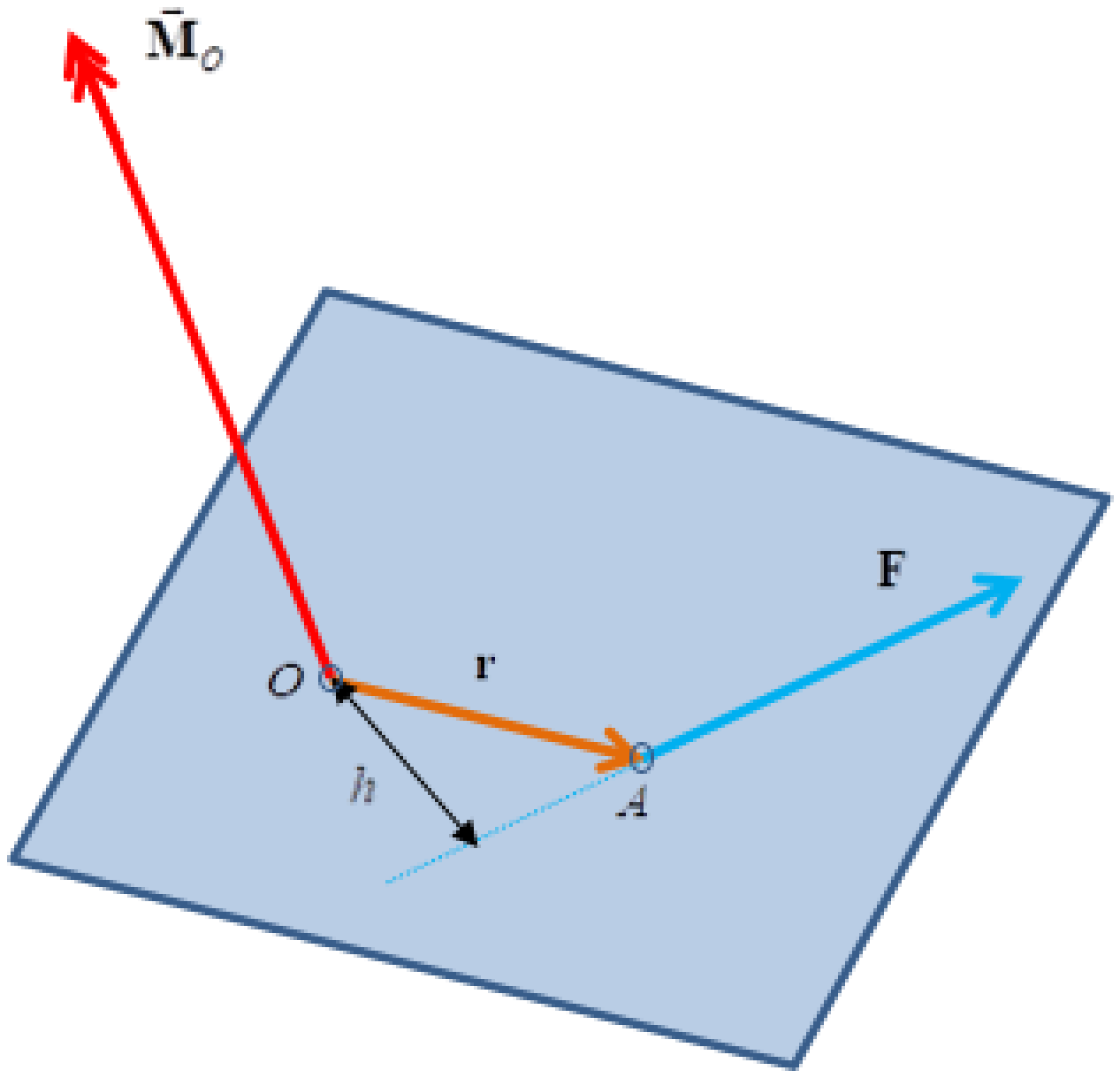

**Figure 1.** Moment of force $\mathbf{F}$ about point *O*.

Although it is possible to define the dyadic product $\mathbf{r} \otimes \mathbf{F}$ ($x_i F_j$) as a general moment, it turns out that the skew-symmetric part of this tensor plays an important role in mechanics. Therefore, the tensorial moment of the force $\mathbf{F}$ at $\mathbf{r}$ about the point *O*, called moment center, is defined as

$$\mathbf{M}_O = \frac{1}{2}(\mathbf{r} \otimes \mathbf{F} - \mathbf{F} \otimes \mathbf{r}) \qquad M_{ij} = \frac{1}{2}(x_i F_j - x_j F_i) \tag{1}$$

where

$$\mathbf{M}_O^T = -\mathbf{M}_O \qquad M_{ji} = -M_{ij} \tag{2}$$

In terms of components, this skew-symmetric moment tensor can be written as

$$\left[M_{ij}\right]=\begin{bmatrix} 0 & M_{12} & M_{13} \\ -M_{12} & 0 & M_{23} \\ -M_{13} & -M_{23} & 0 \end{bmatrix}$$

$$=\begin{bmatrix} 0_1 & \frac{1}{2}\left(x_1F_2-x_2F_1\right) & \frac{1}{2}\left(x_1F_3-x_1F_3\right) \\ -\frac{1}{2}\left(x_1F_2-x_2F_1\right) & 0 & \frac{1}{2}\left(x_2F_3-x_3F_2\right) \\ -\frac{1}{2}\left(x_1F_3-x_1F_3\right) & -\frac{1}{2}\left(x_2F_3-x_3F_2\right) & 0 \end{bmatrix} \tag{3}$$

Since this tensor is specified by three independent components, it looks more convenient to represent it by its dual pseudo- or axial-vector $\vec{\mathbf{M}}_O$ (Figure 1), where

$$\vec{\mathbf{M}}_O=\mathbf{r}\times\mathbf{F} \qquad\qquad M_i=\varepsilon_{ijk}x_jF_k \tag{4}$$

Here $\varepsilon_{ijk}$ is the Levi-Civita alternating symbol. Interestingly, the dual relations are

$$M_i=\varepsilon_{ijk}M_{jk} \qquad\qquad M_{ij}=\frac{1}{2}\varepsilon_{ijk}M_k \tag{5}$$

where

$$M_1=2M_{23}, \qquad M_2=2M_{31}, \qquad M_3=2M_{12} \tag{6}$$

It should be noted that the moment vector $\vec{\mathbf{M}}_O$, defined by vector product (cross product) $\mathbf{r}\times\mathbf{F}$, is perpendicular to the plane containing point $O$ and force vector $\mathbf{F}$. The magnitude of this moment vector is $M_O=Fh$, where $h$ is the perpendicular distance of point $O$ to the line of action of force $\mathbf{F}$. By definition, the pseudo-vector (axial vector) $\vec{\mathbf{M}}_O$ is attributed to the point *O*, as shown in Figure 1. Note that the moment vector $\vec{\mathbf{M}}_O$ has been presented as double headed arrows. It should be also mentioned that the moment of the force vector $\mathbf{F}$ can be considered about many arbitrary points, simultaneously. It is obvious that the vector moment $\vec{\mathbf{M}}_O$ is not applied at the point *O* physically. It is only the force $\mathbf{F}$, which acts the point *A*, not its moment. Although representation of moment of a force by a pseudo-vector has been very convenient in practice, it has created some troubles in the progress of continuum mechanics. This will be illustrated in detail in subsequence sections.

Since the second order moment tensor $\mathbf{M}_O$ is skew-symmetric, it is singular. This means its determinant vanishes

$$\det\left(\mathbf{M}_O\right) = 0 \tag{7}$$

$$\det\left[M_{ij}\right] = \varepsilon_{ijk} M_{1i} M_{2j} M_{3k} = 0 \tag{8}$$

As a result, the tensor $\mathbf{M}_O$ has a zero eigenvalue. Interestingly, the eigenvector corresponding to this zero eigenvalue is in the direction of the pseudo-vector $\vec{\mathbf{M}}_O$, thus

$$\mathbf{M}_O \bullet \vec{\mathbf{M}}_O = 0 \qquad\qquad M_{ij} M_j = 0 \tag{9}$$

The skew-symmetric tensor $\mathbf{M}_O$ or its more popular dual moment vector $\vec{\mathbf{M}}_O$ , which is also preferred in this paper, are very important from a physical point of view. This will be reviewed next, where governing equations of a system of particles are presented.

### 2.2. Fundamental governing equations of motion for a system of particles

Consider the motion of a particle with mass $m$ under the influence of the resultant force $\mathbf{F}$. The location of the particle in space at time $t$ is specified by the position vector $\mathbf{r} = \mathbf{r}(t)$. The velocity and acceleration vectors are defined as $\mathbf{v} = \frac{d\mathbf{r}}{dt}$ and $\mathbf{a} = \frac{d\mathbf{v}}{dt}$, respectively. The governing equations of motion of the particle are given by Newton's second law as

$$\mathbf{F} = m\mathbf{a} \qquad\qquad F_i = ma_i \tag{10}$$

where the vector $m\mathbf{a}$ is called effective force. Therefore, the vector equation (10), which is also called the force equation, states that the resultant force $\mathbf{F}$ acting on the particle equals to the effective force $m\mathbf{a}$.

Now consider a system of particles interacting with each other. This system can represent a continuous body with infinite particles. The equations (10) can be considered for each particle. Note that the resultant force $\mathbf{F}$ for each particle can be decomposed as

$$\mathbf{F} = \mathbf{F}^{ext} + \mathbf{F}^{int} \qquad F_i = F_i^{ext} + F_i^{int} \tag{11}$$

Here $\mathbf{F}^{ext}$ is the external resultant force, whereas $\mathbf{F}^{int}$ represents the internal resultant force on the particle exerted by other particles in the system.

By adding the force and moment equations about point $O$ for all individual particles in the system, one obtains

$$\sum \mathbf{F} = \sum m\mathbf{a} \qquad \sum F_i = \sum m a_i \tag{12}$$

$$\sum \vec{\mathbf{M}}_O = \sum \mathbf{r} \times m\mathbf{a} \qquad \sum M_i = \varepsilon_{ijk} \sum x_j m a_k \tag{13}$$

Note that the total force $\sum \mathbf{F}$ and the total moment $\sum \mathbf{M}_O$ are a combination of the external and internal forces and moments, respectively, where

$$\sum \mathbf{F} = \sum \mathbf{F}^{ext} + \sum \mathbf{F}^{int} \tag{14}$$

$$\sum \vec{\mathbf{M}}_O = \sum \vec{\mathbf{M}}_O^{ext} + \sum \vec{\mathbf{M}}_O^{int} \tag{15}$$

However, due to Newton's third law of action and reaction, the effect of internal forces and moments disappear, that is $\sum \mathbf{F}^{int} = 0$ and $\sum \vec{\mathbf{M}}_O^{int} = 0$ (Goldstein, 1980; Beer and Johnston, 1988; Shames, 1980). Therefore, the force and moment equations for the system of particles become

$$\sum \mathbf{F}^{ext} = \sum m\mathbf{a} \tag{16}$$

$$\sum \vec{\mathbf{M}}_O^{ext} = \sum \mathbf{r} \times m\mathbf{a} \tag{17}$$

These equations state that the resultant external force $\sum \mathbf{F}^{ext}$ and the resultant external moments $\sum \vec{\mathbf{M}}_O^{ext}$ are equal to total effective force $\sum m\mathbf{a}$ and effective moment $\sum \mathbf{r} \times m\mathbf{a}$, respectively.

Interestingly, this result shows that Newton's third law of action and reaction is the reason for defining moment of a force by (4) so that the effect of internal moments disappears for a system of particles.

By introducing the concept of total linear momentum $\mathbf{P}$ and total angular momentum $\mathbf{L}_O$ for the system of particles as

$$\mathbf{P} = \sum m\mathbf{v} \qquad P_i = \sum m v_i \tag{18}$$

$$\mathbf{L}_O = \sum \mathbf{r} \times m\mathbf{v} \qquad L_O = \varepsilon_{ijk} \sum x_j m v_k \tag{19}$$

the force and moment equations (16) and (17) can be expressed as

$$\sum \mathbf{F}^{ext} = \sum m\mathbf{a} = \frac{d\mathbf{P}}{dt} \tag{20}$$

$$\sum \vec{\mathbf{M}}_O^{ext} = \sum \mathbf{r} \times m\mathbf{a} = \frac{d\mathbf{L}_O}{dt} \tag{21}$$

The total linear momentum $\mathbf{P}$ and effective forces $\sum m\mathbf{a}$ can be simplified by the introduction of center of mass (Goldstein, 1980; Beer and Johnston, 1988; Shames, 1980).

Note that although defining higher moments of forces and momentum is possible, they do not have fundamental significance from a physical point of view. For example, defining the dyadic product $\mathbf{r} \otimes \mathbf{F}$ ($x_i F_j$) and its symmetric part $\frac{1}{2}(\mathbf{r} \otimes \mathbf{F} + \mathbf{F} \otimes \mathbf{r})$ ($\frac{1}{2}(x_i F_j + x_j F_i)$) as general moments, do not result in governing equations, because the effect of these moments for internal forces do not vanish for a system of particles. It is only the skew-symmetric part of the tensor $\mathbf{r} \otimes \mathbf{F}$ ($x_i F_j$) that plays a fundamental role in mechanics, as demonstrated.

It is very important to note that two vectorial equations (16) and (17) are the only possible equations for the system of particles, in which the internal forces, and internal moments are cancelled based on Newton's third law of action and reaction. This is the reason why these equations are considered as fundamental governing equations for system of particles or a continuum body. These governing equations (16) and (17) can be used to describe the motion of bodies in integral form. This will be discussed in Section 3. However, note that the governing equations (16) and (17) are not enough to describe the motion of a system of particles or a deformable body. To study of these systems, it is necessary to consider the individual particle or infinitesimal element of matter, which means applying the Newton's second law (10) to individual particles, or applying the governing equations (16) and (17) to all infinitesimal element of matter for a continuum.

Also note that the moment equation about any other arbitrary point can be written as a linear combination of force and moment equations (16) and (17). Therefore, the governing moment equation about point $O'$

$$\sum \vec{\mathbf{M}}_{O'}^{ext} = \sum \mathbf{r}' \times m\mathbf{a} \tag{22}$$

is not a new independent governing equation.

It is also noticed that when the resultant external forces vanish, i.e., $\sum \mathbf{F}^{ext} = 0$ in the force governing equation (20), the total linear momentum of the system $\mathbf{P}$ is conserved. If the external moment about point $O$ is zero, i.e., $\sum \vec{\mathbf{M}}_O^{ext} = 0$ in the moment governing equation (21), the total angular momentum $\mathbf{L}_O$ is conserved about point $O$. Note that based on Noether's theorem (Noether, 1918), the conservation laws are the result of the symmetry properties of nature. Therefore, the conservation laws of linear and angular momentum are the result of the translational and rotational symmetry of space, respectively.

### 2.3. Equipollent system of forces

If two systems of forces have the same resultant force $\sum \mathbf{F}$ and the same resultant moment $\sum \vec{\mathbf{M}}_O$ about an arbitrary point $O$, they are called equipollent (Beer and Johnston, 1988). Interestingly, the system of internal forces is equipollent to zero ($\sum \mathbf{F}^{int} = 0$ and $\sum \vec{\mathbf{M}}_O^{int} = 0$). However, this does not mean the internal forces have no effect on the state of motion of individual particles or elements of matter.

Remarkably, equations (16) and (17) express the fact that the systems of external forces $\sum \mathbf{F}^{ext}$ and effective forces $\sum m\mathbf{a}$ are equipollent. This means these systems have the same resultant and the same resultant moment about any point.

The governing equations (16) and (17) show that if the system of external force system are replaced with an equipollent force system in their left hand sides, their right hand sides do not change. This means the system of effective forces changes to a new equipollent system of effective forces. However, note that the internal forces, and condition of equilibrium and motion of individual particles can change in this replacement. Therefore, the systems of equipollent forces acting on a deformable body are not equivalent, because they create different state of stress and deformation.

## 3. Couple of forces in mechanics

In this section, the important concept of couple is introduced, and its character in continuum mechanics is investigated. It is seen that the effect of a couple cannot be completely represented by its moment vector in continuum mechanics, especially when the deformation and internal stresses in continuum mechanics are studied.

### 3.1. Couple and its moment

The system of two parallel forces $\mathbf{F}$ and $-\mathbf{F}$ that have the same magnitude, but opposite direction form a couple (Figure 2). Let us denote the position vectors of the points of application of $\mathbf{F}$ and $-\mathbf{F}$ with $\mathbf{r}_A$ and $\mathbf{r}_B$, respectively. It turns out that the sum of the moment of forces $\vec{\mathbf{M}}$ about any arbitrary point $O$ is the same

$$\vec{\mathbf{M}} = (\mathbf{r}_A - \mathbf{r}_B) \times \mathbf{F} = \mathbf{r}_{A/B} \times \mathbf{F} \tag{23}$$

where $\mathbf{r}_{A/B} = \mathbf{r}_A - \mathbf{r}_B$ is the vector joining the position vectors $\mathbf{r}_A$ and $\mathbf{r}_B$. The constant vector $\vec{\mathbf{M}}$ is called the moment of the couple, which is perpendicular to the plane of two forces. Note that the magnitude of this constant moment $\vec{\mathbf{M}}$ is $M = Fh$, where $h$ is the perpendicular distance of the line of action of forces (Figure 2).

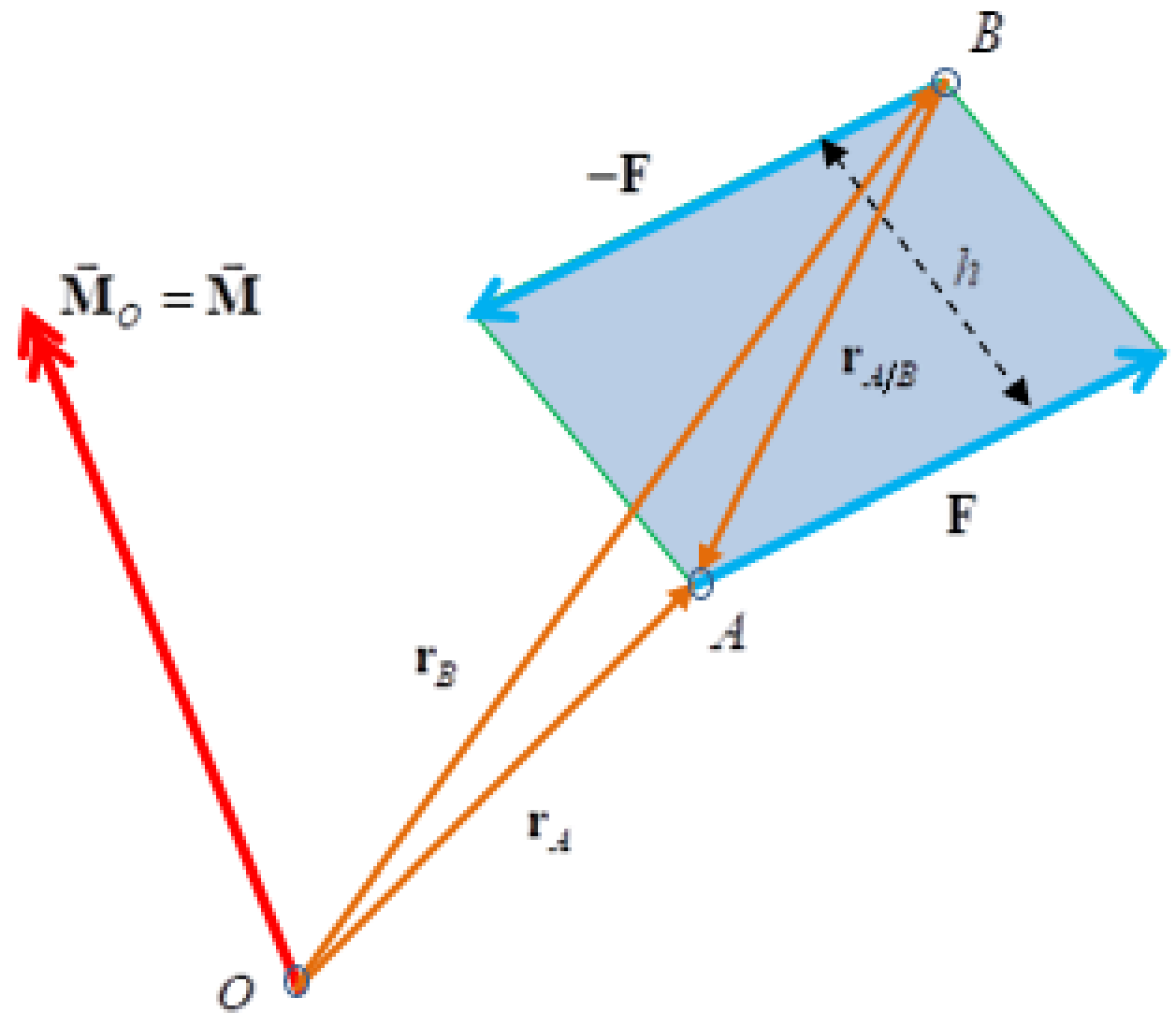

**Figure 2.** Couple $\mathbf{F}$ and $-\mathbf{F}$, and its moment about point *O*.

Note that there are infinite set of equipollent couples with the vector moment $\vec{\mathbf{M}} = \vec{\mathbf{M}}_O$. For example, the system of couple of forces $\mathbf{F}$ and $-\mathbf{F}$ is equipollent to the system of couple of forces $\mathbf{F}'$ and $-\mathbf{F}'$ with the same moment $\vec{\mathbf{M}} = \vec{\mathbf{M}}' = \vec{\mathbf{M}}_O$ (Figure 3). However, these equipollent couples are not equivalent, because they create different state of stress and deformation in the body.

Based on the definition, the pseudo-vector moment $\vec{\mathbf{M}} = \vec{\mathbf{M}}_O$ can be attributed to any arbitrary point. However, this has given the impression that couple is the vector moment $\vec{\mathbf{M}} = \vec{\mathbf{M}}_O$ acting on the body. This incorrect notion seems to have been more convincing for a concentrated couple, as will be discussed next.

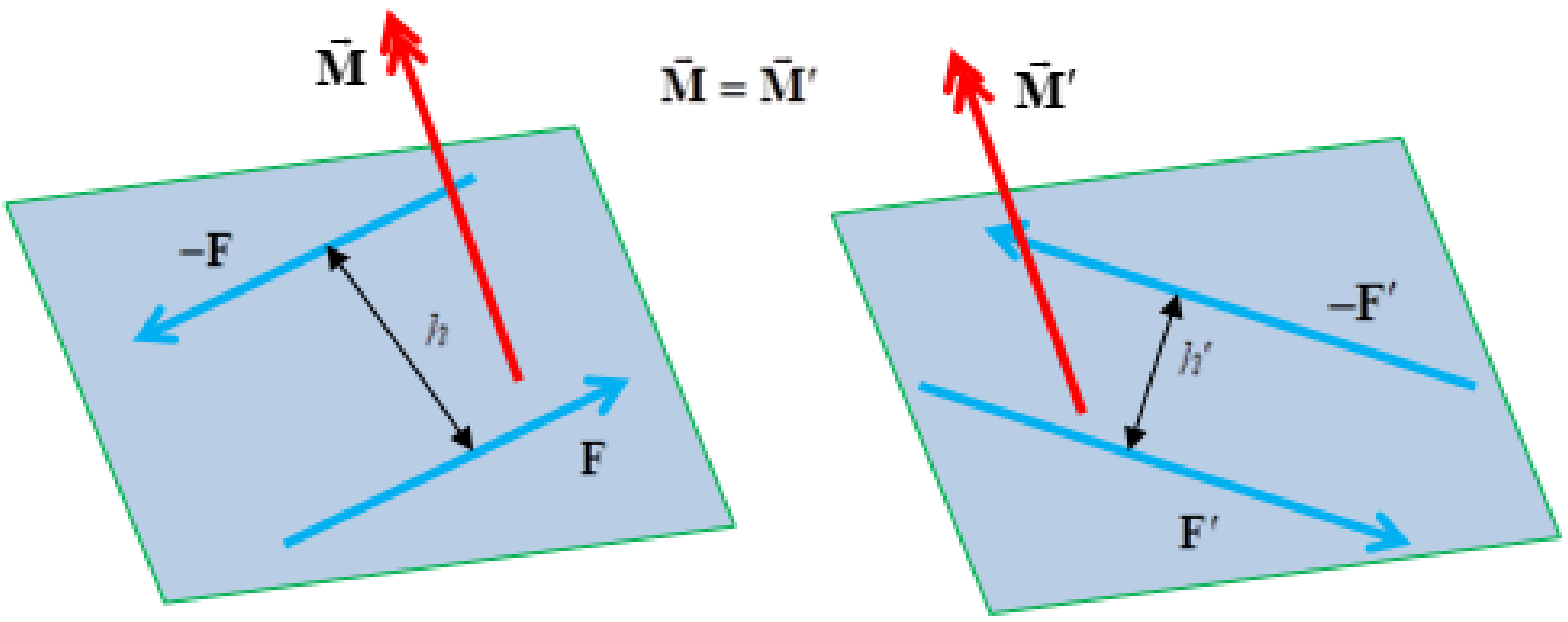

**Figure 3.** Equipollent couples on the body $M = Fh = F'h'$.

### 3.2. Concentrated couple

In mechanics of deformable bodies, the concept of concentrated couple is very useful. A concentrated couple with moment $\vec{\mathbf{M}}$ acting at point $A$ can be considered as the limit of system of parallel forces $\mathbf{F}$ and $-\mathbf{F}$ when $B$ approaches $A$, such that the moment vector $\vec{\mathbf{M}}$ remains constant (Figure 4). This means the perpendicular distance $h$ approaches zero whereas $F$ (the magnitude of $\mathbf{F}$) approaches infinity, such that

$$Fh = M \qquad \text{for } h \to 0 \text{ and } F \to \infty \tag{24}$$

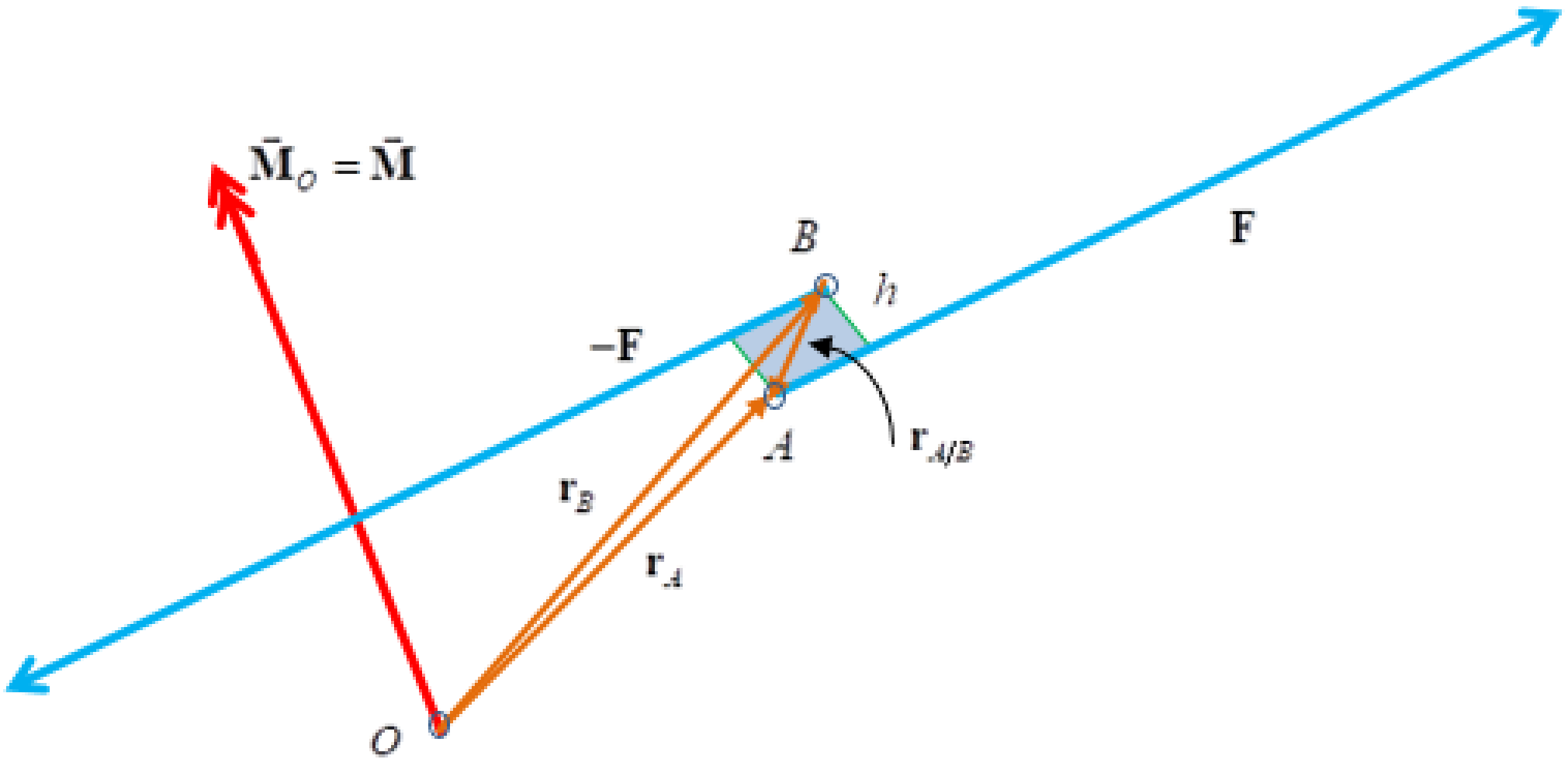

**Figure 4.** Couple $\mathbf{F}$ and $-\mathbf{F}$, and its moment, when approaching to a concentrated couple.

Although the moment of concentrated couple $\vec{\mathbf{M}}$ can still be attributed to any arbitrary point, it is usually attributed to the limiting point at $A$ (Figure 5). However, this has been very misleading because:

1. It has given the false notion that the moment vector $\vec{\mathbf{M}}$ is a real vector exerted to the point $A$, analogous to application of a force to a point;

2. It has given the false impression that the couple moment vector $\vec{\mathbf{M}}$ by itself completely describe the effect of concentrated couple at $A$.

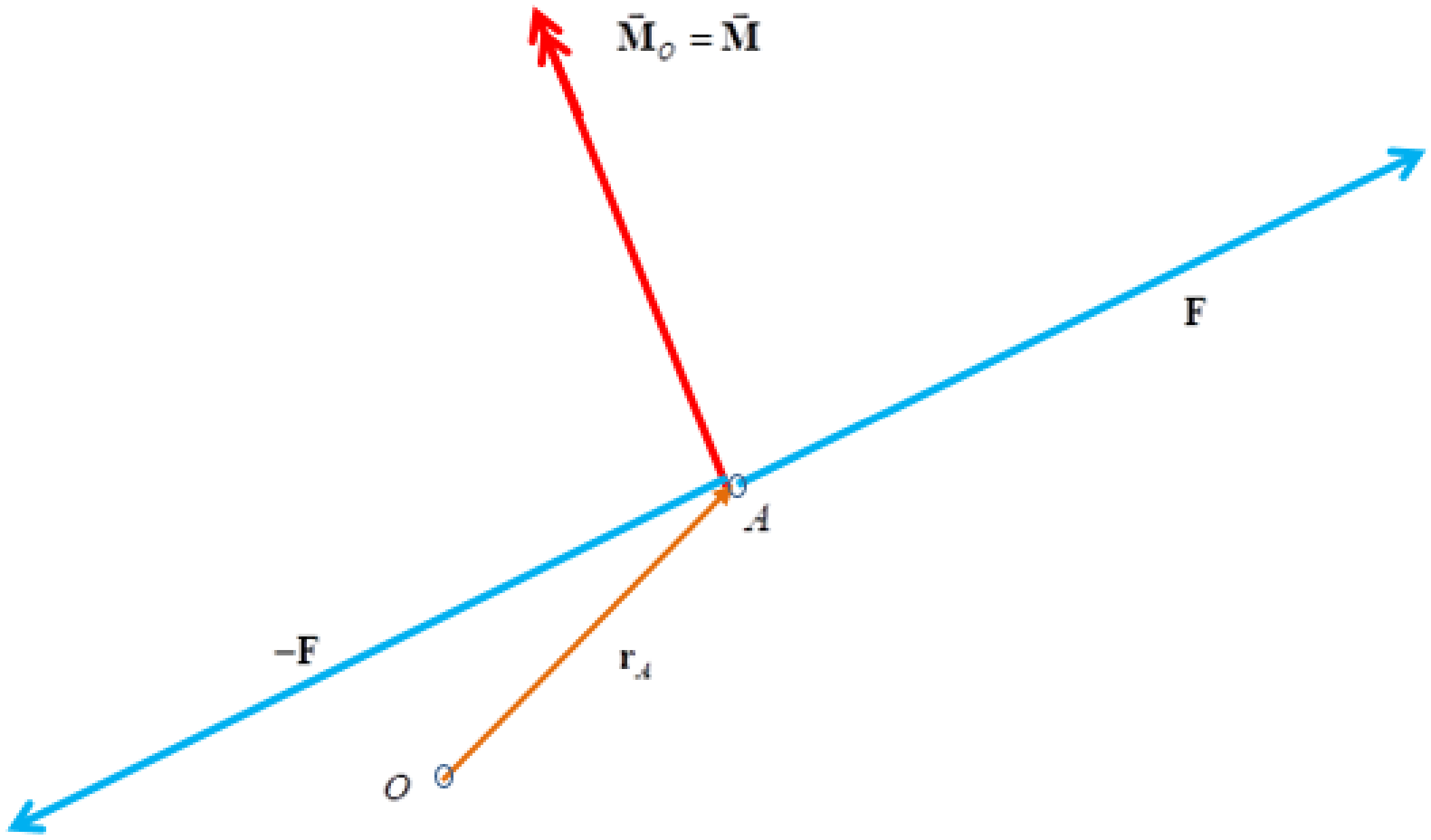

**Figure 5.** Concentrated couple $\mathbf{F}$ and $-\mathbf{F}$, and its moment $\vec{\mathbf{M}}$ at point $A$.

However, it is only the forces $\mathbf{F}$ and $-\mathbf{F}$ that are acting on the body, not the moment vector $\vec{\mathbf{M}}$. Note that the if $\vec{\mathbf{M}}$ has been defined based on a left hand rule in (23), its direction would have been opposite. Furthermore, the effect of concentrated couple with moment $\vec{\mathbf{M}}$ acting at point $A$ is not completely specified by the moment $\vec{\mathbf{M}} = \vec{\mathbf{M}}_O$. The vector $\vec{\mathbf{M}}$ does not uniquely specify the force system $\mathbf{F}$ and $-\mathbf{F}$ at $A$. Note that there are countless sets of infinitely long parallel concentrated forces at point $A$ with the same moment $\vec{\mathbf{M}} = \vec{\mathbf{M}}_O$. For example, the concentrated couple with moment $\vec{\mathbf{M}}$ acting at $A$ can be the result of either of equipollent concentrated couples with concentrated forces $\mathbf{F}$ and $-\mathbf{F}$, and concentrated forces $\mathbf{F}'$ and $-\mathbf{F}'$ in Figure 6. However, the deformation and state of stresses in the body are different even for these two equipollent concentrated couples. Interestingly, if the material is isotropic and infinitely extended, the deformation and state of stresses from the effect of couple with concentrated forces $\mathbf{F}'$ and $-\mathbf{F}'$ can be obtained by rotating the deformation and state of stresses from the effect of couple with concentrated forces $\mathbf{F}$ and $-\mathbf{F}$ with the same angle of rotation of $\mathbf{F}$ to $\mathbf{F}'$ (Figure 6).

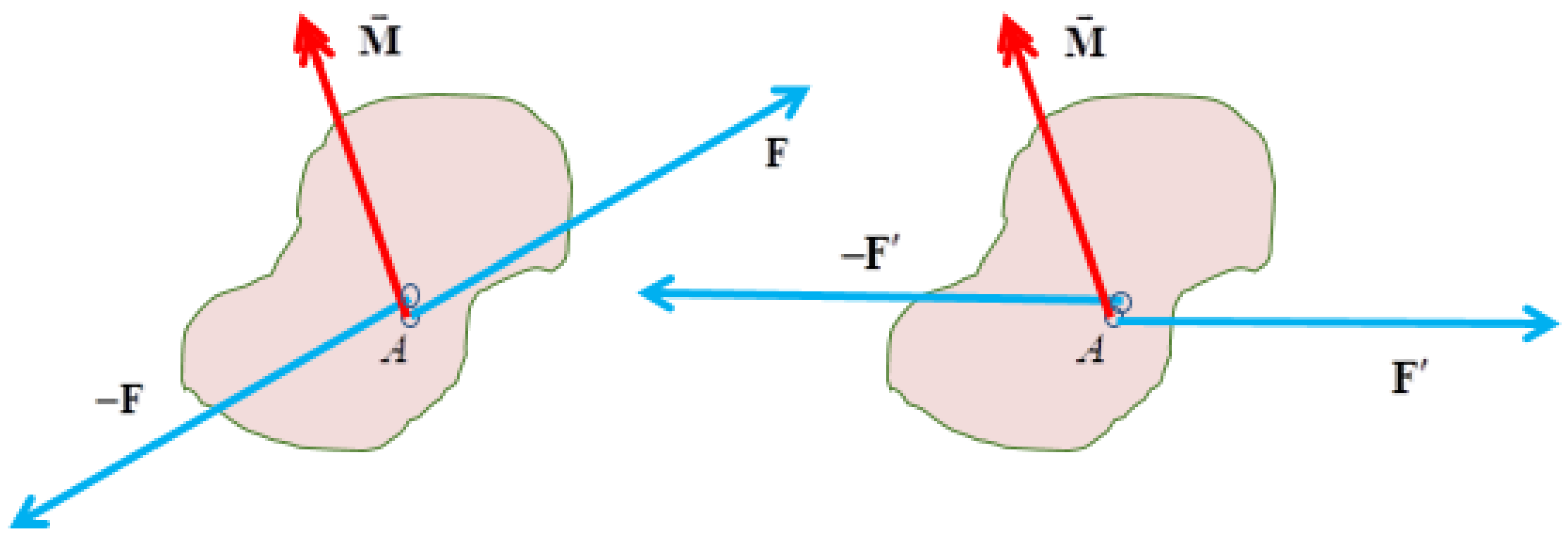

**Figure 6.** Equipollent concentrated couples at point $A$.

Note that the specification of concentrated couple moment $\vec{\mathbf{M}}$ at point $A$ specifies the plane of parallel couple forces $\mathbf{F}$ and $-\mathbf{F}$, not their directions. This clearly shows that the effect of concentrated couple cannot be completely defined by concentrating the moment vector $\vec{\mathbf{M}}$ at point $A$.

Notice that the effect of a system of concentrated couples applied at a point $A$ cannot be replaced by a single resultant couple. For example, two concentrated couples with couple moments $\vec{\mathbf{M}}_1$ and $\vec{\mathbf{M}}_2$ acting at a point cannot be generally replaced with a couple with resultant moment couple $\vec{\mathbf{M}}^R = \vec{\mathbf{M}}_1 + \vec{\mathbf{M}}_2$ in Figure 7. This means the parallelogram law for addition of moment of concentrated couples is not generally valid for concentrated couples. Interestingly, the resultant of concentrated couples with moments $\vec{\mathbf{M}}_1$ and $\vec{\mathbf{M}}_2$ represents a concentrated quadrupole not a concentrated couple or dipole.

It is only when the concentrated forces of couples act on the same points *A* and *B* before limiting process, that the parallelogram law can be used for forces and couple moments. As a result, the system of two couples $\vec{\mathbf{M}}_1$ and $\vec{\mathbf{M}}_2$ can be replaced by a resultant couple moment of $\vec{\mathbf{M}}^R = \vec{\mathbf{M}}_1 + \vec{\mathbf{M}}_2$. However, the moments $\vec{\mathbf{M}}_1$, $\vec{\mathbf{M}}_2$ and $\vec{\mathbf{M}}^R = \vec{\mathbf{M}}_1 + \vec{\mathbf{M}}_2$ cannot completely describe the effect of these couples. Interestingly, the converse is true, that is a given concentrated couple with moment $\vec{\mathbf{M}}$ at point *A* can be decomposed to some components by using parallelogram law, as long as the corresponding forces follow the parallelogram law.

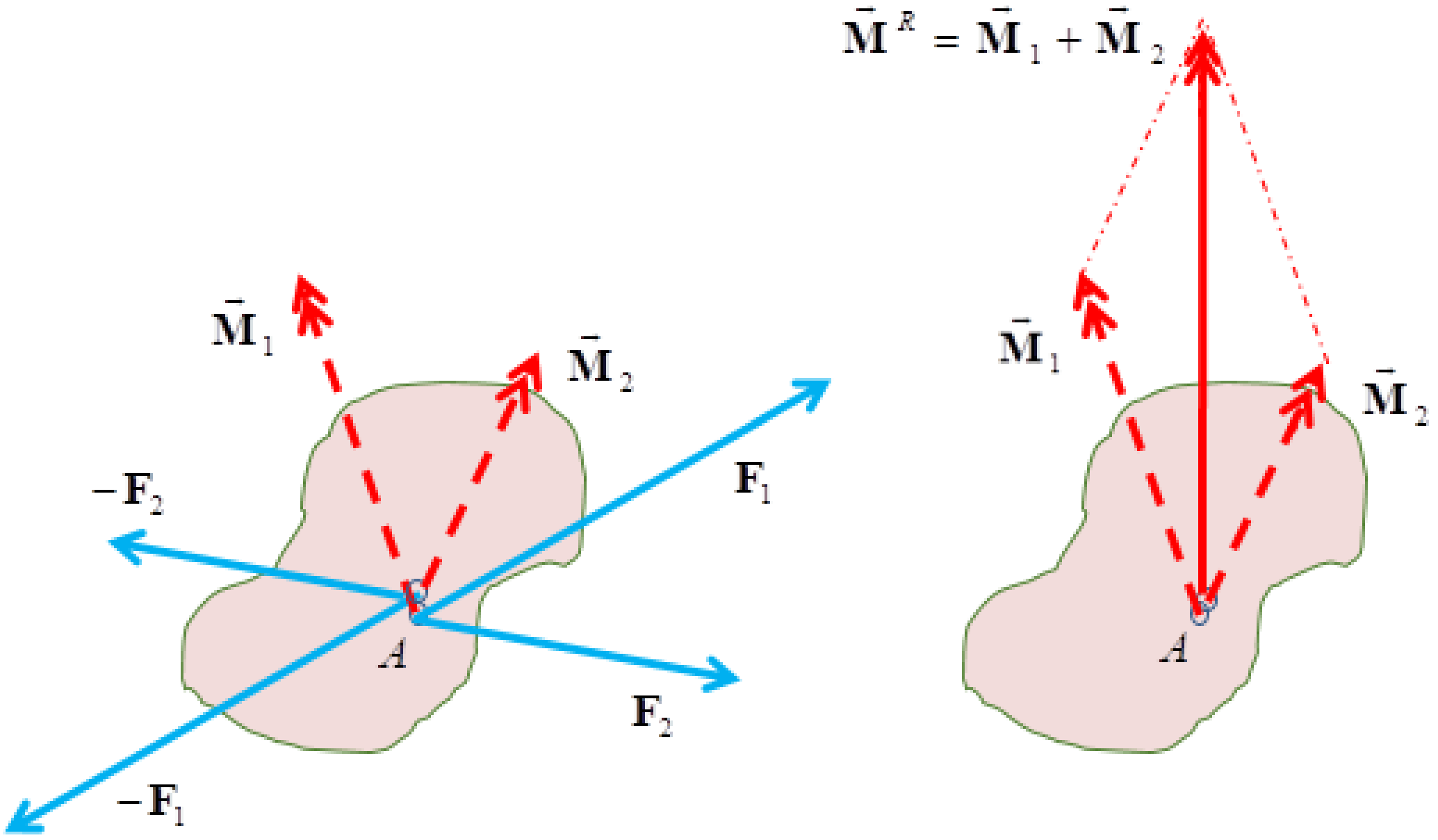

**Figure 7.** System of two concentrated couples with moments $\vec{\mathbf{M}}_1$ and $\vec{\mathbf{M}}_2$, and equipollent resultant couple with moment $\vec{\mathbf{M}}^R = \vec{\mathbf{M}}_1 + \vec{\mathbf{M}}_2$ at point *A*.

### 3.3. System of forces equipollent to a system of one force and one couple

The introduction of couple shows that a system of forces acting on a body is equipollent to a system of one resultant force $\mathbf{F}^R$ and one resultant couple with moment $\vec{\mathbf{M}}_O^R$ at point $O$ (Beer and Johnston, 1988), where

$$\mathbf{F}^R = \sum \mathbf{F} \tag{25}$$

$$\vec{\mathbf{M}}_O^R = \sum \vec{\mathbf{M}}_O = \sum \mathbf{r} \times \mathbf{F} \tag{26}$$

Figure 8 shows reduction of the system of forces $\mathbf{F}_1$, $\mathbf{F}_2$ and $\mathbf{F}_3$ acting on a body has been replaced with an equipollent system of one resultant force $\mathbf{F}^R$ and one resultant couple with moment $\vec{\mathbf{M}}_O^R$ at point $O$.

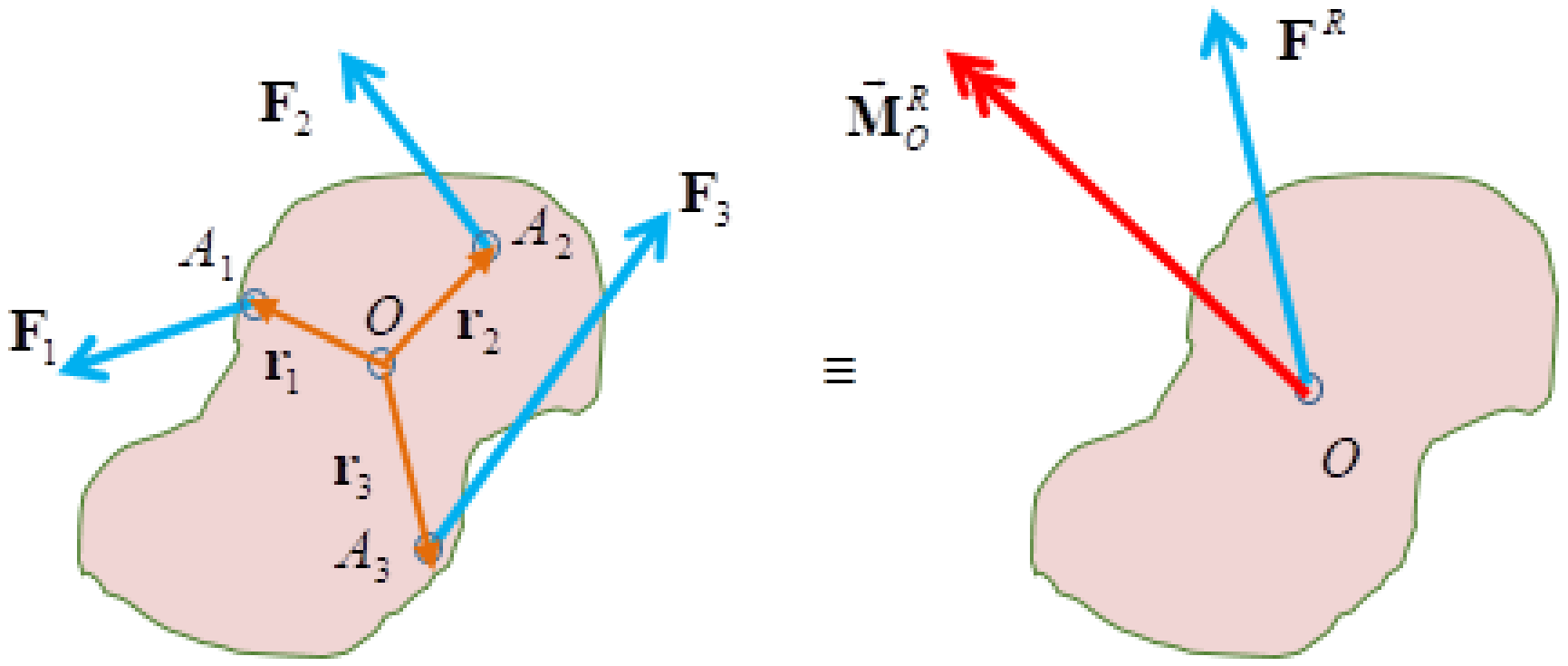

**Figure 8.** System of one resultant force $\mathbf{F}^R$ and one resultant couple with moment $\vec{\mathbf{M}}_O^R$ at point $O$ equipollent to a system of forces.

However, it is noticed that the effect of this new simple equipollent system is not generally equivalent to the original system of forces, because it creates a different state of stress and deformation. As discussed previously, equipollent systems have different effect on the internal interactions and deformation of the body.

Despite this character, the concept of equipollent system of one resultant force $\mathbf{F}^R$ and one resultant couple with moment $\vec{\mathbf{M}}_O^R$ still plays an important role in mechanics. For example, for investigating the condition of equilibrium or motion of rigid bodies, or determining the deformation of beams, plates and shells in structural mechanics and strength of material, the reduction of forces to an equipollent system of one force and one couple is very useful. This will be discussed in the following sections.

### 3.4. Equipollent system of forces in rigid body mechanics

The motion of a rigid body is specified by motion of one point describing its translation and rotation about this point. Interestingly, the force and moment governing equations

$$\mathbf{F}^R = \sum \mathbf{F} = \sum m\mathbf{a} \tag{27}$$

$$\vec{\mathbf{M}}_O^R = \sum \vec{\mathbf{M}}_O = \sum \mathbf{r} \times m\mathbf{a} \tag{28}$$

are enough to describe the motion of a rigid body (Goldstein, 1980; Beer and Johnston, 1988; Shames, 1980), where each scalar equation of (27) and (28) describes the motion corresponding to a degree of freedom of the rigid body. Consequently, equipollent system of external forces create the same condition of equilibrium or motion for a rigid body. This means although the internal forces might not remain the same, the motion of individual particles remain the same. Therefore, in rigid body mechanics, systems of equipollent forces are considered equivalent.

Note that the effective forces $\sum m\mathbf{a}$ and $\sum \mathbf{r} \times m\mathbf{a}$ are usually simplified for rigid body by introducing the center of mass and the moment of inertia tensor (Goldstein, 1980; Beer and Johnston, 1988; Shames, 1980).

Figure 9 shows reduction of the system of forces $\mathbf{F}_1$, $\mathbf{F}_2$ and $\mathbf{F}_3$ acting on a rigid body has been replaced with an equivalent (equipollent) system of one resultant force $\mathbf{F}^R$ and one resultant couple $\vec{\mathbf{M}}_O^R$ at point *O*, which in turn is equivalent to system of effective force $\sum m\mathbf{a}$ and couple with moment $\sum \mathbf{r} \times m\mathbf{a}$.

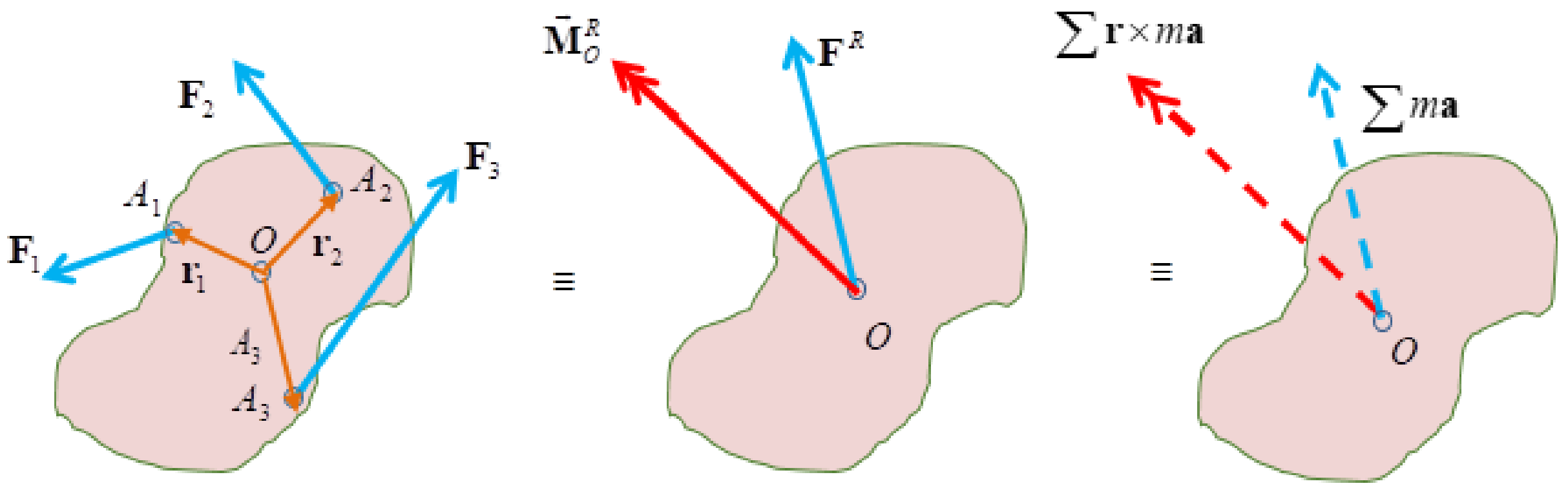

**Figure 9.** System of one resultant force $\mathbf{F}^R$ and one resultant couple with moment $\vec{\mathbf{M}}_O^R$ at point $O$ equivalent to system of external forces and effective forces.

Interestingly, in mechanics of rigid bodies very important results are obtained as follows.

### 3.4.1. Sliding force (Transmissibility principle)

It is noted that by sliding an external force vector $\mathbf{F}$ along its line of action the state of equilibrium or motion of a rigid body does not change. In other words, in rigid body mechanics the forces with the same magnitude and line of action are not only equipollent, but also equivalent. This is called the principle of transmissibility for rigid body, which states that the action of a force $\mathbf{F}$ on a rigid body may be transmitted along its line of action, as shown in Figure 10.

However, transmissibility principle is not valid when considering state of internal stresses and deformation, where forces are concentrated and cannot be transmitted on their line of action.

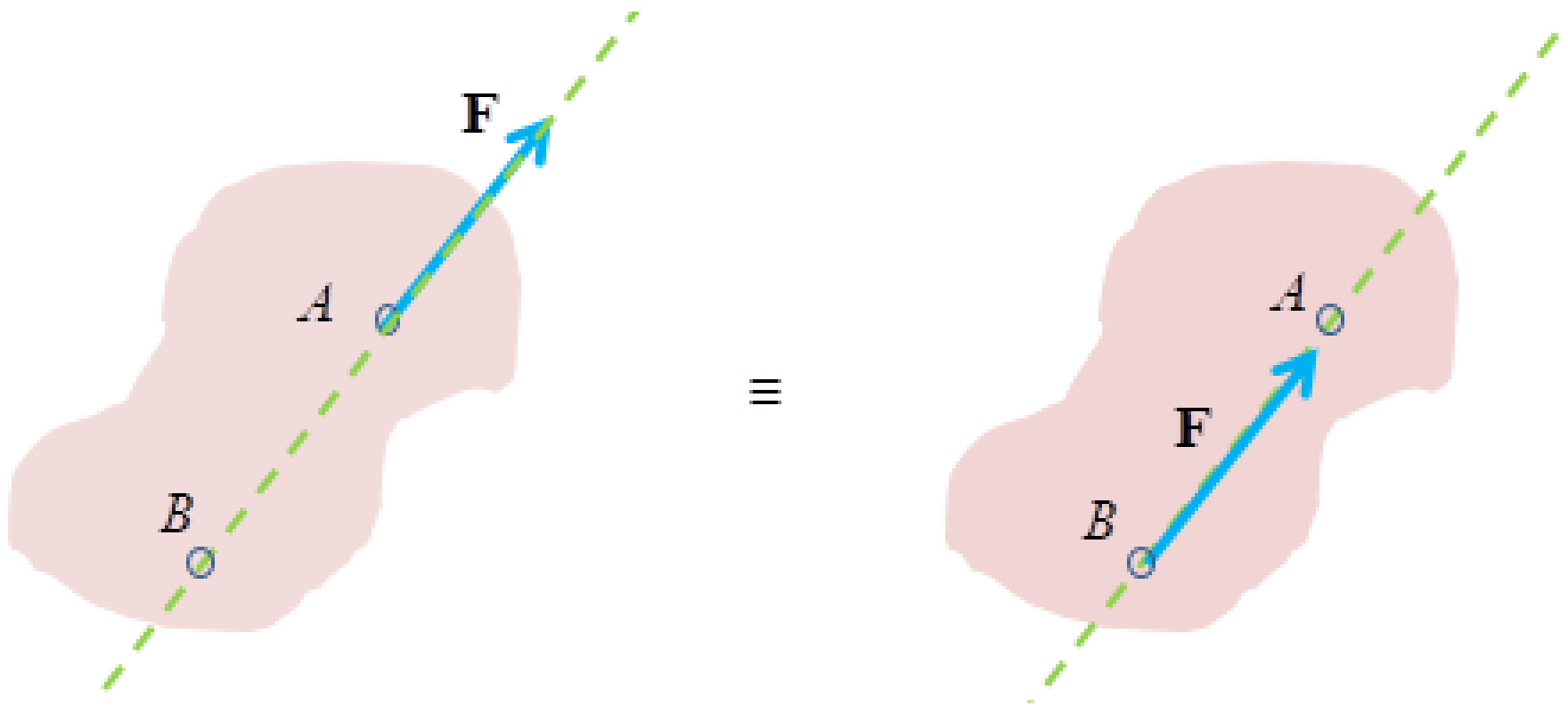

**Figure 10.** Transmissibility of force $\mathbf{F}$ on a rigid body.

### 3.4.2. Freedom of couple

Note that replacing a couple with an equipollent couple does not change the state of equilibrium or motion of a rigid body. This means in rigid body mechanics, equipollent couples are equivalent (Beer and Johnston, 1988; Shames, 1980). For example, the couple of forces $\mathbf{F}$ and $-\mathbf{F}$ can be replaced by the equipollent couple of forces $\mathbf{F}'$ and $-\mathbf{F}'$, as shown in Figure 11. As a result, in rigid body mechanics two couples are equivalent if they have the same moment $\vec{\mathbf{M}}$, no matter they act in the same plane or in parallel planes (Beer and Johnston, 1988; Shames, 1980).

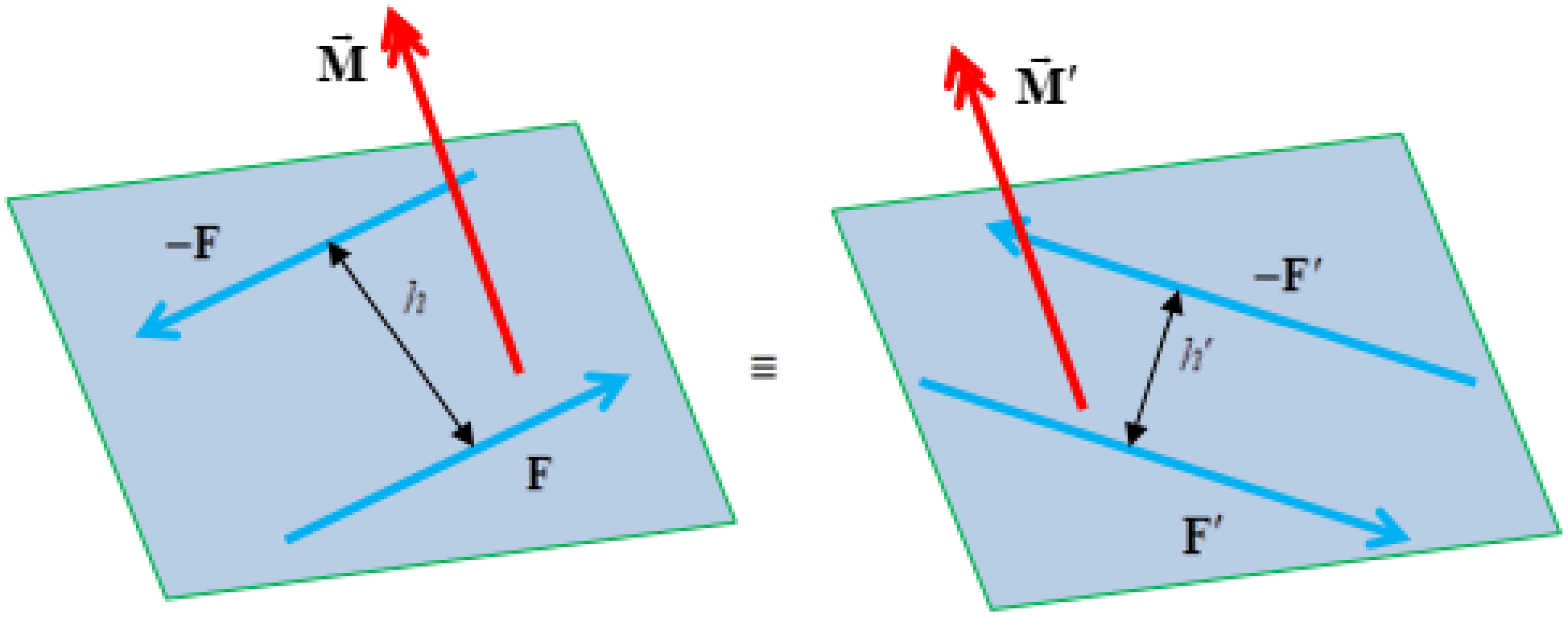

**Figure 11.** Equipollent couples on a rigid body are equivalent.

Also notice that the effect of two couples with moments $\vec{\mathbf{M}}_1$ and $\vec{\mathbf{M}}_2$ can be replaced by a couple with moment equal to the sum $\vec{\mathbf{M}}_1 + \vec{\mathbf{M}}_2$. This result suggests that couple can be completely represented by its moment in rigid body mechanics.

Accordingly, in rigid body mechanics the effect of couple of forces $\mathbf{F}$ and $-\mathbf{F}$ can be completely represented by its pseudo-vector moment $\vec{\mathbf{M}}$ applied to any arbitrary point. Therefore, the couple is usually denoted with its moment, and instead of the term “couple with moment $\vec{\mathbf{M}}$”. However, one can simply use “couple $\vec{\mathbf{M}}$”. Note that only for a rigid body, a couple can be represented by its free moment vector $\vec{\mathbf{M}}$ and attribute it to any arbitrary point without affecting its condition of equilibrium or motion, as shown in Figure 12.

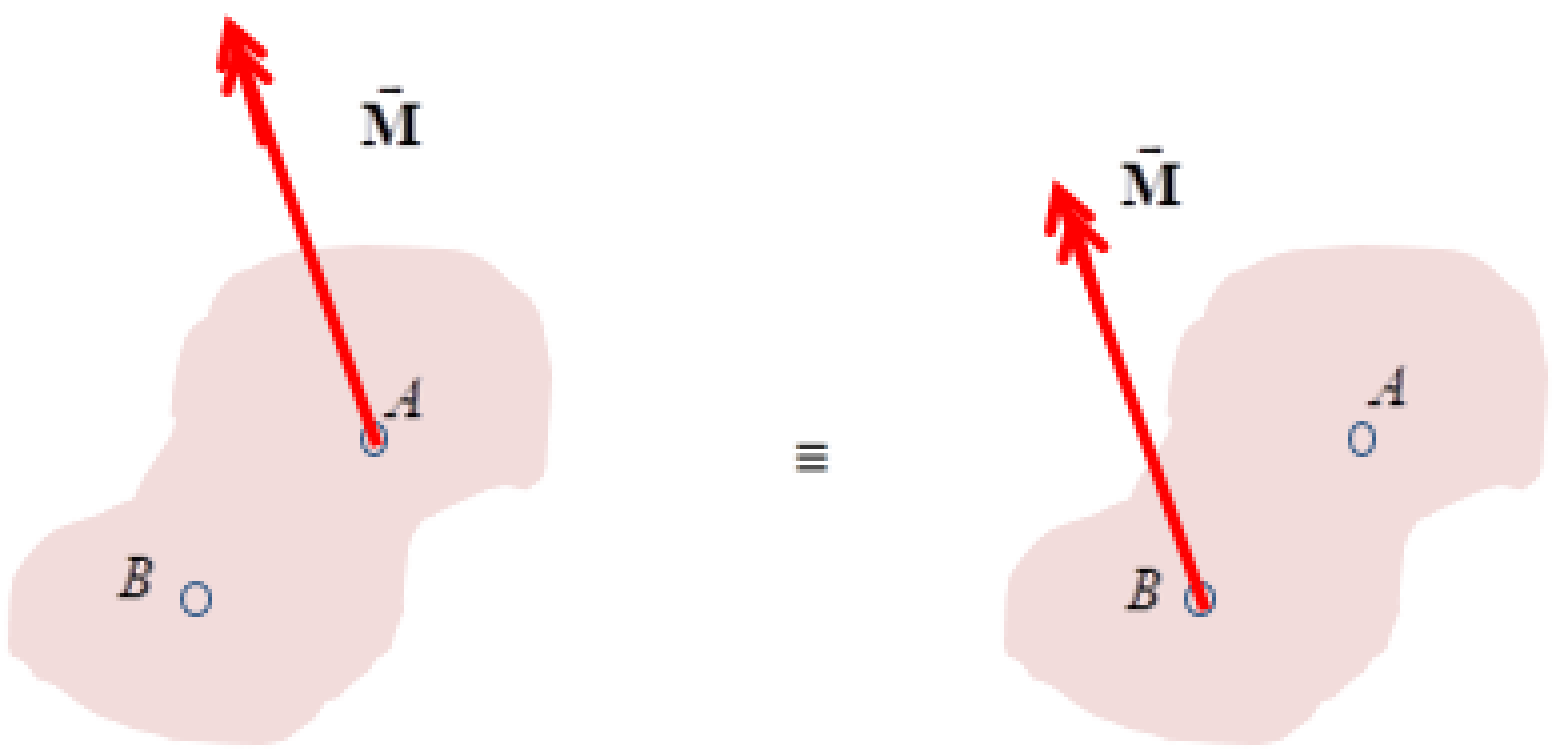

**Figure 12.** Freedom of a couple with moment $\vec{\mathbf{M}}$ on a rigid body.

Couple freedom is not valid when considering the state of internal stresses and deformation. In this case, a couple cannot be completely represented by its moment. Let it be a free vector.

### 3.5. Equipollent system of forces in continuum mechanics and strength of material

In continuum mechanics, where the state of internal forces and deformation are studied, the equipollent systems of forces are not equivalent. This means the system of forces $\mathbf{F}_1$, $\mathbf{F}_2$ and $\mathbf{F}_3$ acting on a deformable body, shown in Figure 9, cannot be replaced with an equipollent system of one resultant force $\mathbf{F}^R$ and one resultant couple $\vec{\mathbf{M}}_O^R$ at point *O*. Therefore, forces are not sliding (transmissible) vectors, and couples cannot be completely represented by their moments as free vectors. This is more essential for concentrated couples. There are infinite sets of concentrated couples at point *A* with the same moment $\vec{\mathbf{M}}$. Figure 13 shows two of these equipollent concentrated couples, one with concentrated forces $\mathbf{F}$ and $-\mathbf{F}$, and the other with concentrated forces $\mathbf{F}'$ and $-\mathbf{F}'$. Note that the deformation and state of stresses of these two equipollent concentrated couples are different.

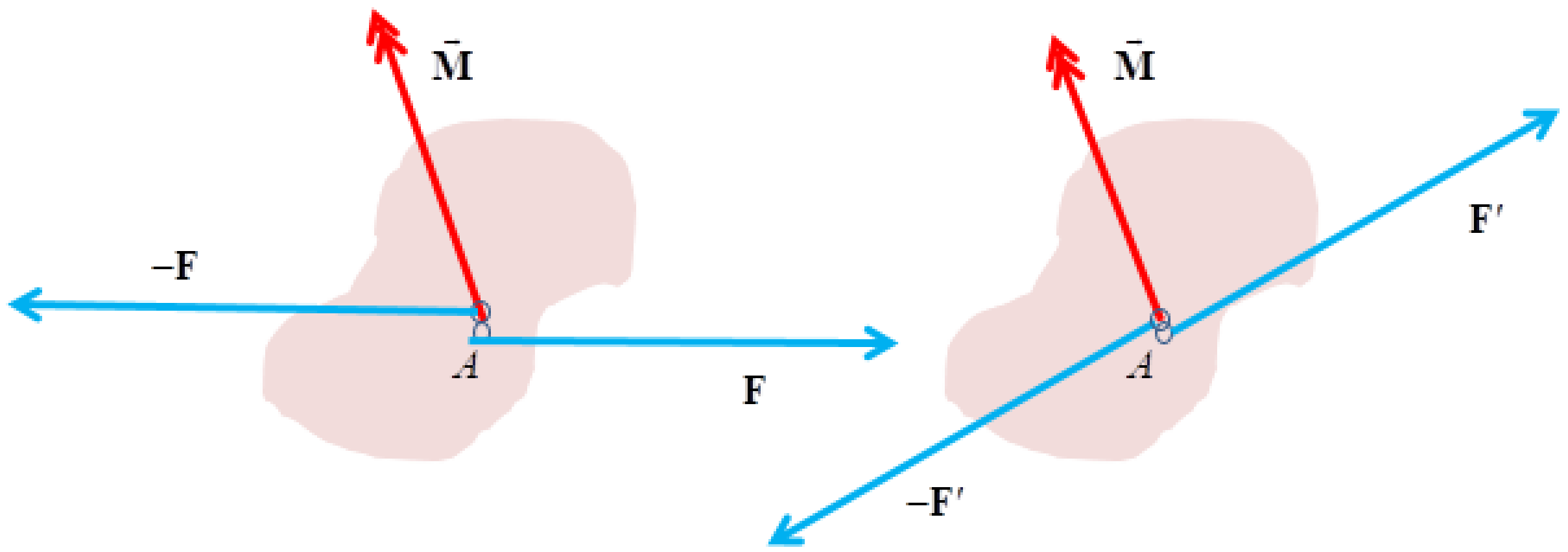

**Figure 13.** Equipollent concentrated couples at point $A$ are not equivalent in continuum mechanics.

Consequently, even a concentrated couple cannot be completely represented by its moment $\vec{\mathbf{M}}$ as a free vector. Unfortunately, representing a couple by its corresponding pseudo-vector moment carelessly has created some fundamental misunderstanding in the development of couple stress continuum mechanics. It seems this is the result of confusion of continuum mechanics with structural mechanics and strength of material, which is explained as follows.

Although equipollent systems of forces are not equivalent, it turns out that the state of stresses and deformation are approximately the same for some cases. This is the result of Saint-Venants's principle in elastostatics, which states that the deformation and state of stresses for static equipollent system of forces are approximately the same for parts of continuum far away from loading points. Saint-Venant's principle allows us to replace boundary loadings by an equipollent system of forces to find analytical solutions far away from loadings. This principle is generally used in the semi-inverse method to solve elasticity problems, such as extension, pure bending, torsion and flexure of elastic bars (Sadd, 2009). It turns out that the Saint-Venant's principle is an important part of the strength of materials, such as beam, plate and shell bending, where the loading on boundary cross sections is replaced by the equipollent loading that the far field deformation predicts. For example, the equipollent couples $(\mathbf{F}, -\mathbf{F})$ and $(\mathbf{F}', -\mathbf{F}')$ on the end cross sections of the beam create almost the same deformation and stresses far away from these ends. Therefore, their loadings are represented by their moment $\vec{\mathbf{M}}_B$ at the ends of the beam, as shown in Figure 14.

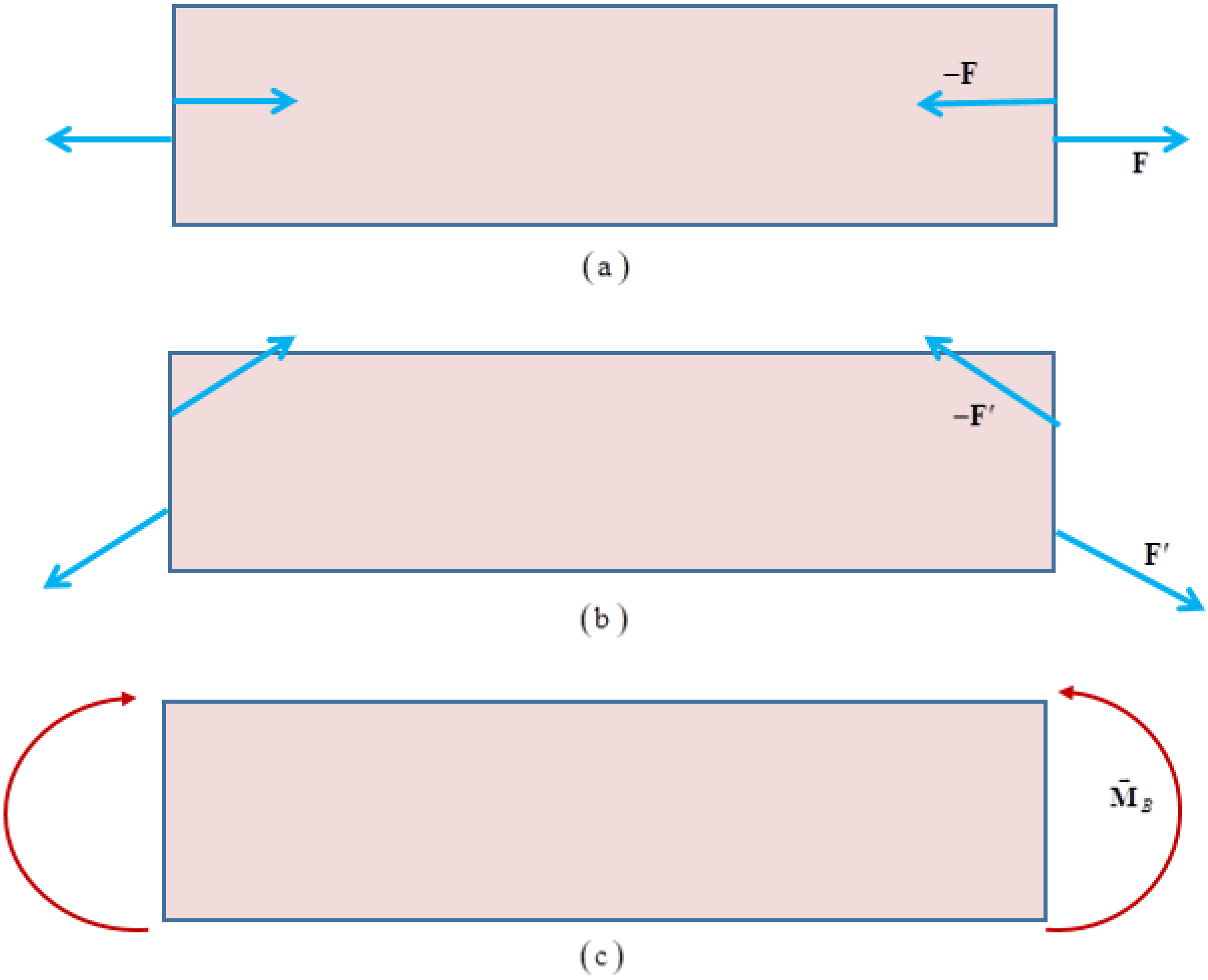

**Figure 14.** Equipollent couples (a) and (b) on the end cross sections are represented by their moment $\vec{\mathbf{M}}_B$ in (c).

In strength of materials, the couple $\vec{\mathbf{M}}_B$ is replaced with the triangular normal force-stress distribution predicted by beam theory, as shown in Figure 15. However, note that on and near the end cross sections the deformations and stresses are different, and beam theory solution is not accurate and can be even misleading.

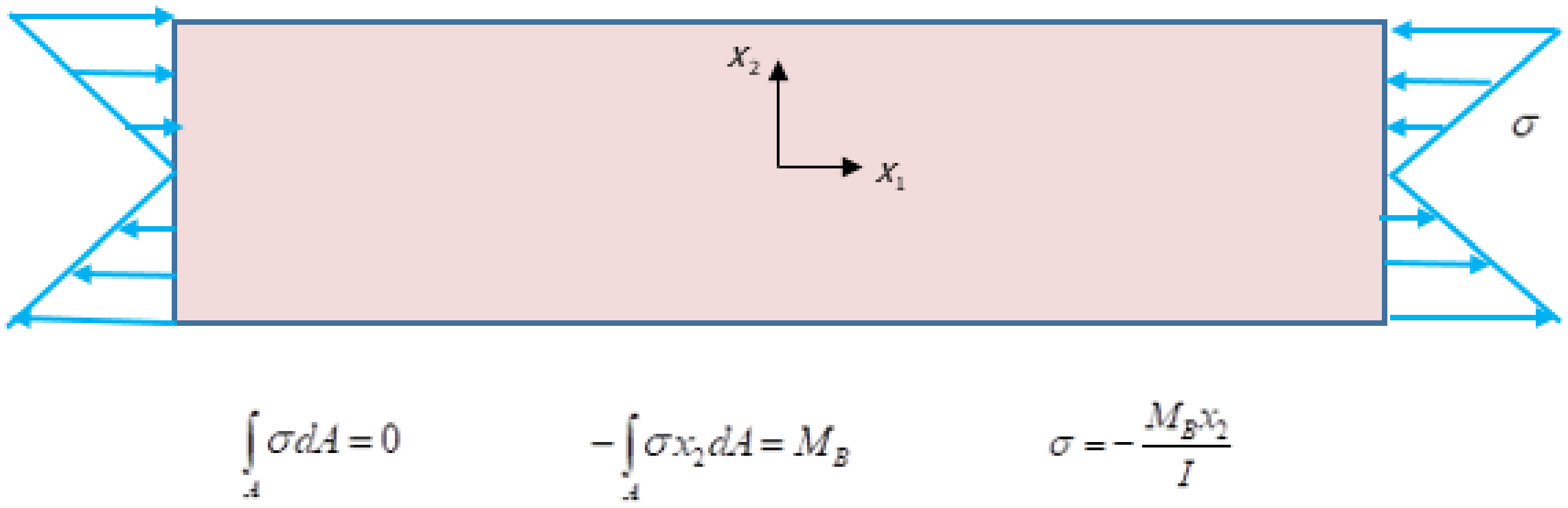

**Figure 15.** Moment $\vec{\mathbf{M}}_B$ on the end cross sections are replaced with equipollent triangular normal force-stress distribution predicted from beam theory.

Based on Saint-Venant's principle, the deflection and stresses from beam and plate theories are accurate enough far away from the loaded regions. However, at the vicinity of loading regions these structural mechanics results are not accurate and can be even misleading. This fundamental fact is what has been missed by Cosserat and Cosserat (1909), Mindlin and Tiersten (1962) and Koiter (1964) in their developments in couple stress continuum mechanics. Interestingly, the Cosserats came from structural mechanics to develop their formulation for the continuum.

## 4. Loading in continuum mechanics

Consider a continuous body occupying a volume $V$ bounded by a surface $S$, as shown in Figure 15. Also for future reference, an arbitrary subdomain volume $V_a$ having surface $S_a$ is considered, as shown in Figure 16.

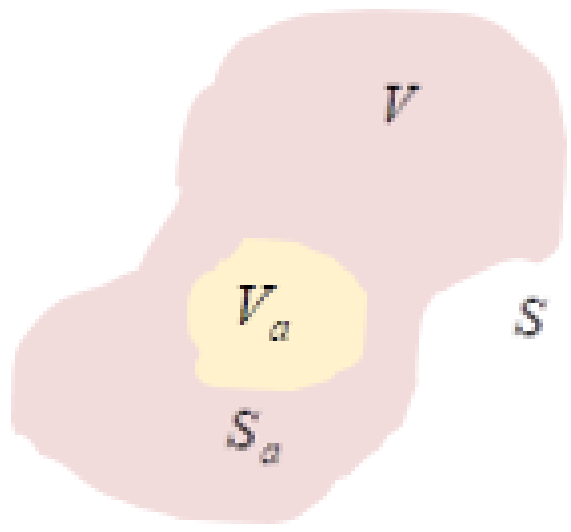

**Figure 16.** The body configuration.

It is assumed the body is under the influence of external loading, such as body and surface loading. These loadings may be concentrated force and couple or distributed force and couple in the volume or on the surface.

### 4.1. Distributed loads

In continuum mechanics, it is postulated that the distributed quantities, such a body force, surface force, body couple, and surface couple are piecewise continuous, if they exist.

#### 4.1.1. Body force and body couple

Body loads are those external distributed forces and couples that act on the elements of volume or mass in the body. Therefore, the system of forces exerting to a volume element $dV$ at a point $P$ is equipollent (not equivalent) to the general force-couple system with force $d\mathbf{F} = \mathbf{f}dV$ and couple with moment $d\vec{\mathbf{M}} = \vec{\mathbf{c}}dV$, as shown in Figure 17. Here, $\mathbf{f}$ and $\vec{\mathbf{c}}$ are the body force and moment of body couple, respectively, which are considered piecewise continuous in $V$.

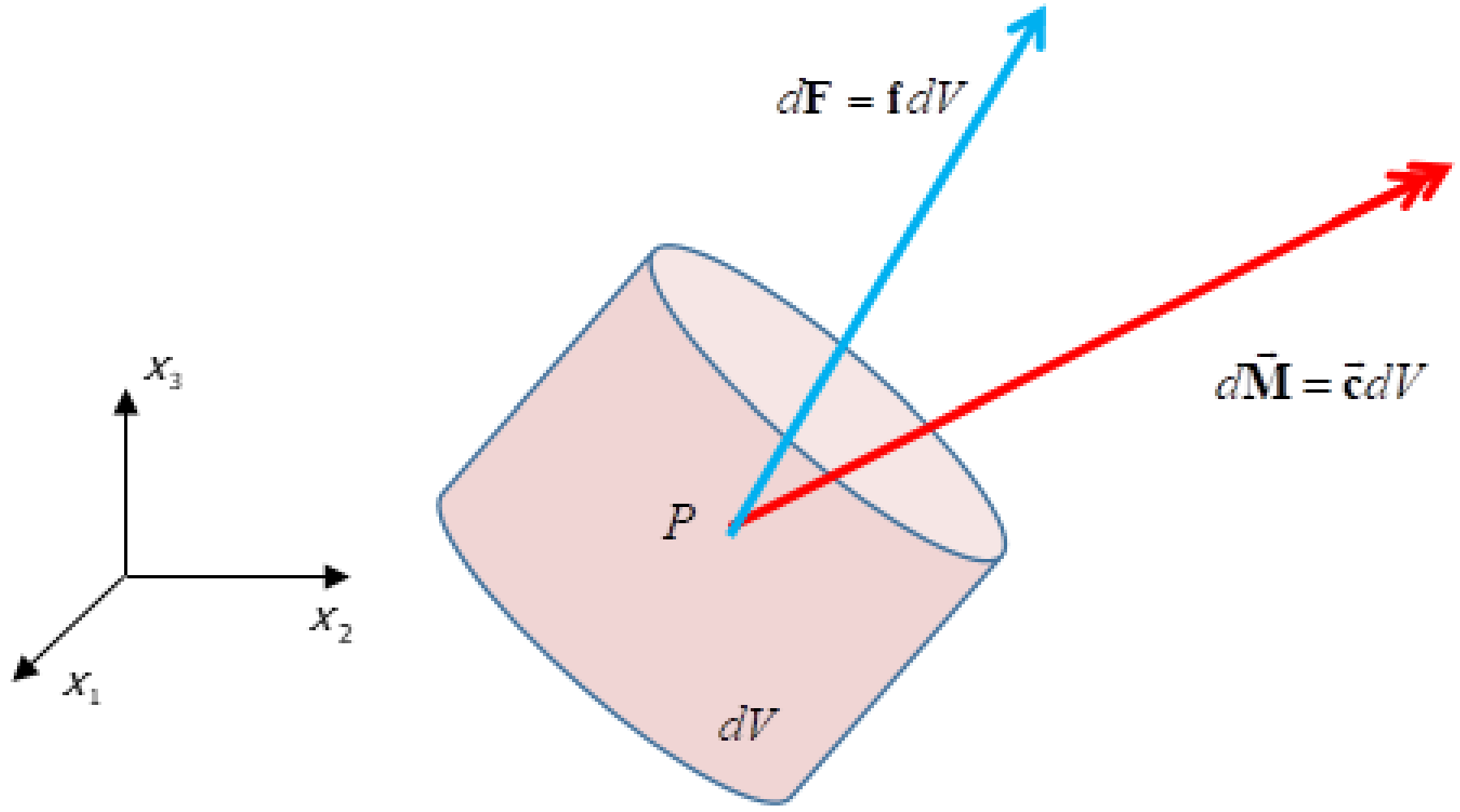

**Figure 17.** Force $d\mathbf{F} = \mathbf{f}dV$ and couple with moment $d\vec{\mathbf{M}} = \vec{\mathbf{c}}dV$ in the volume element $dV$.

However, note that the pseudo-vector moment $d\vec{\mathbf{M}} = \vec{\mathbf{c}}dV$ does not completely describe the effect of couple in the volume element $dV$. It is also necessary to specify the line of action of forces creating the couple. This obviously imposes some restriction on the form of body couple, which will be examined in more details later.

### 4.1.2. Force- and couple-tractions

Surface loads are those forces and couples acting on the elements of the surface of the body. Interestingly, the state of interactions inside the body among elements of matter can also be represented with surface forces and couples by using the method of section.

Let us consider the state of interactions inside a continuous body. For this the interactions on a surface element $dS$ with unit normal vector $\mathbf{n}$ at point $P$ on an arbitrary surface $S_a$ in the body is considered as follows (Figure 16). It is generally assumed that the system of forces through this surface element $dS$ is equipollent (not equivalent) to the force-couple system with force $d\mathbf{F} = \mathbf{t}^{(n)} dS$ and couple moment $d\vec{\mathbf{M}} = \vec{\mathbf{m}}^{(n)} dS$, as shown in Figure 18. Here, $\mathbf{t}^{(n)}$ and $\vec{\mathbf{m}}^{(n)}$ are the force-traction and moment of couple-traction, respectively, which are considered piecewise continuous on $S_a$.

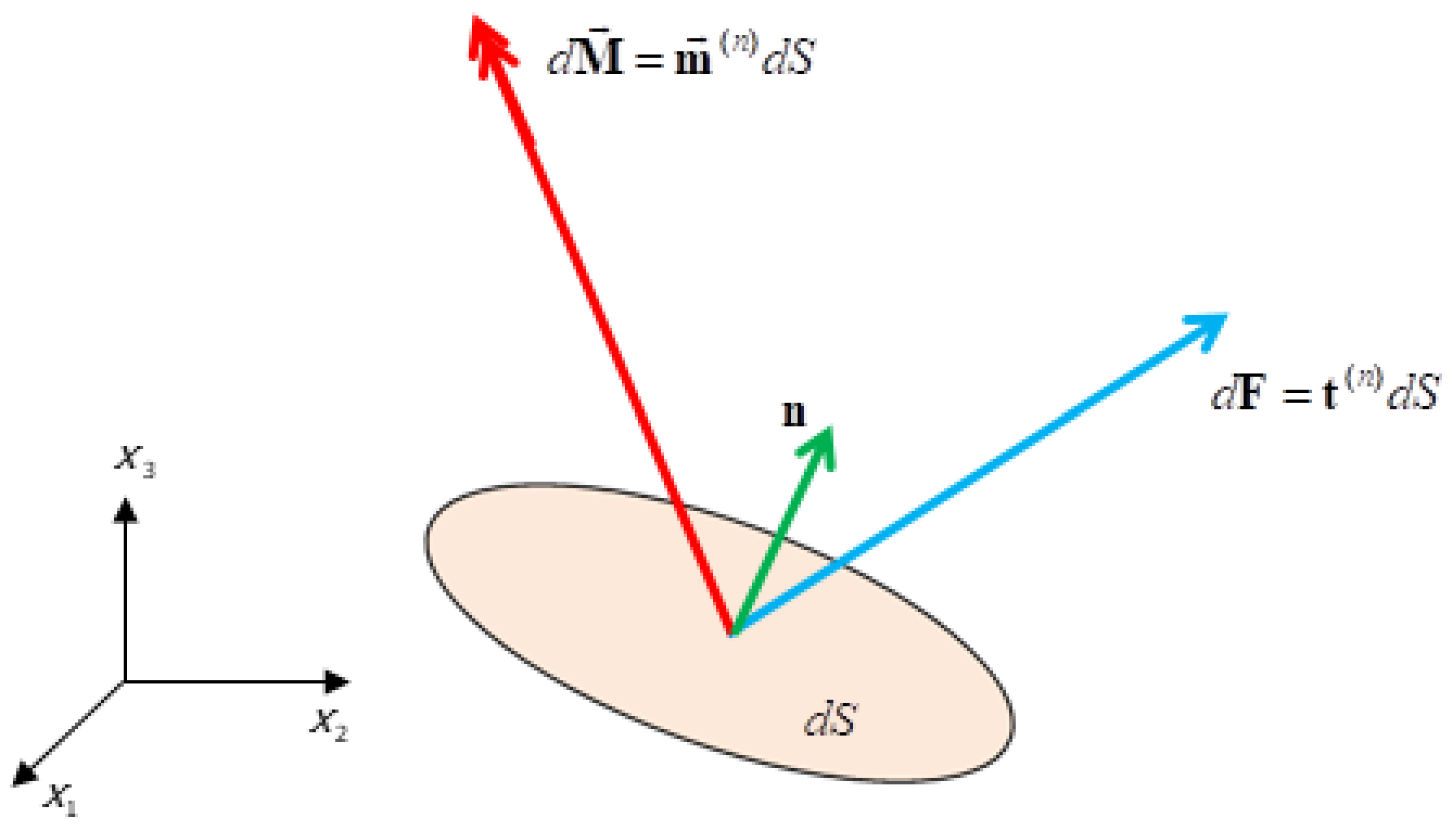

**Figure 18.** Force $d\mathbf{F} = \mathbf{t}^{(n)} dS$ and couple with moment $d\vec{\mathbf{M}} = \vec{\mathbf{m}}^{(n)} dS$ on the surface element $dS$.

Figure 19 shows the force-traction vector $\mathbf{t}^{(n)}$ and couple-traction moment pseudo vector $\vec{\mathbf{m}}^{(n)}$, which are considered continuous on the arbitrary surface $S_a$.

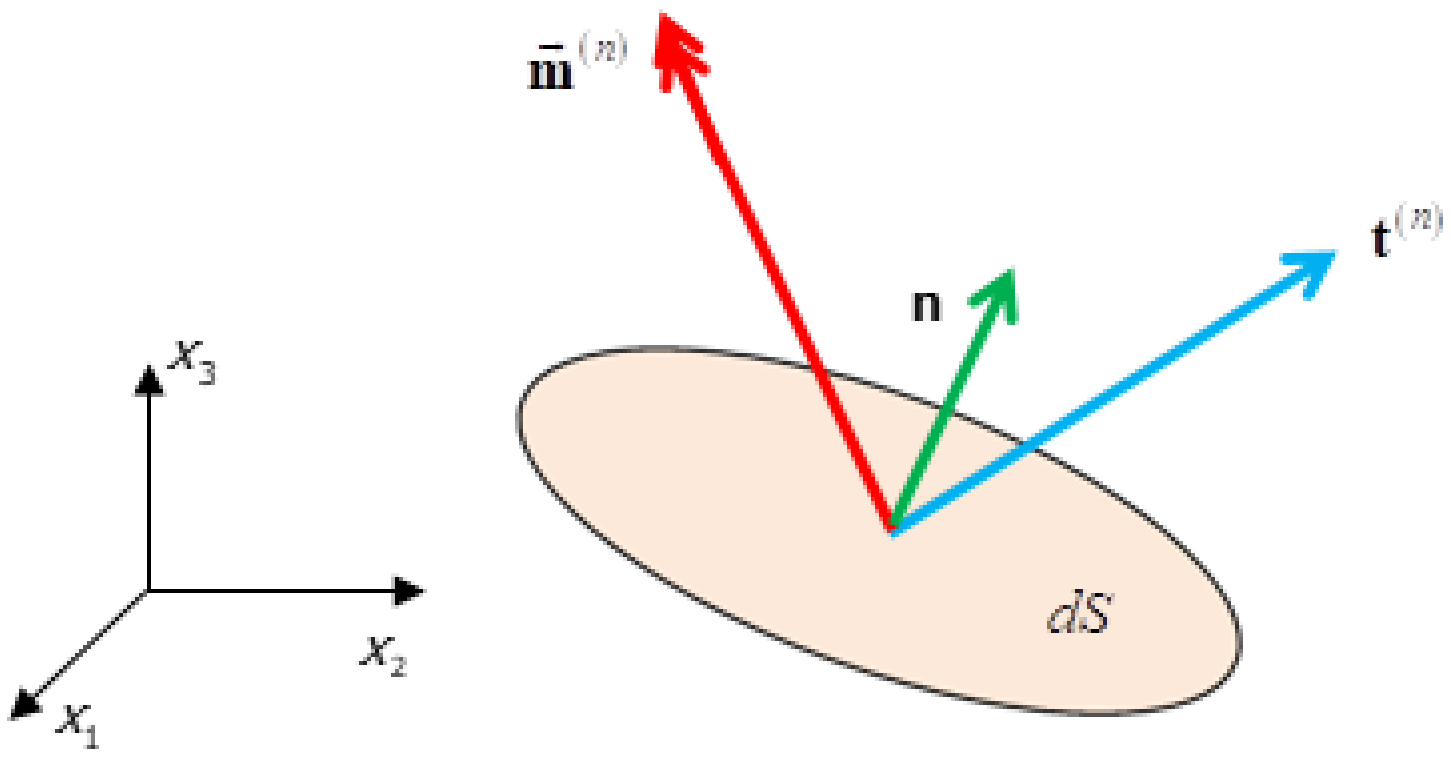

**Figure 19.** Force-traction $\mathbf{t}^{(n)}$ and couple-traction moment $\vec{\mathbf{m}}^{(n)}$ on the surface element $dS$ .

However, the couple with pseudo-vector moment $d\vec{\mathbf{M}} = \vec{\mathbf{m}}^{(n)} dS$ does not completely describe the effect of this surface couple on the surface element $dS$ . Obviously, this character imposes some restrictions on the form of couple-traction, which will be also discussed in more details later.

The tractions $\mathbf{t}^{(n)}$ and $\vec{\mathbf{m}}^{(n)}$ can be decomposed into their normal and tangential components as

$$\mathbf{t}^{(n)} = \mathbf{t}^{(nn)} + \mathbf{t}^{(nt)} \qquad t_i^{(n)} = t_i^{(nn)} + t_i^{(nt)} \tag{29}$$

$$\vec{\mathbf{m}}^{(n)} = \vec{\mathbf{m}}^{(nn)} + \vec{\mathbf{m}}^{(nt)} \qquad m_i^{(n)} = m_i^{(nn)} + m_i^{(nt)} \tag{30}$$

Note that the normal force-traction $\mathbf{t}^{(nn)}$ pulls or pushes material, whereas the tangential transverse force-traction component $\mathbf{t}^{(nt)}$ creates shear, as shown in Figure 20.

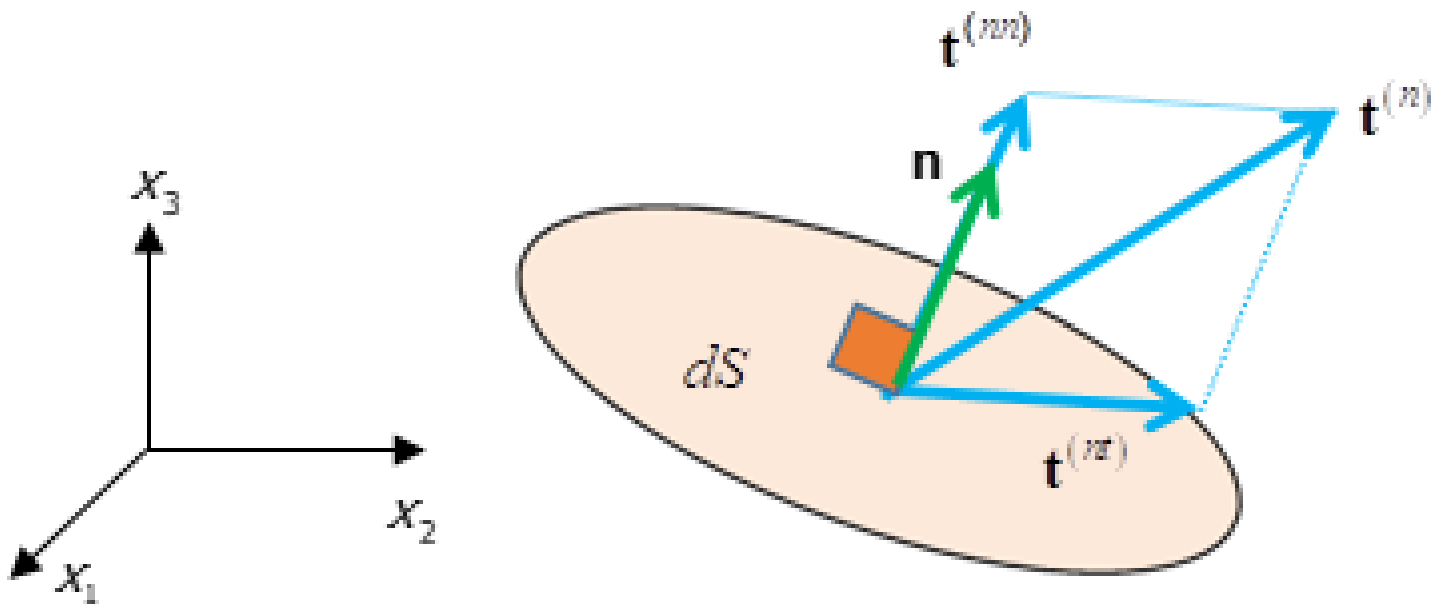

**Figure 20.** Normal and tangential components of surface force-traction $\mathbf{t}^{(n)}$ .

Note that the normal moment of couple-traction $\vec{\mathbf{m}}^{(nn)}$ causes torsion or twisting, whereas the tangential moment $\vec{\mathbf{m}}^{(nt)}$ of couple-traction creates bending, as shown in Figure 21.

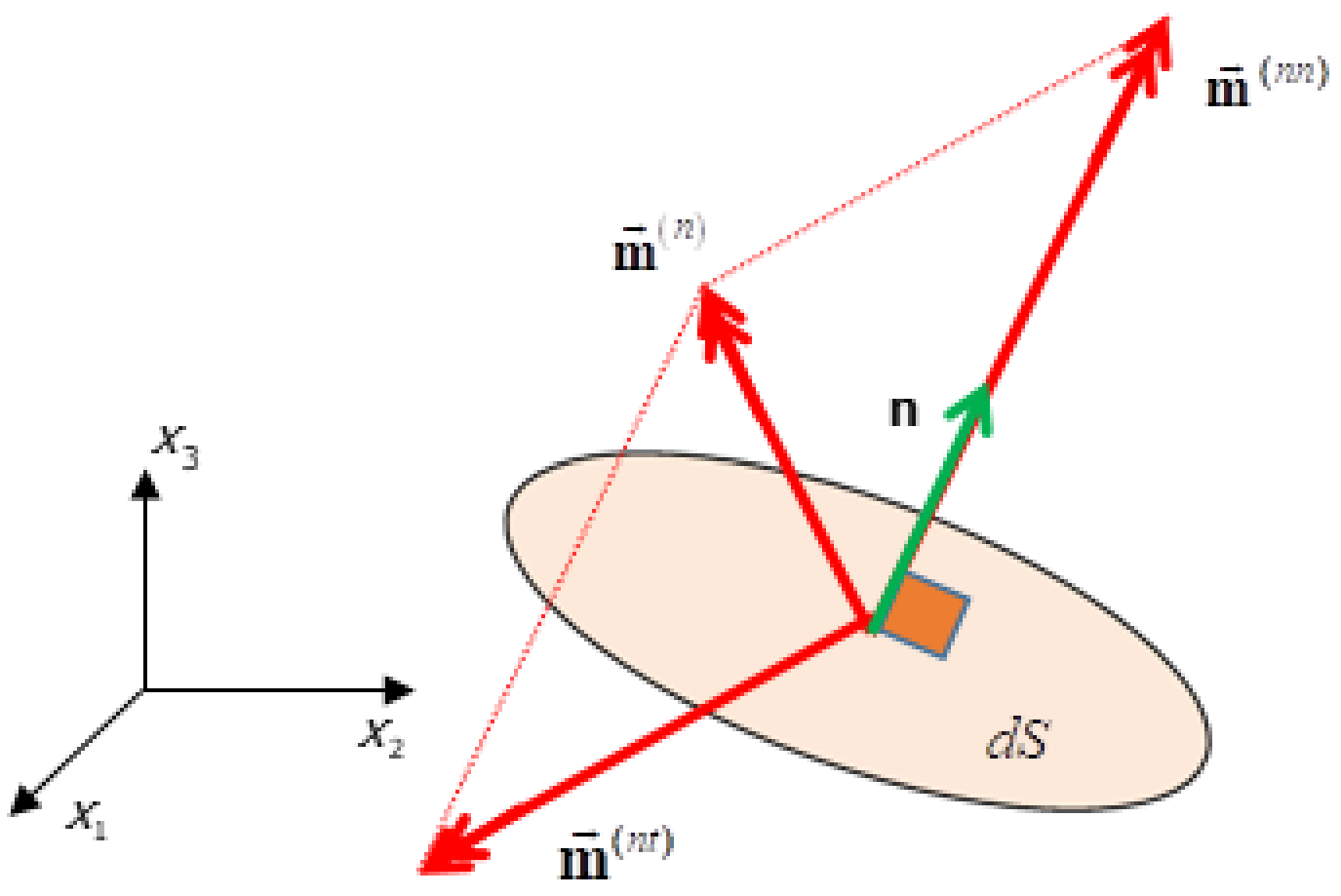

**Figure 21.** Normal and tangential components of surface couple-traction moment $\vec{\mathbf{m}}^{(n)}$.

The magnitude of normal components are

$$t^{(nn)} = \mathbf{t}^{(n)} \bullet \mathbf{n} = t_i^{(n)} n_i \tag{31}$$

$$m^{(nn)} = \vec{\mathbf{m}}^{(n)} \bullet \mathbf{n} = m_i^{(n)} n_i \tag{32}$$

Therefore, these components are

$$\mathbf{t}^{(nn)} = \left(\mathbf{t}^{(n)} \bullet \mathbf{n}\right)\mathbf{n} \qquad t_i^{(nn)} = t_j^{(n)} n_j n_i \tag{33}$$

$$\begin{aligned} \mathbf{t}^{(nt)} &= \mathbf{t}^{(n)} - \left(\mathbf{t}^{(n)} \bullet \mathbf{n}\right)\mathbf{n} & t_i^{(nt)} &= t_i^{(n)} - t_j^{(n)} n_j n_i \\ &= \left(1 - \mathbf{n} \otimes \mathbf{n}\right) \bullet \mathbf{t}^{(n)} & &= \left(\delta_{ij} - n_i n_j\right) t_j^{(n)} \end{aligned} \tag{34}$$

$$\vec{\mathbf{m}}^{(nn)} = \left(\vec{\mathbf{m}}^{(n)} \bullet \mathbf{n}\right)\mathbf{n} \qquad m_i^{(nn)} = m_j^{(n)} n_j n_i \tag{35}$$

$$\begin{aligned} \vec{\mathbf{m}}^{(nt)} &= \vec{\mathbf{m}}^{(n)} - \left(\vec{\mathbf{m}}^{(n)} \bullet \mathbf{n}\right)\mathbf{n} & m_i^{(nt)} &= m_i^{(n)} - m_j^{(n)} n_j n_i \\ &= \left(1 - \mathbf{n} \otimes \mathbf{n}\right) \bullet \vec{\mathbf{m}}^{(n)} & &= \left(\delta_{ij} - n_i n_j\right) m_j^{(n)} \end{aligned} \tag{36}$$

It seems instructive if the distribution of the tangential shear force-traction vector $\mathbf{t}^{(nt)}$ is considered as a single layer of shear-force traction. This is the vectorial analogy to a single layer of electric charge in electrostatic. This analogy is helpful in revealing the character of the bending couple-traction with moment $\vec{\mathbf{m}}^{(nt)}$, as will be seen later.

## 5. Fundamental governing equations for continuum

The fundamental governing equations in continuum mechanics are based on the force and moment equations (16) and (17) for system of particles. Therefore, the force and moment equations for the arbitrary part of the material continuum occupying the volume $V_a$ enclosed by the boundary surface $S_a$, shown in Figure 16, are

$$\int_{S_a} \mathbf{t}^{(n)} dS + \int_{V_a} \mathbf{f} dV = \int_{V_a} \rho \mathbf{a} dV \tag{37}$$

$$\int_{S_a} \left[ \mathbf{r} \times \mathbf{t}^{(n)} + \vec{\mathbf{m}}^{(n)} \right] dS + \int_{V_a} \left( \mathbf{r} \times \mathbf{f} + \vec{\mathbf{c}} \right) dV = \int_{V_a} \mathbf{r} \times \rho \mathbf{a} dV \tag{38}$$

where $\rho$ is the mass density. In terms of components, these equations become

$$\int_{S_a} t_i^{(n)} dS + \int_{V_a} f_i dV = \int_{V_a} \rho a_i dV \tag{39}$$

$$\int_{S_a} [\varepsilon_{ijk} x_j t_k^{(n)} + m_i^{(n)}]\, dS + \int_{V_a} \left( \varepsilon_{ijk} x_j f_k + c_i \right) dV = \int_{V_a} \varepsilon_{ijk} x_j \rho a_k dV \tag{40}$$

Note that these equations are the integral form of governing equations of motion in continuum mechanics. To obtain the differential form of the governing equations, force-stress tensor and couple-stress moment tensor are introduced as follows.

### 5.1. Force- and couple- stress tensors

The state of internal stress at point $P$ is known, if the force-traction vector $\mathbf{t}^{(n)}$ and couple-traction with moment $\vec{\mathbf{m}}^{(n)}$ on arbitrary surfaces at that point are known. This requires knowledge of only the force-traction and couple-traction on three mutually independent planes passing the point. When these planes are taken perpendicular to coordinate axes, the force-traction vectors are $\mathbf{t}^{(1)}$, $\mathbf{t}^{(2)}$ and $\mathbf{t}^{(3)}$, and the couple-traction moment vectors are $\vec{\mathbf{m}}^{(1)}$, $\vec{\mathbf{m}}^{(2)}$, and $\vec{\mathbf{m}}^{(3)}$ as shown in Figure 22.

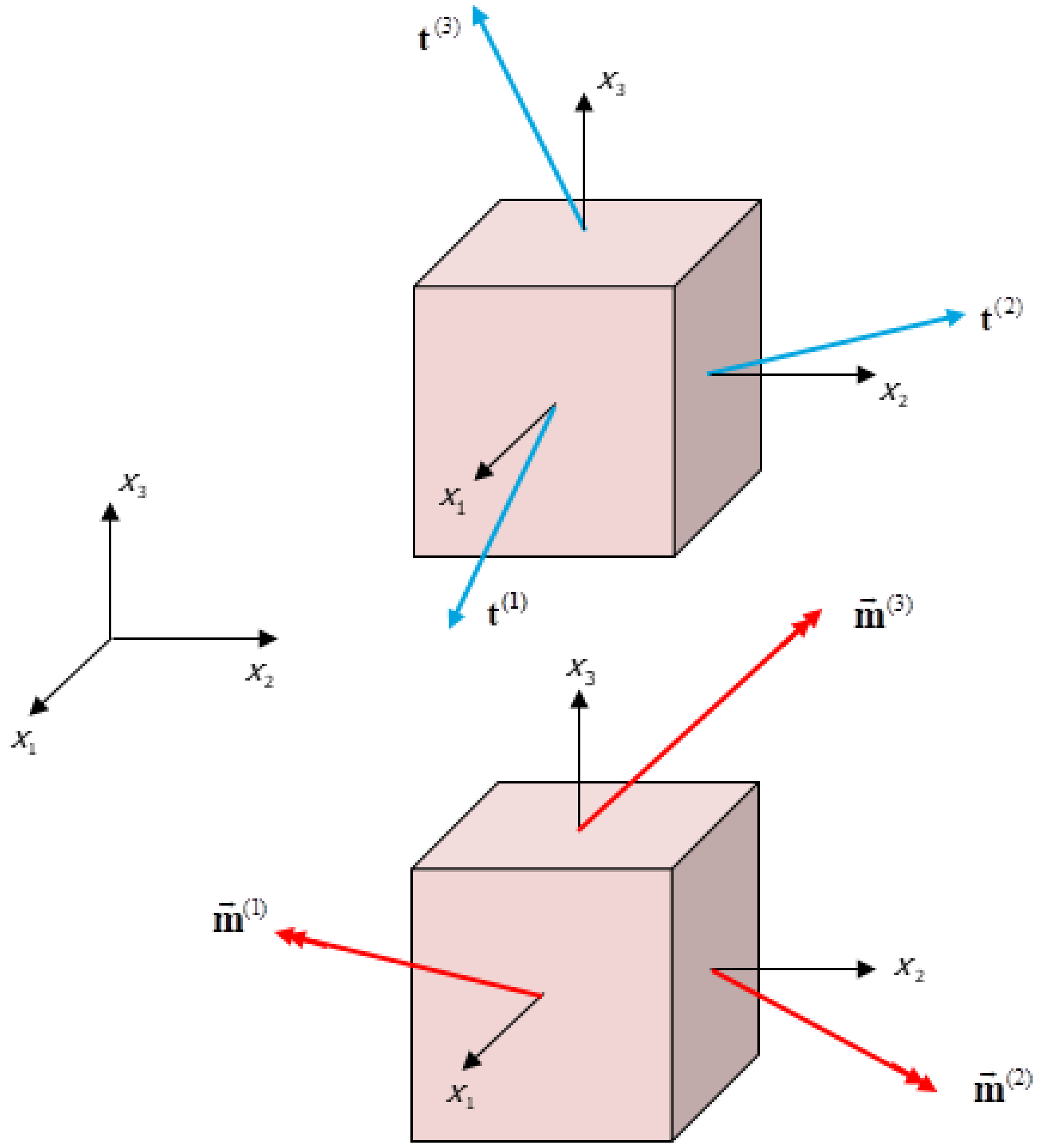

**Figure 22.** Force- and couple-traction vectors on planes perpendicular to coordinate axes.

Consequently, the internal stresses are represented by the second order force-stress tensor $\sigma_{ij}$ and couple-stress system with second order moment tensor $\mu_{ij}$, where the following relations hold:

$$\sigma_{ij} = t_j^{(i)} \tag{41}$$

$$\mu_{ij} = m_j^{(i)} \tag{42}$$

Note that the tensors $\sigma_{ij}$ and $\mu_{ij}$ each can have at most nine components, respectively. In terms of components, these tensors are represented as

$$\left[\sigma_{ij}\right]=\begin{bmatrix}\sigma_{11} & \sigma_{12} & \sigma_{13}\\ \sigma_{21} & \sigma_{22} & \sigma_{23}\\ \sigma_{31} & \sigma_{32} & \sigma_{33}\end{bmatrix}=\begin{bmatrix}t_1^{(1)} & t_2^{(1)} & t_3^{(1)}\\ t_1^{(2)} & t_2^{(2)} & t_3^{(2)}\\ t_1^{(3)} & t_2^{(3)} & t_3^{(3)}\end{bmatrix} \tag{43}$$

$$\left[\mu_{ij}\right]=\begin{bmatrix}\mu_{11} & \mu_{12} & \mu_{13}\\ \mu_{21} & \mu_{22} & \mu_{23}\\ \mu_{31} & \mu_{32} & \mu_{33}\end{bmatrix}=\begin{bmatrix}m_1^{(1)} & m_2^{(1)} & m_3^{(1)}\\ m_1^{(2)} & m_2^{(2)} & m_3^{(2)}\\ m_1^{(3)} & m_2^{(3)} & m_3^{(3)}\end{bmatrix} \tag{44}$$

The components of the force-stress $\sigma_{ij}$ and couple-stress moment $\mu_{ij}$ tensors are shown in Figure 23.

The force-traction vector $\mathbf{t}^{(n)}$ and couple-traction moment vector $\vec{\mathbf{m}}^{(n)}$ at point $P$ on a surface element $dS$ with unit normal vector $\mathbf{n}$ are obtained by applying the governing equations (39) and (40) for a Cauchy tetrahedron (Malvern, 1969) at given point $P$ as

$$\mathbf{t}^{(n)}=\mathbf{n}\bullet\boldsymbol{\sigma} \qquad t_i^{(n)}=\sigma_{ji}n_j \tag{45}$$

$$\vec{\mathbf{m}}^{(n)}=\mathbf{n}\bullet\boldsymbol{\mu} \qquad m_i^{(n)}=\mu_{ji}n_j \tag{46}$$

It is important to note that the couple-stress moment pseudo-tensor $\boldsymbol{\mu}$ cannot completely describe the effect of couple-stresses, because the moment pseudo-vectors $\vec{\mathbf{m}}^{(1)}$, $\vec{\mathbf{m}}^{(2)}$, and $\vec{\mathbf{m}}^{(3)}$ do not completely describe their corresponding couple-tractions. It should be noted that this is the fundamental source of the indeterminacy in Mindlin, Tiersten and Koiter couple-stress theory. Obviously, this character imposes some restrictions on the form of couple-stresses.

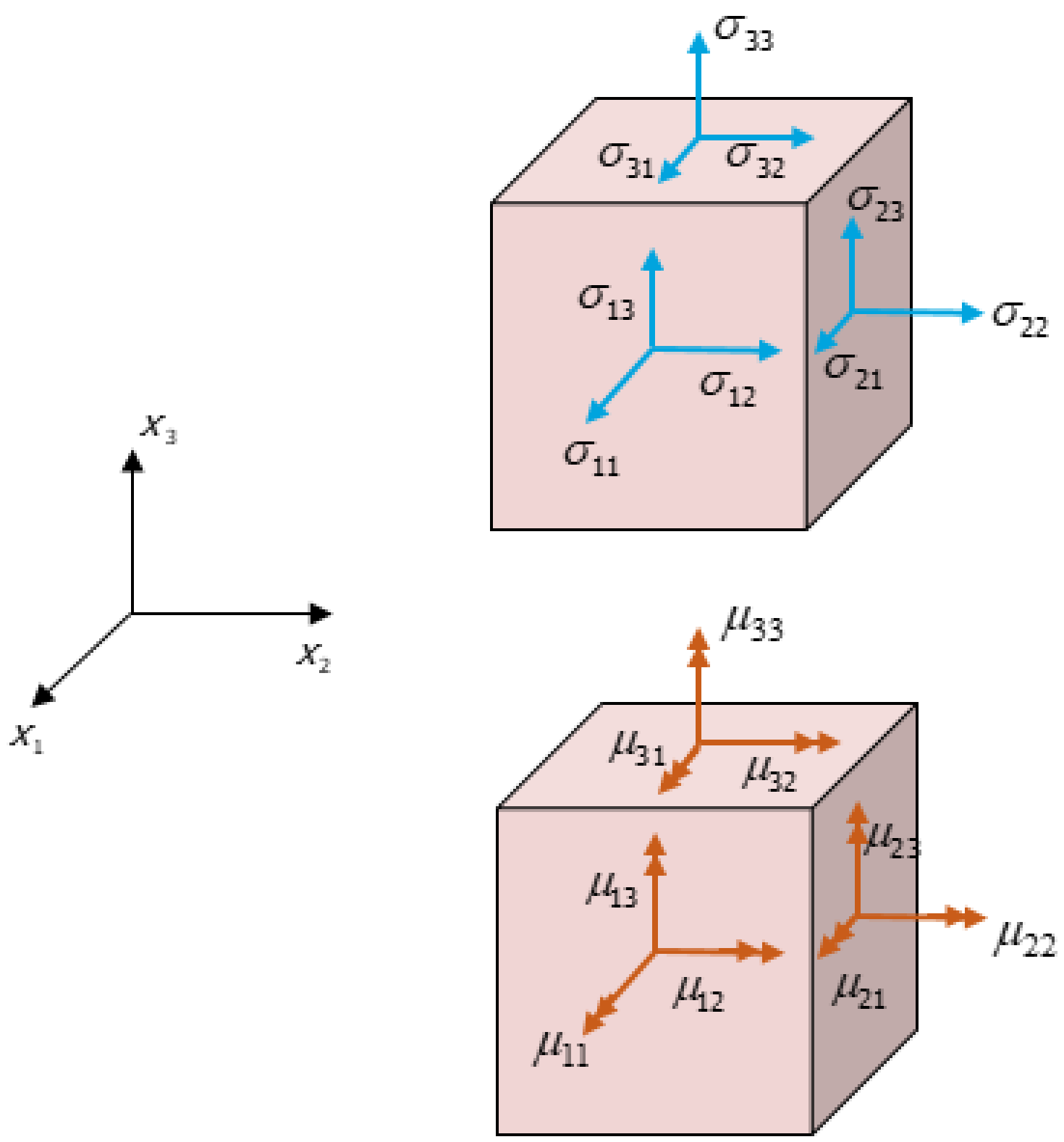

**Figure 23.** Components of force-stress and indeterminate couple-stress moment tensors.

### 5.2. Fundamental governing equations of motion in differential form

The differential form of the governing equations of motion for an infinitesimal element of matter are obtained by using the relations (45) and (46) for tractions in (39) and (40), along with the divergence theorem, and noticing the arbitrariness of volume $V_a$ as

$$\nabla \bullet \boldsymbol{\sigma} + \mathbf{f} = \rho \mathbf{a} \qquad \sigma_{ji,j} + f_i = \rho a_i \tag{47}$$

$$\nabla \bullet \boldsymbol{\mu} + \boldsymbol{\varepsilon} : \boldsymbol{\sigma} + \vec{\mathbf{c}} = 0 \qquad \mu_{ji,j} + \varepsilon_{ijk}\sigma_{jk} + c_i = 0 \tag{48}$$

These equations were derived by Cosserat and Cosserat (1909), Mindlin and Tiersten (1962) and Koiter (1964). However, since they confused the body couple and couple-stresses by their moments $\vec{\mathbf{c}}$ and $\boldsymbol{\mu}$, they did not realize the possible restrictions on the form of body couple and couple-stress distributions. As a result, they end up with the ill-posed indeterminate couple stress theory.

Note that the moment governing equation (48) and the traction relation (46) involve only with body couple moment density $\vec{\mathbf{c}}$, the couple-traction moment $\vec{\mathbf{m}}^{(n)}$, and couple-stress moment tensor $\boldsymbol{\mu}$. This means the line of action of opposite parallel couple forces does not affect these equations. However, the general moment pseudo-vectors $\vec{\mathbf{c}}$, and $\vec{\mathbf{m}}^{(n)}$, and the general moment pseudo-tensor $\boldsymbol{\mu}$, respectively, cannot completely represent the effect of body couple, couple-traction and couple-stresses in continuum. This is the result of the fact that the moments of couples do not specify their corresponding couples uniquely. There are infinitely possible equipollent system of forces, which represent the body couple with moment $\vec{\mathbf{c}}$, the couple-traction with moment $\vec{\mathbf{m}}^{(n)}$, and couple-stress with moment $\boldsymbol{\mu}$. These systems of equipollent couples have different effect on the deformation and internal interactions (stresses) of the continuum. However, the state of stresses in the body is physical and unique. This means the line of action of these couple forces is unique if they exist. This requires that the effect of body couple, couple-traction and couple-stress be completely described by their moments $\vec{\mathbf{c}}$, $\vec{\mathbf{m}}^{(n)}$ and $\boldsymbol{\mu}$, respectively, without requiring the specification of the line of action of opposite parallel couple forces. This is the statement of the determinacy or uniqueness of interactions in the continuum mechanics, which imposes some restrictions on the form of body couple, couple-traction and couple-stress distributions as following:

1. *If body couple with moment vector* $\vec{\mathbf{c}}$ *exists in a volume, its effect on the continuum must be completely represented by its moment vector* $\vec{\mathbf{c}}$*;*

2. *If couple-traction with moment vector* $\vec{\mathbf{m}}^{(n)}$ *exists on an arbitrary surface with unit normal vector* $\mathbf{n}$*, its effect on the continuum must be completely represented by its moment vector* $\vec{\mathbf{m}}^{(n)}$.

In the next section, the consequences of the determinacy of interactions in the continuum is investigated.

## 6. Fundamental character of body couple, couple-traction and couple-stresses in continuum

Now the consistent form of body couple, couple-traction and couple-stress based on the determinacy of interactions in the continuum, such that their effect is completely described by their moments $\vec{\mathbf{c}}$, $\vec{\mathbf{m}}^{(n)}$ and $\boldsymbol{\mu}$, respectively, is examined without requiring the specification of the line of action of opposite parallel couple forces.

### 6.1. Body couple distribution does not exist in continuum mechanics

Note that the pseudo-vector moment $d\vec{\mathbf{M}} = \vec{\mathbf{c}}dV$ does not completely describe the effect of external body couple in volume $dV$ (Figure 24).

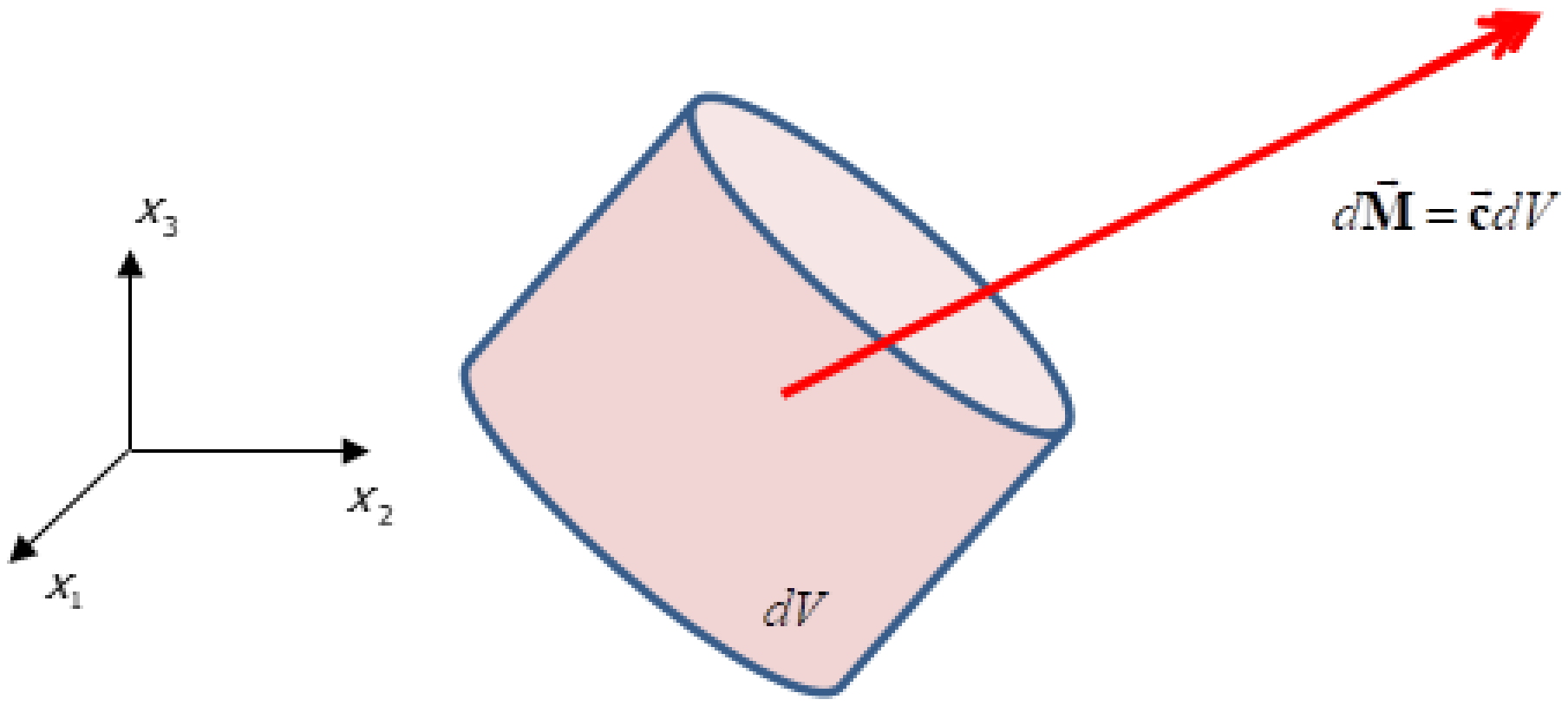

**Figure 24.** Couple with moment $d\vec{\mathbf{M}} = \vec{\mathbf{c}}dV$ in the volume $dV$.

One needs to know the line of action of couple forces $d\mathbf{F}^c$ and $-d\mathbf{F}^c$ creating this couple. However, the system of couple forces $d\mathbf{F}^c$ and $-d\mathbf{F}^c$ is not unique. The couple can be represented by forces $d\mathbf{F}^c$ and $-d\mathbf{F}^c$ in $dV$ in infinite number of ways, where $dM^{(n)} = cdV = dF^c dh$. Here $dh$ is the relative perpendicular distance between the line of action of forces $d\mathbf{F}^c$ and $-d\mathbf{F}^c$. For example, this system of couple tangential forces $d\mathbf{F}^c$ and $-d\mathbf{F}^c$ can be arbitrarily chosen either of the systems in Figure 25. However, these system of equipollent couples create different state of stresses and deformation. Therefore, the couple with moment $d\vec{\mathbf{M}} = \vec{\mathbf{c}}dV$ can never be completely represented by its moment. This contradicts the uniqueness of interactions in the continuum. This contradiction indicates that the volume couple with moment $d\vec{\mathbf{M}} = \vec{\mathbf{c}}dV$ cannot exist, which results in $\vec{\mathbf{c}} = 0$. Therefore, body couple distribution does not exist in continuum mechanics.

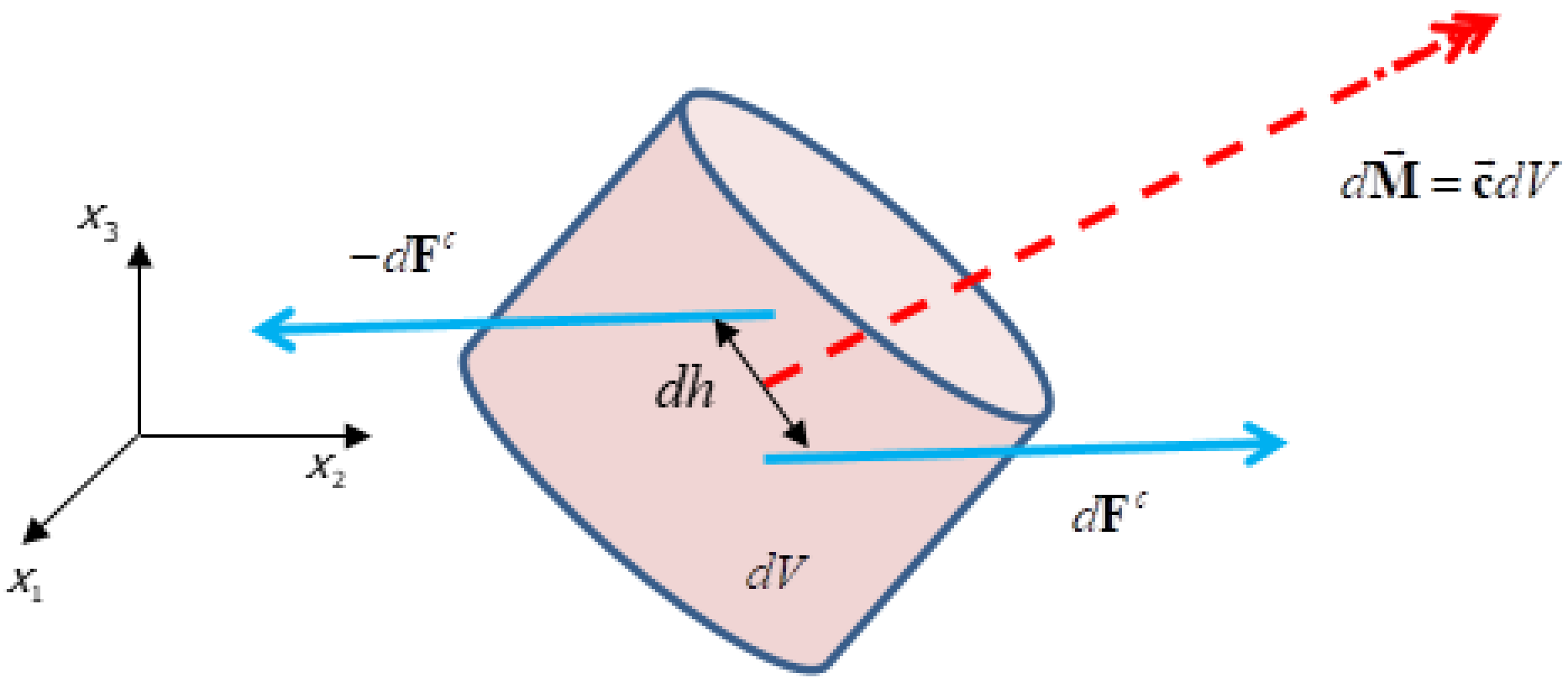

(a)

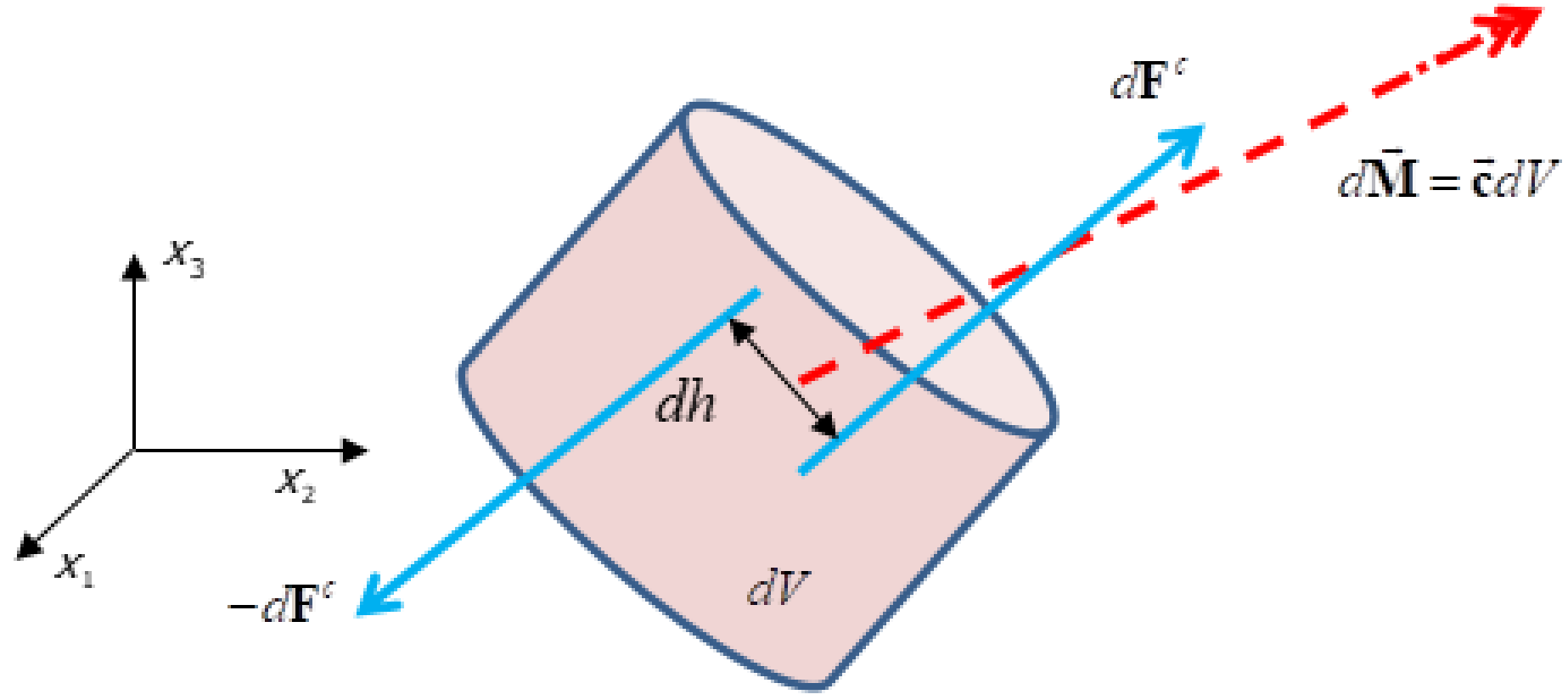

(b)

**Figure 25.** Different equipollent couple force systems with moment $d\vec{\mathbf{M}} = \vec{\mathbf{c}}dV$ in the volume element $dV$.

One might claim the possibility of a distribution of body couple with a specified piecewise continuous distribution of external couple forces $d\mathbf{F}^c$ and $-d\mathbf{F}^c$ creating the moment $d\vec{\mathbf{M}} = \vec{\mathbf{c}}dV$ in the volume $dV$. It can be always shown that this body couple can be replaced with an equivalent body force and a surface shear force-traction (Hadjesfandiari and Dargush, 2011). This means the supposed body couple distribution with moment density $\vec{\mathbf{c}}$ is not distinguishable from a body force $\mathbf{f}$ in continuum mechanics and its effect is simply equivalent to a system of body force and surface traction.

## 6.2. Fundamental character of the couple-traction

Based on the uniqueness of interactions in the continuum, the couple-traction with moment $\vec{\mathbf{m}}^{(n)}$ must be in such a form that its effect on a surface element $dS$ of any arbitrary surface $S_a$ is completely described by its pseudo vector moment $\vec{\mathbf{m}}^{(n)}$. Interestingly, this requirement does not impose the condition $\vec{\mathbf{m}}^{(n)} = 0$ inside the body. Instead, it imposes some restriction on the form of couple-traction and its moment $\vec{\mathbf{m}}^{(n)}$ as follows.

For more insight, let us decompose the moment vector $d\vec{\mathbf{M}} = \vec{\mathbf{m}}^{(n)}dS$ into its normal and tangential components on the surface element $dS$ as

$$d\vec{\mathbf{M}} = d\vec{\mathbf{M}}^{(n)} + d\vec{\mathbf{M}}^{(t)} \qquad (49)$$

where

$$d\vec{\mathbf{M}}^{(n)} = \vec{\mathbf{m}}^{(nn)}dS \qquad\qquad d\vec{\mathbf{M}}^{(t)} = \vec{\mathbf{m}}^{(nt)}dS \qquad (50)$$

These components have been shown in Figure 26.

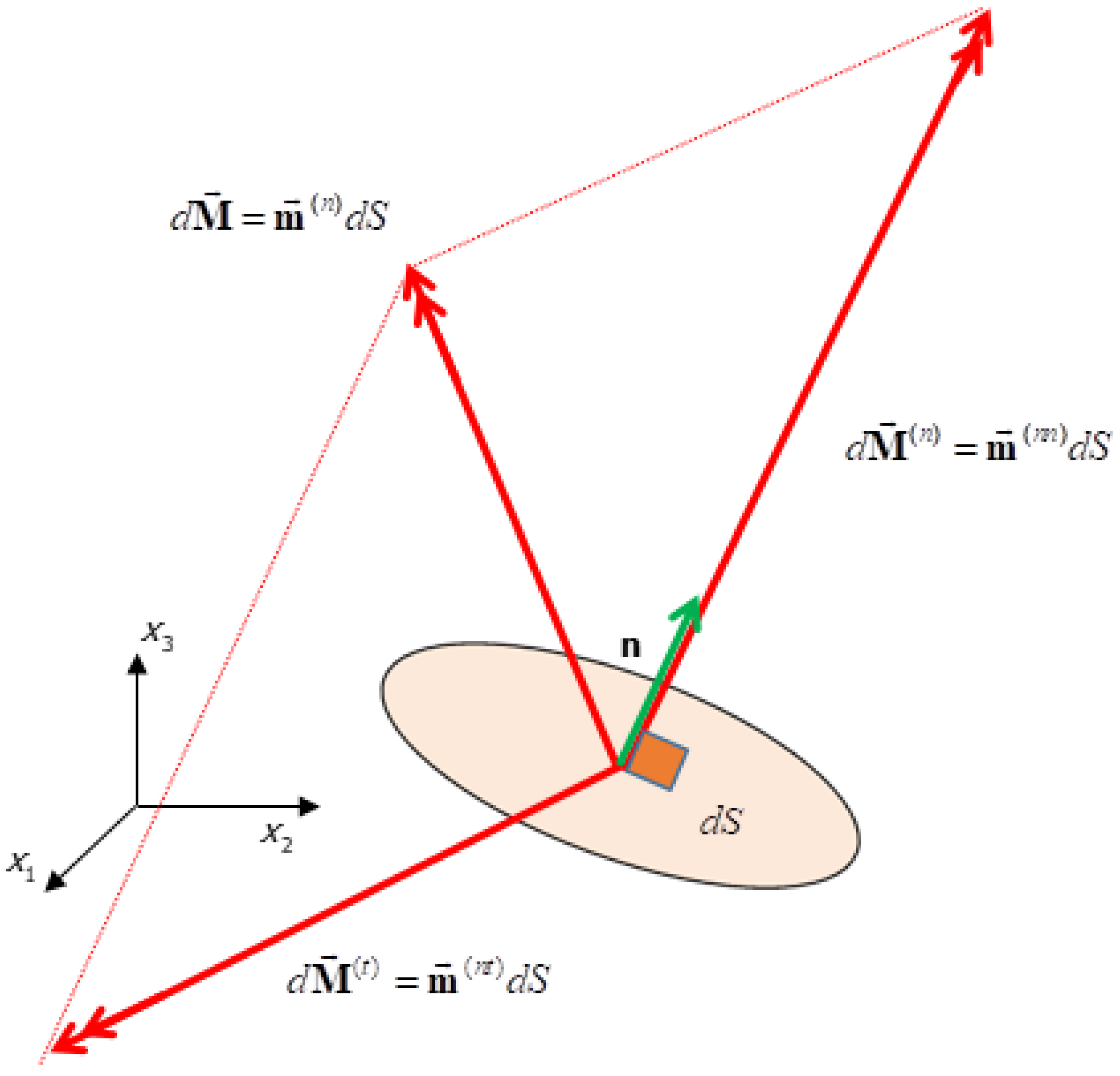

**Figure 26.** Normal and tangential components of moment $d\vec{\mathbf{M}} = \vec{\mathbf{m}}^{(n)}dS$ of the surface couple.

Note that the normal moment $d\vec{\mathbf{M}}^{(n)} = \vec{\mathbf{m}}^{(nn)} dS$ is the moment of the couple component that causes twisting or torsion on the surface element $dS$, whereas the tangential moment $d\vec{\mathbf{M}}^{(t)} = \vec{\mathbf{m}}^{(nt)} dS$ is the moment of the couple component that causes bending.

It seems also instructive if the distribution of the tangential shear force-traction vector $\mathbf{t}^{(nt)}$ is considered as a single layer. This is the vectorial analogy to a single layer of electric charge in electrostatic.

Now, the character of normal and tangential couple-traction components with moments $\vec{\mathbf{m}}^{(nn)}$ and $\vec{\mathbf{m}}^{(nt)}$, respectively are investigated.

### 6.2.1. Twisting couple-traction does not exist in continuum mechanics

The possible surface twisting couple-traction with normal moment $d\vec{\mathbf{M}}^{(n)} = \vec{\mathbf{m}}^{(nn)} dS$ , as shown in Figure 27, creates torsion on the surface element $dS$ .

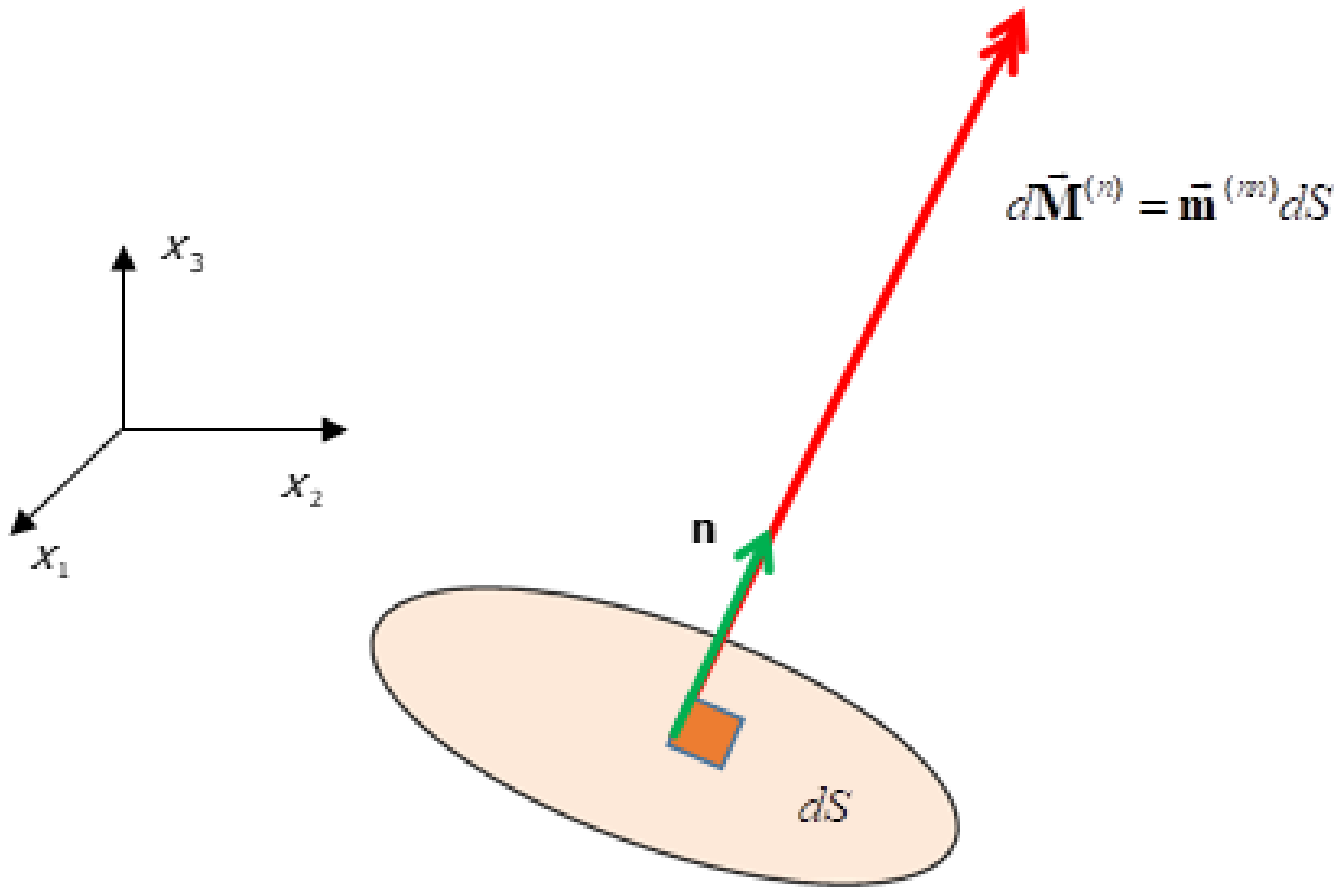

**Figure 27.** Twisting couple with normal moment $d\vec{\mathbf{M}}^{(n)} = \vec{\mathbf{m}}^{(nn)} dS$ on the surface element $dS$ .

**Theorem.** Twisting couple-traction with normal moment vanishes on any surface in the continuum, that is $\vec{\mathbf{m}}^{(nn)} = 0$.

**Proof**. If the surface twisting couple with moment $d\vec{\mathbf{M}}^{(n)} = \vec{\mathbf{m}}^{(nn)} dS$ exists on the surface element $dS$ of any surface $S_a$, it represents a system of couple tangential forces $d\mathbf{F}^{s}$ and $-d\mathbf{F}^{s}$ on the surface creating twisting. However, this system of couple of tangential forces $d\mathbf{F}^{s}$ and $-d\mathbf{F}^{s}$ is not unique for a given $\vec{\mathbf{m}}^{(nn)}$. Note that the twisting couple with moment $d\vec{\mathbf{M}}^{(n)} = \vec{\mathbf{m}}^{(nn)} dS$ can be represented by tangential forces $d\mathbf{F}^{s}$ and $-d\mathbf{F}^{s}$ on $dS$ in infinite number of ways. For example, this system of couple tangential forces $d\mathbf{F}^{s}$ and $-d\mathbf{F}^{s}$ can be arbitrarily chosen either of the equipollent systems in Figure 28. Therefore, the normal couple with moment $d\vec{\mathbf{M}}^{(n)} = \vec{\mathbf{m}}^{(nn)} dS$ cannot be completely represented by its moment $d\vec{\mathbf{M}}^{(n)} = \vec{\mathbf{m}}^{(nn)} dS$. This contradicts the uniqueness of interactions in the continuum that the effect of normal couple can be completely represented by its moment $d\vec{\mathbf{M}}^{(n)} = \vec{\mathbf{m}}^{(nn)} dS$. This contradiction shows that the surface normal twisting couple with moment $d\vec{\mathbf{M}}^{(n)} = \vec{\mathbf{m}}^{(nn)} dS$ cannot exist in continuum, which results in $\vec{\mathbf{m}}^{(nn)} = 0$. This means continuum does not support normal twisting couple-traction on any surface.

Although it has been rigorously established that twisting couple-traction does not exist on any surface in the continuum in continuum mechanics, one might claim the possibility of a distribution of twisting couple-traction with a specified piecewise continuous distribution of external couple forces $d\mathbf{F}^{s}$ and $-d\mathbf{F}^{s}$ creating the normal moment $d\vec{\mathbf{M}}^{(n)} = \vec{\mathbf{m}}^{(nn)} dS$ on the surface element $dS$ of the physical bounding surface $S$. It can be always shown that this twisting surface couple-traction can be replaced with an equivalent shear force and some line force on the physical surface. Therefore, the possible distribution of normal twisting couple-traction with moment $\vec{\mathbf{m}}^{(nn)}$ on physical bounding surface $S$ is not distinguishable from a shear force distribution in continuum mechanics. However, there is no need for such replacement on arbitrary surfaces inside the continuum, where $\vec{\mathbf{m}}^{(nn)}$ does not exist.

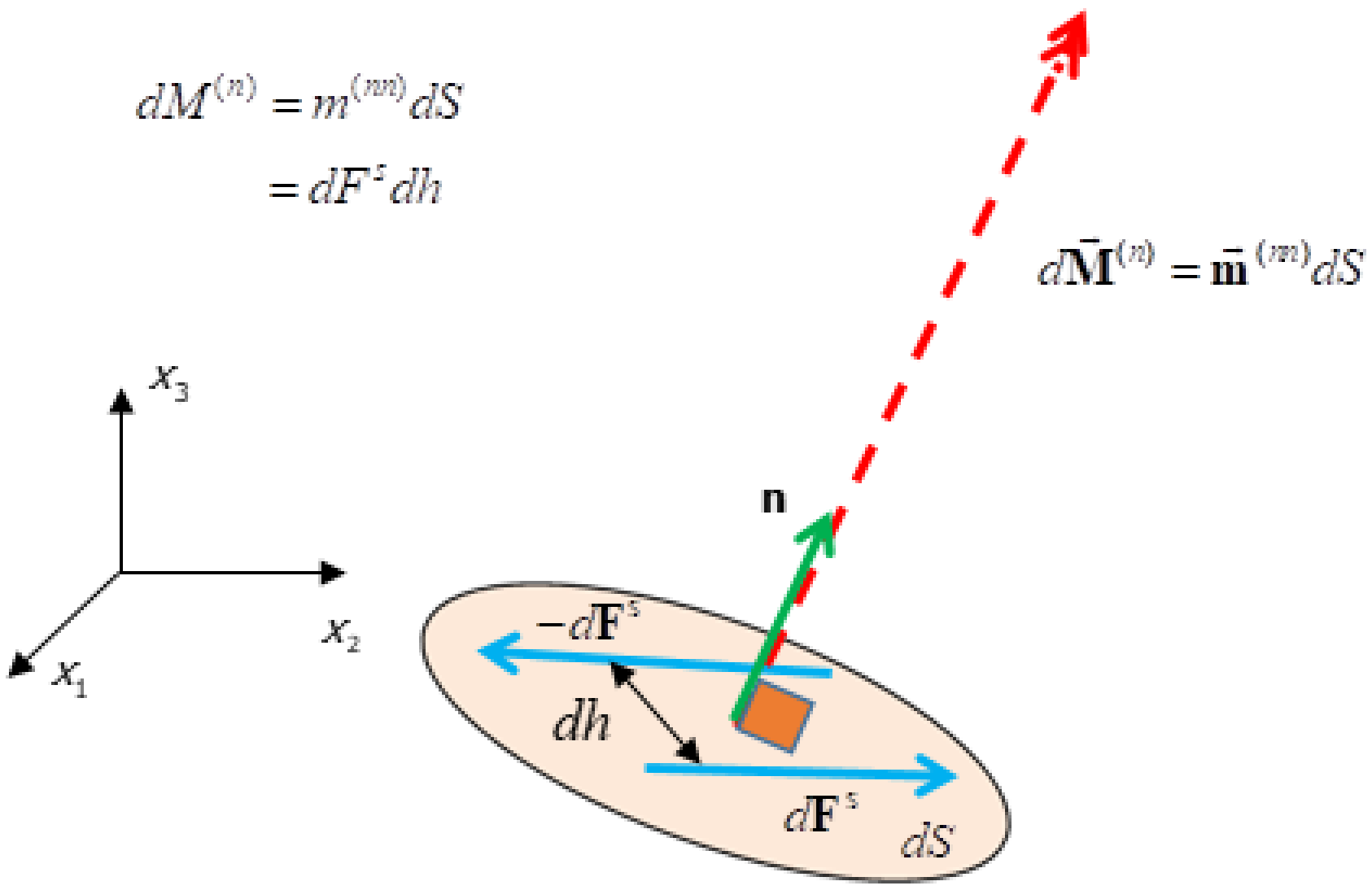

(a)

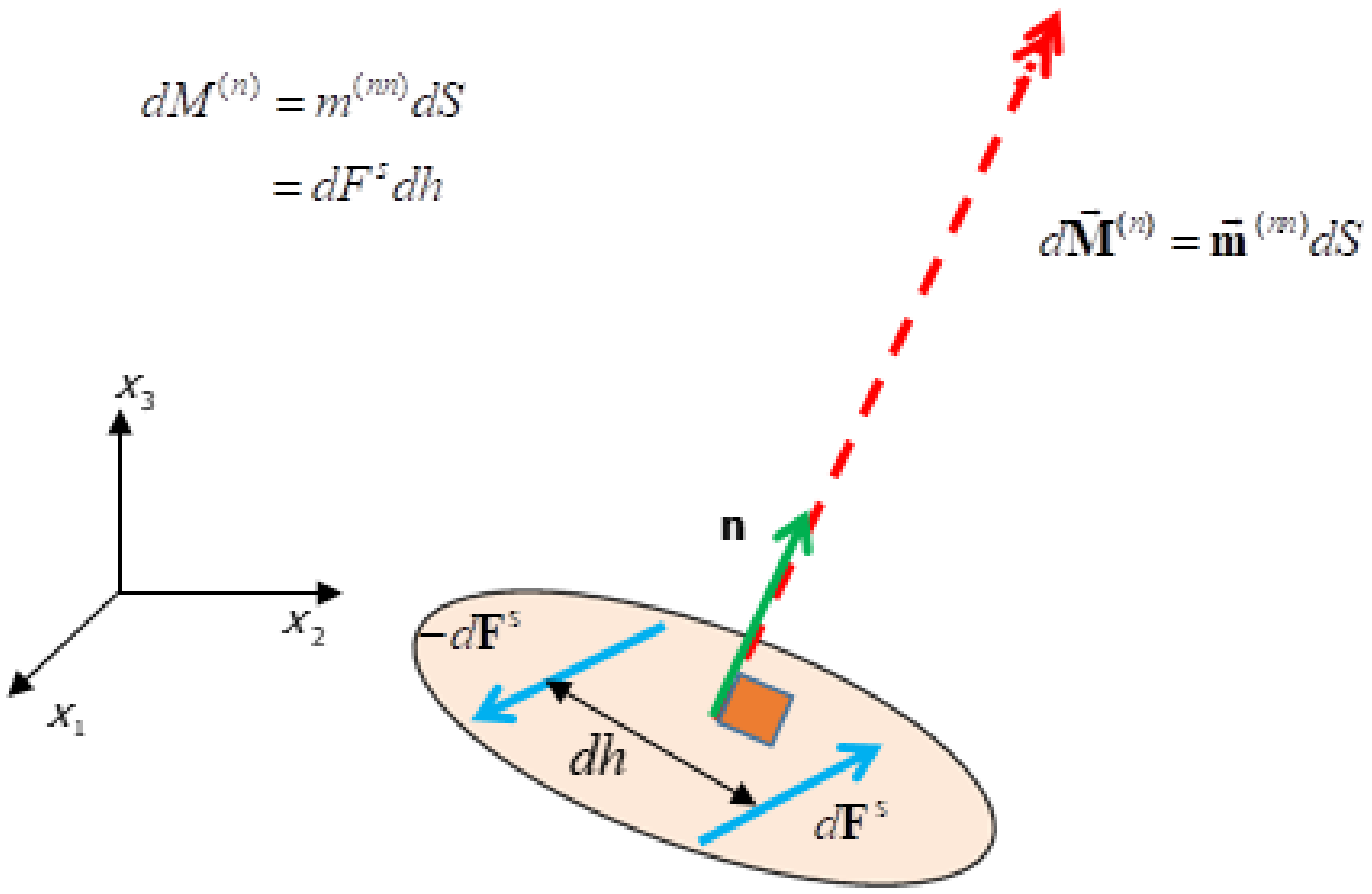

(b)

**Figure 28.** Equipollent surface twisting couples with moment $d\vec{\mathbf{M}}^{(n)} = \vec{\mathbf{m}}^{(nn)} dS$ .

### 6.2.2. Bending couple-traction can exist in continuum mechanics

The possible surface bending couple with tangential moment $d\vec{\mathbf{M}}^{(t)} = \vec{\mathbf{m}}^{(nt)} dS$, as shown in Figure 29, creates bending on the surface element $dS$.

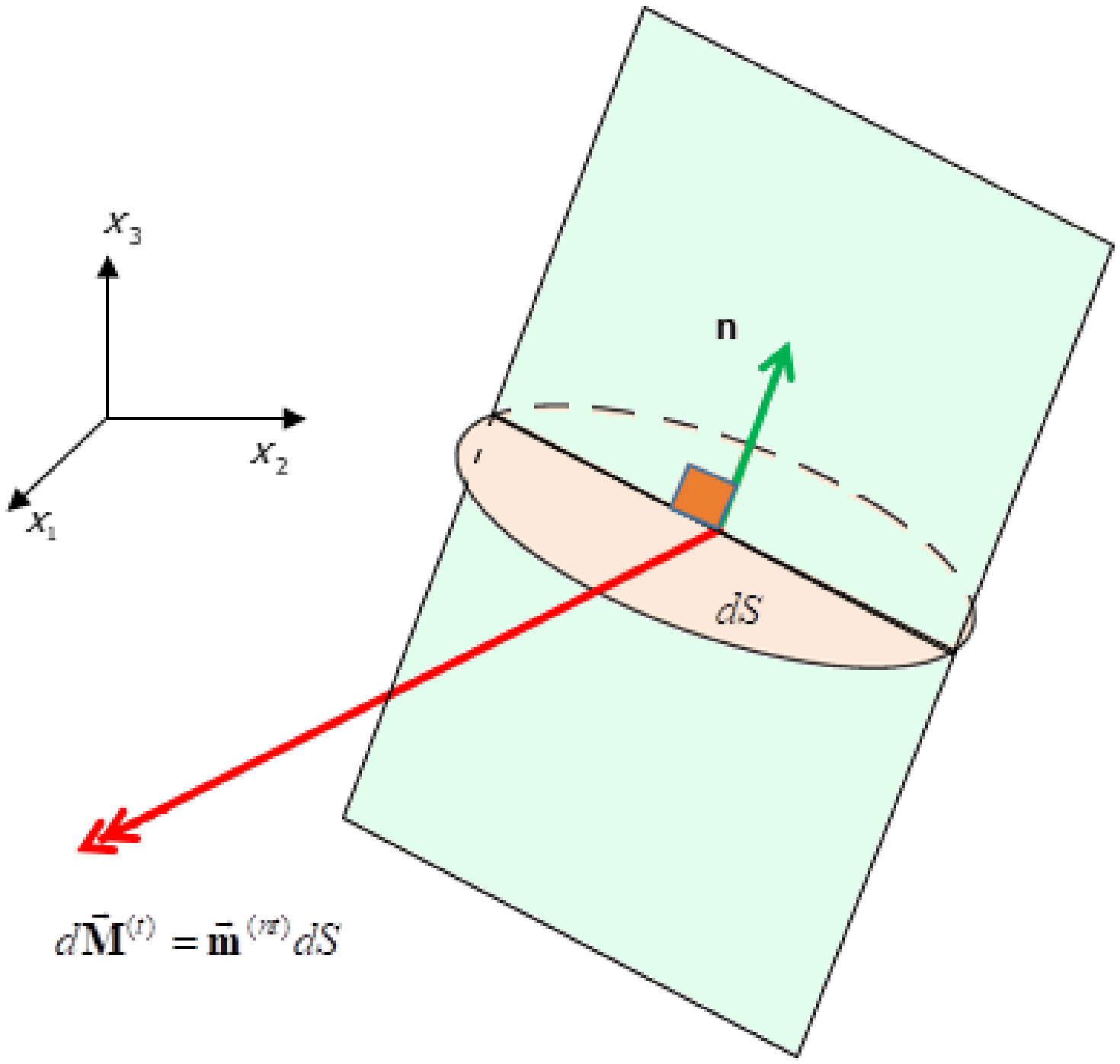

**Figure 29.** Bending couple with tangential moment $d\vec{\mathbf{M}}^{(t)} = \vec{\mathbf{m}}^{(nt)} dS$ on the surface element $dS$.

**Theorem.** The bending couple-traction with tangential moment can exist on any surface as a double layer of shear force-tractions.

**Proof**. If the surface bending couple with tangential moment $d\vec{\mathbf{M}}^{(t)} = \vec{\mathbf{m}}^{(nt)} dS$ exists on the surface element $dS$, it represents the effect of the system of couple forces $d\mathbf{F}^b$ and $-d\mathbf{F}^b$ in the plane normal to the surface element $dS$ and normal to the moment vector $d\vec{\mathbf{M}}^{(t)} = \vec{\mathbf{m}}^{(nt)} dS$ (Figure 29). Two possible cases are examined as follows:

I. The couple forces $d\mathbf{F}^b$ and $-d\mathbf{F}^b$ are not parallel to the surface element $dS$.

Note that the system of couple $d\mathbf{F}^b$ and $-d\mathbf{F}^b$ is not unique for this case. This means the bending couple with moment $d\vec{\mathbf{M}}^{(t)} = \vec{\mathbf{m}}^{(nt)} dS$ can be represented by system of equipollent forces $d\mathbf{F}^b$ and $-d\mathbf{F}^b$ in infinite number of ways. For example, this system of couple forces $d\mathbf{F}^b$ and $-d\mathbf{F}^b$ can be arbitrarily chosen either of equipollent systems of forces nonparallel to the surface element $dS$ in Figure 30. Therefore, if the tangential couple exist, it cannot be represented by a unique two force system $d\mathbf{F}^b$ and $-d\mathbf{F}^b$ non-parallel to the surface element $dS$. This is in contradiction with the uniqueness of interactions in the continuum. Therefore, the couple forces $d\mathbf{F}^b$ and $-d\mathbf{F}^b$ cannot be non-parallel to the surface element $dS$.

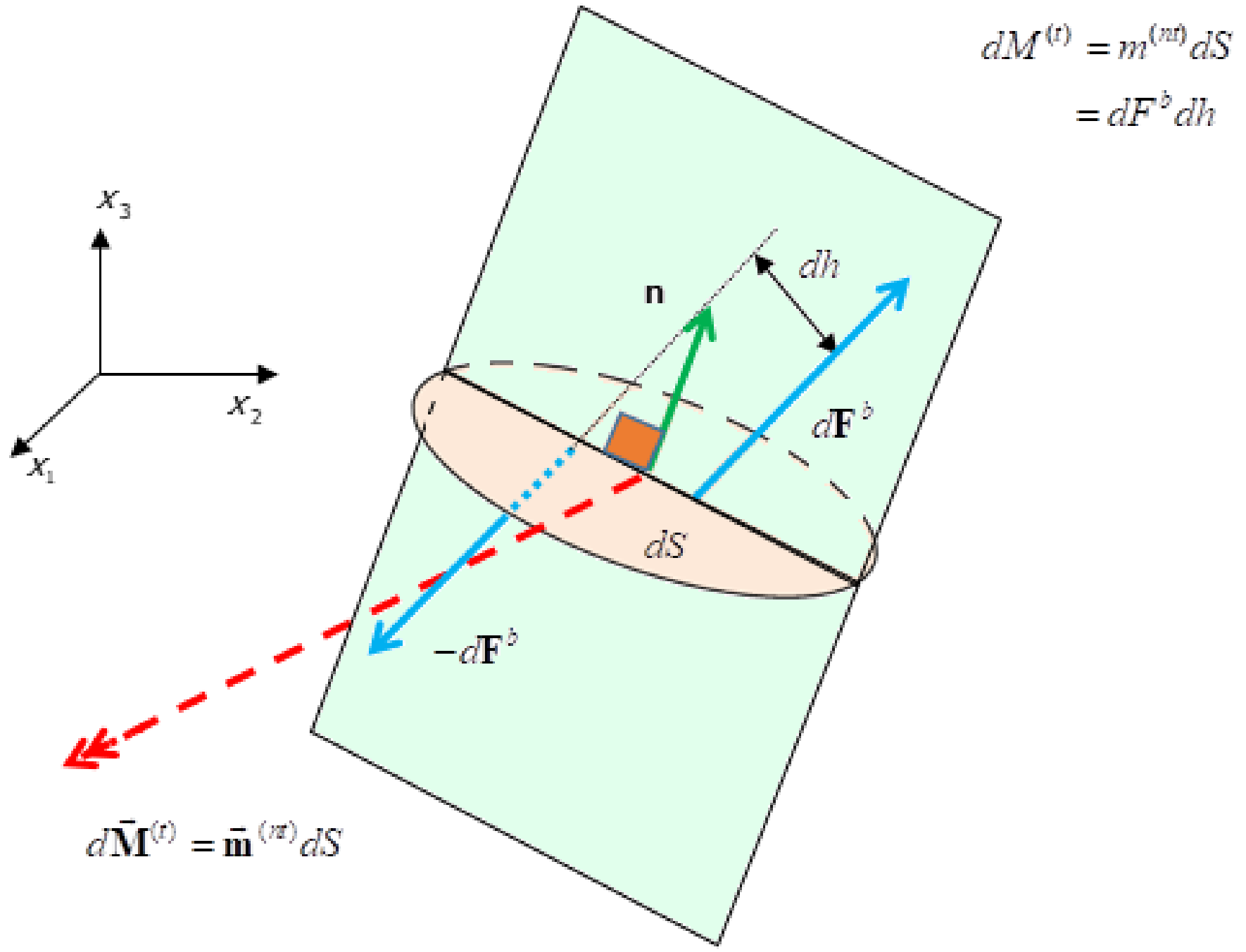

(a)

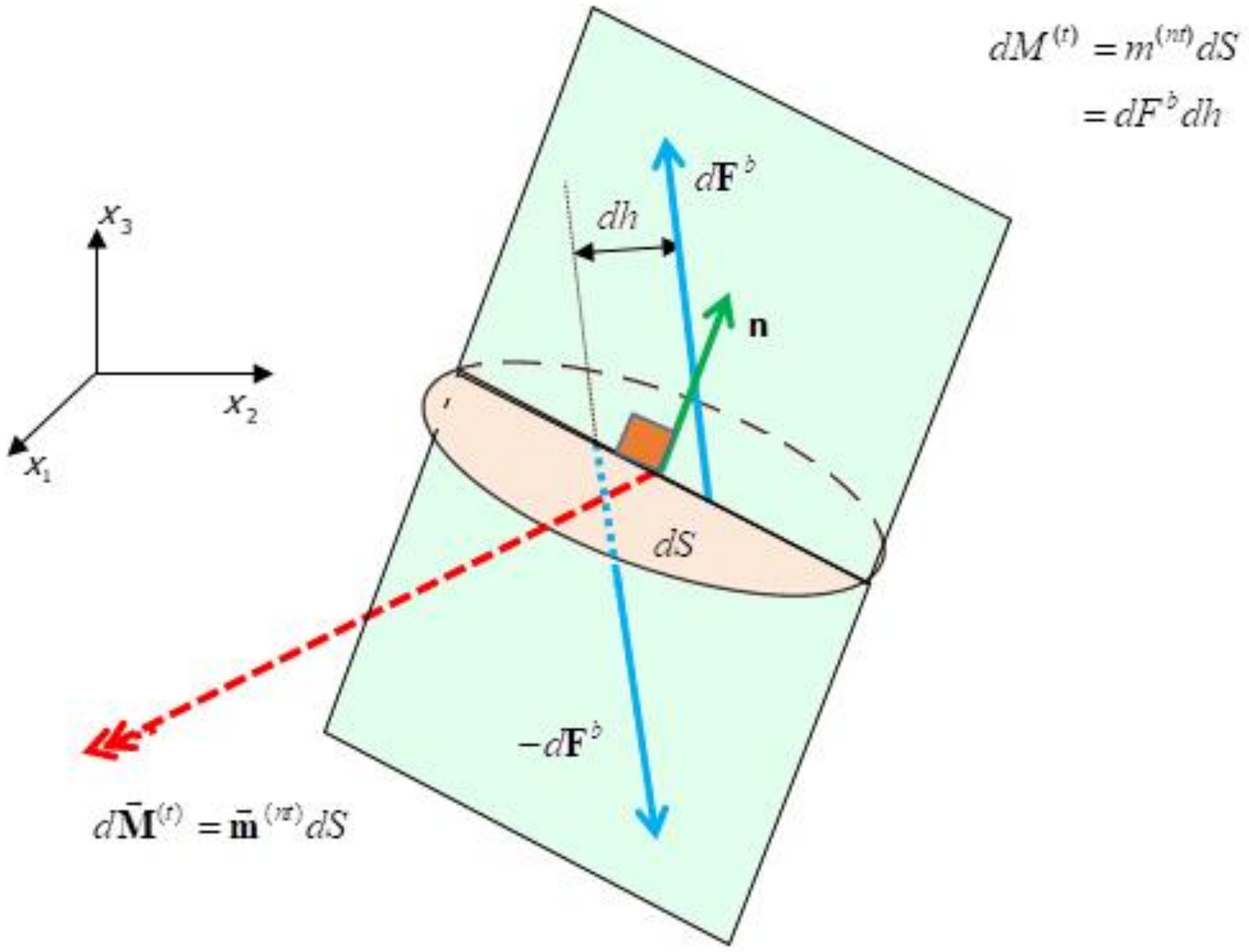

(b)

**Figure 30.** Equipollent bending couples with moment $d\vec{\mathbf{M}}^{(t)} = \vec{\mathbf{m}}^{(nt)} dS$ represented by forces non-parallel to surface element $dS$ in normal plane.

II. The couple forces $d\mathbf{F}^b$ and $-d\mathbf{F}^b$ are parallel to the surface element $dS$.

Figure 31 shows the unique system of couple forces $d\mathbf{F}^b$ and $-d\mathbf{F}^b$ parallel to the surface element $dS$. Therefore, the tangential moment $d\vec{\mathbf{M}}^{(t)} = \vec{\mathbf{m}}^{(nt)}dS$ completely describes the effect of bending couple of forces $d\mathbf{F}^b$ and $-d\mathbf{F}^b$ parallel to the surface element $dS$. This is in accordance with the uniqueness of interactions in the continuum, which not only shows that the bending couple $d\vec{\mathbf{M}}^{(t)} = \vec{\mathbf{m}}^{(nt)}dS$ can exist, but also reveals its structure as a system of couple forces $d\mathbf{F}^b$ and $-d\mathbf{F}^b$ parallel to the surface element $dS$. Therefore, the bending couple-traction with moment $\vec{\mathbf{m}}^{(nt)}$ can exist as a double layer of shear force-tractions $\mathbf{t}^{(b)} = \frac{d\mathbf{F}^b}{dS}$ and $-\mathbf{t}^{(b)} = -\frac{d\mathbf{F}^b}{dS}$, where $t^{(b)} = \frac{dF^b}{dS} = \frac{m^{(nt)}}{dh}$, on the arbitrary surface $S_a$. Interestingly, this is the vectorial analogy of double layer in electrostatics, where the tangential shear traction-force distribution is analogous to a single layer of electric charge.

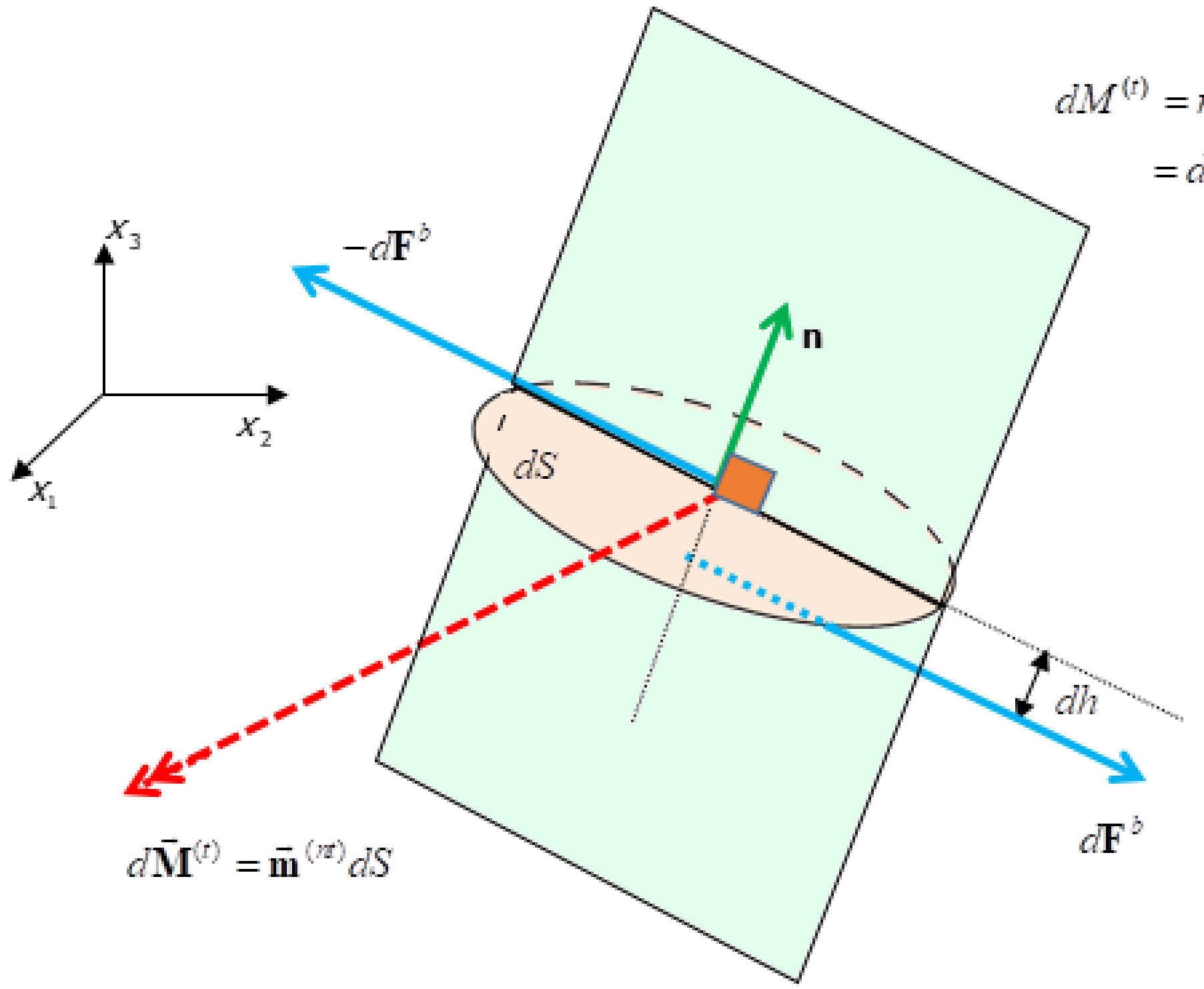

**Figure 31.** Couple forces parallel to surface element $dS$ in normal plane with tangential moment $d\vec{\mathbf{M}}^{(t)} = \vec{\mathbf{m}}^{(nt)}dS$.

Therefore, continuum supports piecewise continuous distributions of bending couple-traction with tangential moment $\vec{\mathbf{m}}^{(nt)}$ on any arbitrary surface $S_a$ as a double layer of shear-force tractions.

Since the bending couple-traction is completely described by its tangential moment vector $\vec{\mathbf{m}}^{(nt)}$, it can be denoted by $\vec{\mathbf{m}}^{(nt)}$. Therefore, in consistent continuum mechanics the load density acting at point $P$ on an arbitrary surface with unit normal vector $\mathbf{n}$ is equivalent to a system of force-traction vector $\mathbf{t}^{(n)}$ and bending couple-traction with tangential moment $\vec{\mathbf{m}}^{(n)} = \vec{\mathbf{m}}^{(nt)}$, as shown in Figure 32.

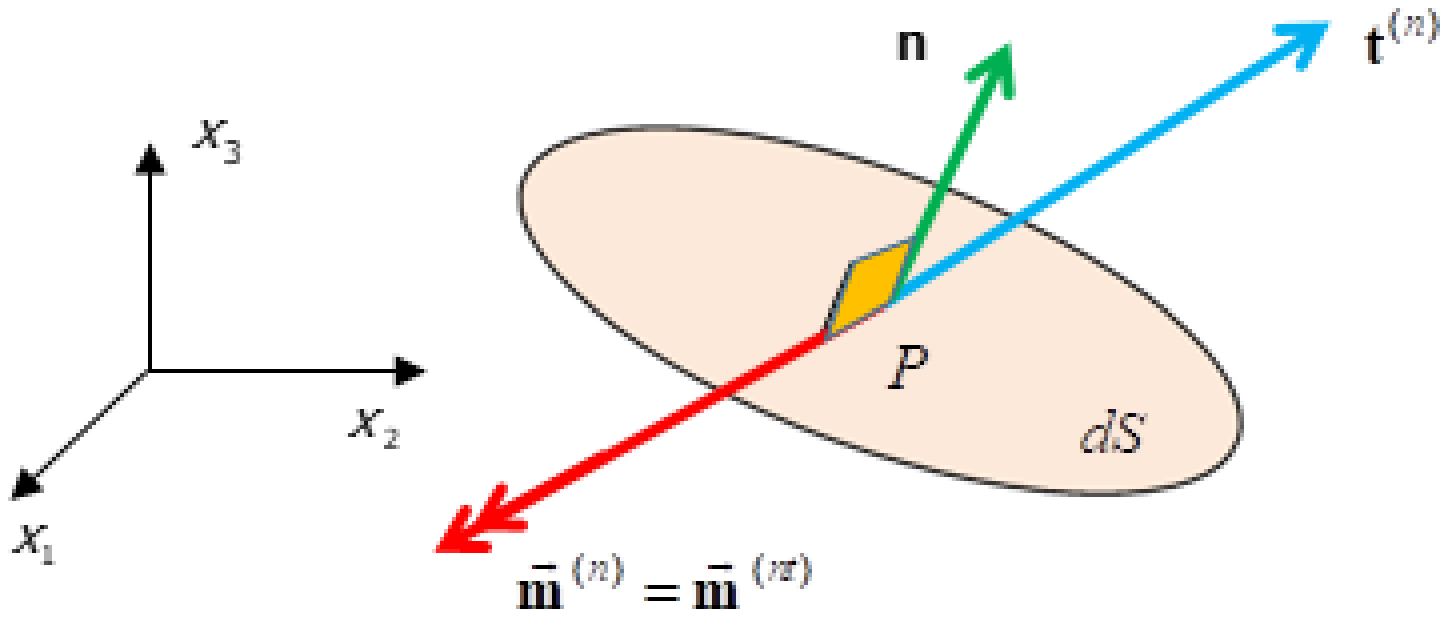

**Figure 32.** Force-traction $\mathbf{t}^{(n)}$ and the consistent couple-traction $\vec{\mathbf{m}}^{(n)} = \vec{\mathbf{m}}^{(nt)}$ system.

### 6.3. Skew-symmetric character of couple- stress tensor and its consequences

Since there is no twisting couple-traction with normal moment component $\vec{\mathbf{m}}^{(nn)}$ on any surface element $dS$ on the arbitrary $S_a$

$$m^{(nn)} = \vec{\mathbf{m}} \bullet \mathbf{n} = m_i^{(n)} n_i = 0 \qquad \text{on} \quad S_a \tag{51}$$

This means

$$m^{(nn)} = \mu_{ji} n_j n_i = 0 \qquad \text{on} \quad S_a \tag{52}$$

However, in this relation, $n_i$ is arbitrary at each point; one may construct subdomains with any surface normal orientation at a point. Consequently, in (52), $n_i n_j$ is an arbitrary symmetric second order tensor of rank one at each point. Therefore, for (52) to hold in general, the moment stress pseudo tensor $\mu_{ij}$ must be skew-symmetric, that is

$$\boldsymbol{\mu}^T = -\boldsymbol{\mu} \qquad\qquad \mu_{ji} = -\mu_{ij} \tag{53}$$

This is the fundamental character of the couple-stress moment tensor in continuum mechanics, which guarantees the couple-traction moment vector $\vec{\mathbf{m}}^{(n)}$ is tangent to the surface, thus creating a bending effect. It should be emphasized that there is no mention of constitutive relations in any of this development, so that these results are in no way limited to linear elastic materials or to isotropic response. In this development, there are no additional assumptions beyond that of the continuum as a domain-based concept having no special characteristics associated with the actual bounding surface $S$ over any arbitrary internal surface $S_a$.

Since the couple-stress system is completely described by its skew-symmetric moment tensor $\mu_{ij}$, it can be denoted by $\mu_{ij}$ and the term couple-stress tensor $\mu_{ij}$ can be used. Therefore, in consistent continuum mechanics the state of stresses at any arbitrary point is specified by force-stress tensor $\sigma_{ij}$ and skew-symmetric couple-stress tensor $\mu_{ij}$. The skew-symmetric character immediately resolves the indeterminacy problem in original couple stress theory developed by Mindlin and Tiersten (1962) and Koiter (1964). Since the diagonal components of the couple-stress tensor vanish, it is noticed that the couple-stress tensor automatically is determinate in this consistent couple stress theory.

The components of the force-stress $\sigma_{ij}$ and couple-stress $\mu_{ij}$ tensors in this consistent theory are shown in Figure 33.

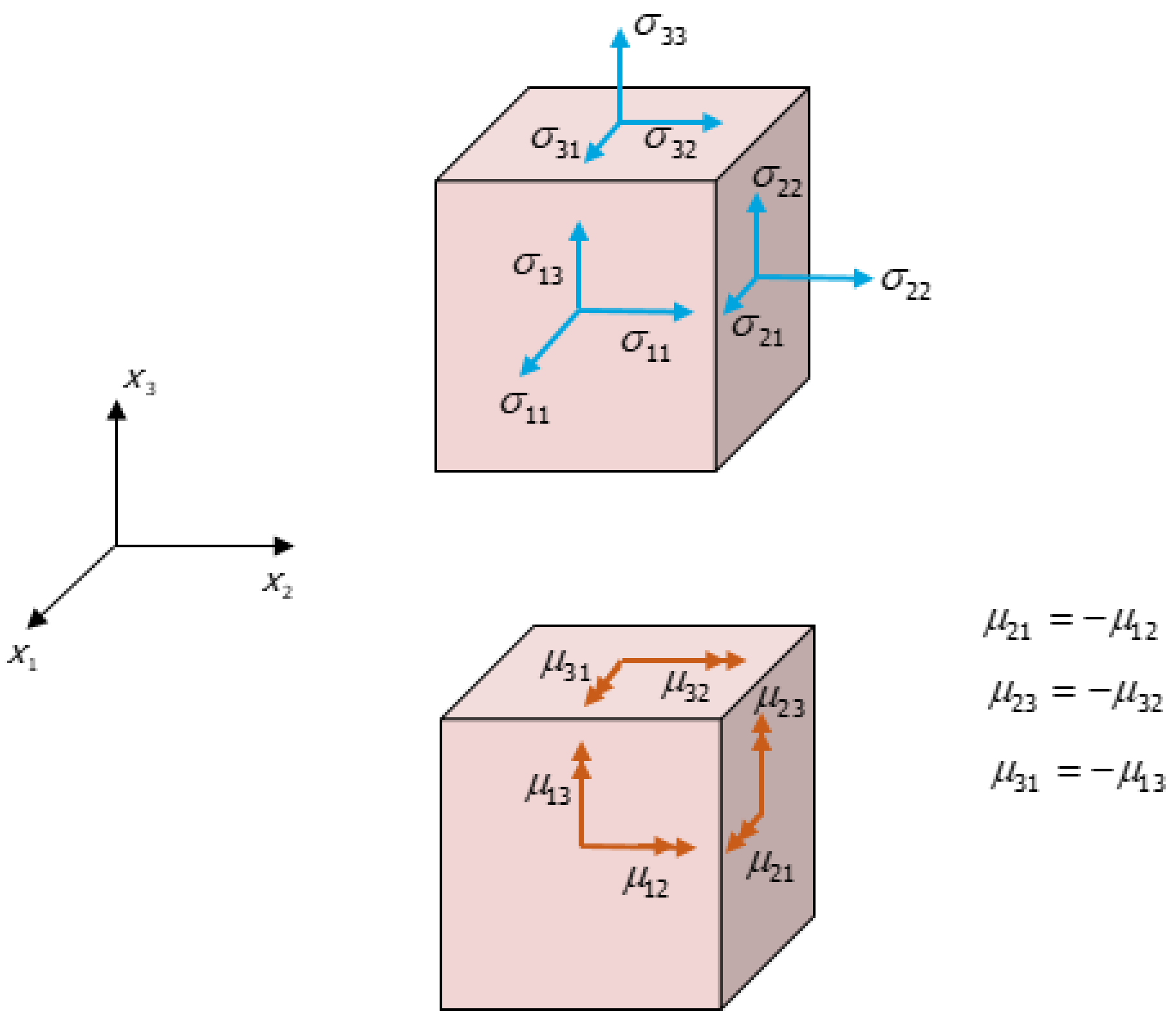

**Figure 33.** Components of force- and couple-stress tensors in consistent couple stress theory.

In terms of components, the skew-symmetric couple-stress tensor (couple-stress moment tensor) can be written as

$$[\mu_{ij}] = \begin{bmatrix} 0 & \mu_{12} & \mu_{13} \\ -\mu_{12} & 0 & \mu_{23} \\ -\mu_{13} & -\mu_{23} & 0 \end{bmatrix} \tag{54}$$

As a result, the couple-traction vectors $\vec{\mathbf{m}}^{(1)}$, $\vec{\mathbf{m}}^{(2)}$ and $\vec{\mathbf{m}}^{(3)}$ reduce to

$$\{m_i^{(1)}\} = [0 \quad \mu_{12} \quad \mu_{13}]^T \tag{55a}$$

$$\{m_i^{(2)}\} = [-\mu_{12} \quad 0 \quad \mu_{23}]^T \tag{55b}$$

$$\{m_i^{(3)}\} = [-\mu_{13} \quad -\mu_{23} \quad 0]^T \tag{55c}$$

Since the couple-stress tensor $\boldsymbol{\mu}$ is skew-symmetric, it is singular. This means its determinant vanishes

$$\det\left[\mu_{ij}\right] = \varepsilon_{ijk}\mu_{1i}\mu_{2j}\mu_{3k} = 0 \tag{56}$$

and its rank is two. Interestingly, the determinant (56) can also be expressed as

$$\det\left[\mu_{ij}\right] = \varepsilon_{ijk}m_i^{(1)}m_j^{(2)}m_k^{(3)} = 0 \tag{57}$$

or in vectorial form

$$\det\left[\mu_{ij}\right] = \left[\vec{\mathbf{m}}^{(1)} \times \vec{\mathbf{m}}^{(2)}\right] \bullet \vec{\mathbf{m}}^{(3)} = 0 \tag{58}$$

Remarkably, the relation (58) shows that the three couple traction vectors $\vec{\mathbf{m}}^{(1)}$, $\vec{\mathbf{m}}^{(2)}$ and $\vec{\mathbf{m}}^{(3)}$ are linearly dependent. This interestingly means that these three vectors are coplanar. Then any two of these vector tractions span a plane $\Gamma$, the third traction vector is a linear combination of the first two.

Also notice that the three independent components $\mu_{12}$, $\mu_{13}$, $\mu_{23}$ of the couple-stress tensor $\mu_{ij}$ are specified by the components of only two rows or columns of the matrix representation (54). Therefore, the couple-stress tensor $\mu_{ij}$ is specified by only two of the coplanar couple-traction vectors $\vec{\mathbf{m}}^{(1)}$, $\vec{\mathbf{m}}^{(2)}$ and $\vec{\mathbf{m}}^{(3)}$ ( $m_i^{(1)}$, $m_i^{(2)}$ and $m_i^{(3)}$). This means that the state of couple-stress in the continuum is completely specified by two of the couple-traction vectors, e.g., $m_i^{(1)}$, $m_i^{(2)}$. Note that this is an important implication of the skew-symmetric couple-stress tensor.

Interestingly, the skew-symmetric tensors $\boldsymbol{\mu}$ can be represented by its dual true couple-stress vector $\vec{\boldsymbol{\mu}}$ (Hadjesfandiari and Dargush, 2011), where

$$\mu_i = \frac{1}{2}\varepsilon_{ijk}\mu_{kj} \tag{59}$$

This relation can also be written in the form

$$\varepsilon_{ijk}\mu_k = \mu_{ji} \tag{60}$$

which simply shows

$$\mu_1 = \mu_{32}, \qquad \mu_2 = \mu_{13}, \qquad \mu_3 = \mu_{21} \tag{61}$$

Components of couple-stress tensor $\boldsymbol{\mu}$ and its dual couple-stress vector $\vec{\boldsymbol{\mu}}$ are shown in Figure 34. Note that the components of couple-stress vector $\vec{\boldsymbol{\mu}}$ have been presented as single headed arrows.

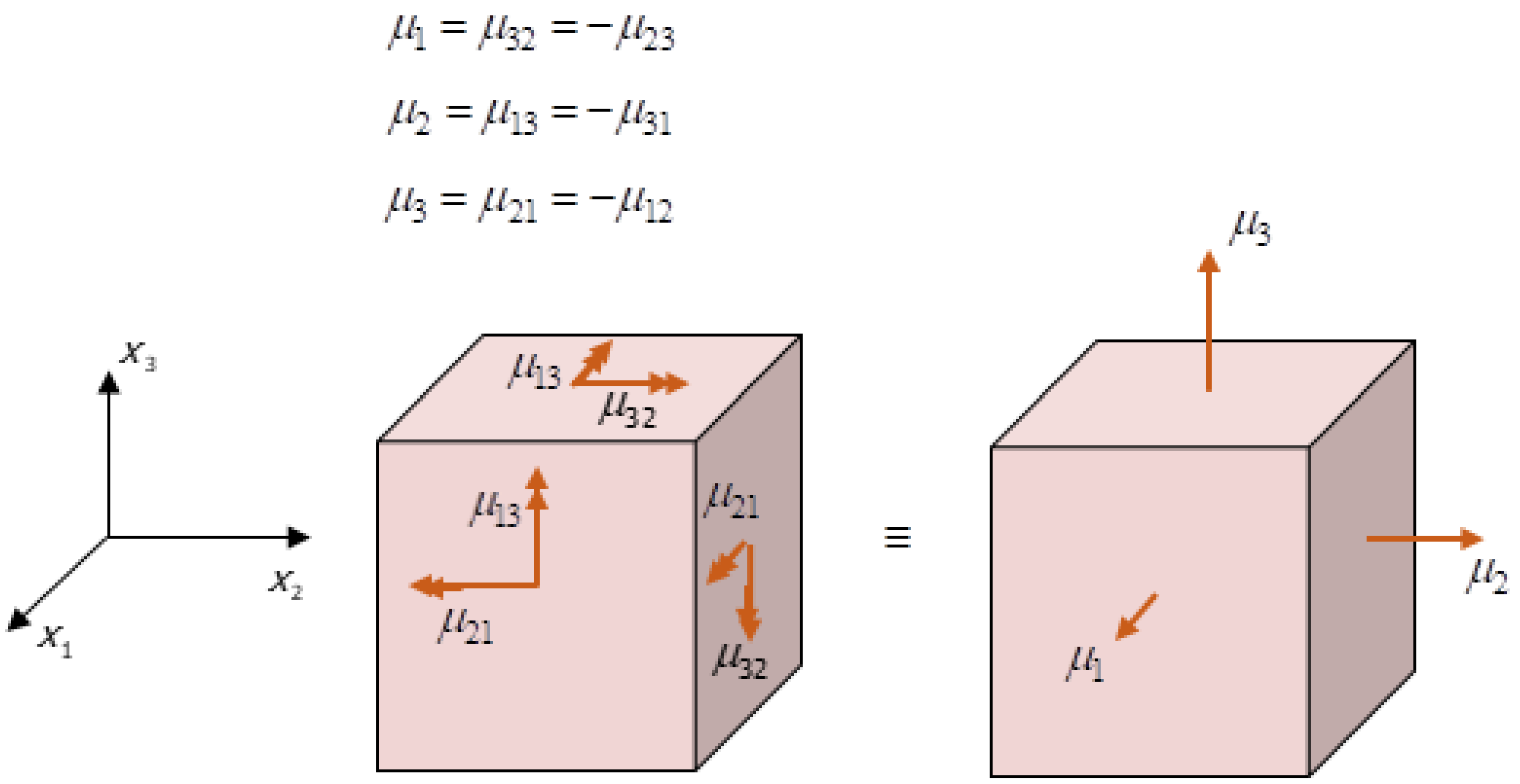

**Figure 34.** Components of couple-stress tensor $\boldsymbol{\mu}$ and couple-stress vector $\vec{\boldsymbol{\mu}}$.

Therefore, the couple-stress pseudo-tensor $\mu_{ij}$ and couple-traction true-vector $\mu_i$ can be represented as

$$[\mu_{ij}] = \begin{bmatrix} 0 & -\mu_3 & \mu_2 \\ \mu_3 & 0 & -\mu_1 \\ -\mu_2 & \mu_1 & 0 \end{bmatrix}, \qquad \{\mu_i\} = \begin{Bmatrix} \mu_1 \\ \mu_2 \\ \mu_3 \end{Bmatrix} \tag{62}$$

The magnitude of the couple-stress vector $\mu_i$ is

$$\mu = |\vec{\boldsymbol{\mu}}| = \sqrt{\mu_1^2 + \mu_2^2 + \mu_3^2} = \sqrt{\mu_{32}^2 + \mu_{13}^2 + \mu_{12}^2} \tag{63}$$

Note that the couple-stress components $\mu_{32}$, $\mu_{13}$ and $\mu_{21}$ represent the bending effect of double layer of tangential shear-force stresses on planes normal to coordinate axis $x_1$, $x_2$ and $x_3$, respectively. Interestingly, their dual couple-stress vector components $\mu_1$, $\mu_2$ and $\mu_3$ are in the direction of double layers of shear force-stresses associated with the couple-stress components $\mu_{32}$

, $\mu_{13}$ and $\mu_{21}$, respectively. For example, Figure 35 shows this double layer of shear force-stresses for the couple stress component $\mu_3$ corresponding to $\mu_{21}$.

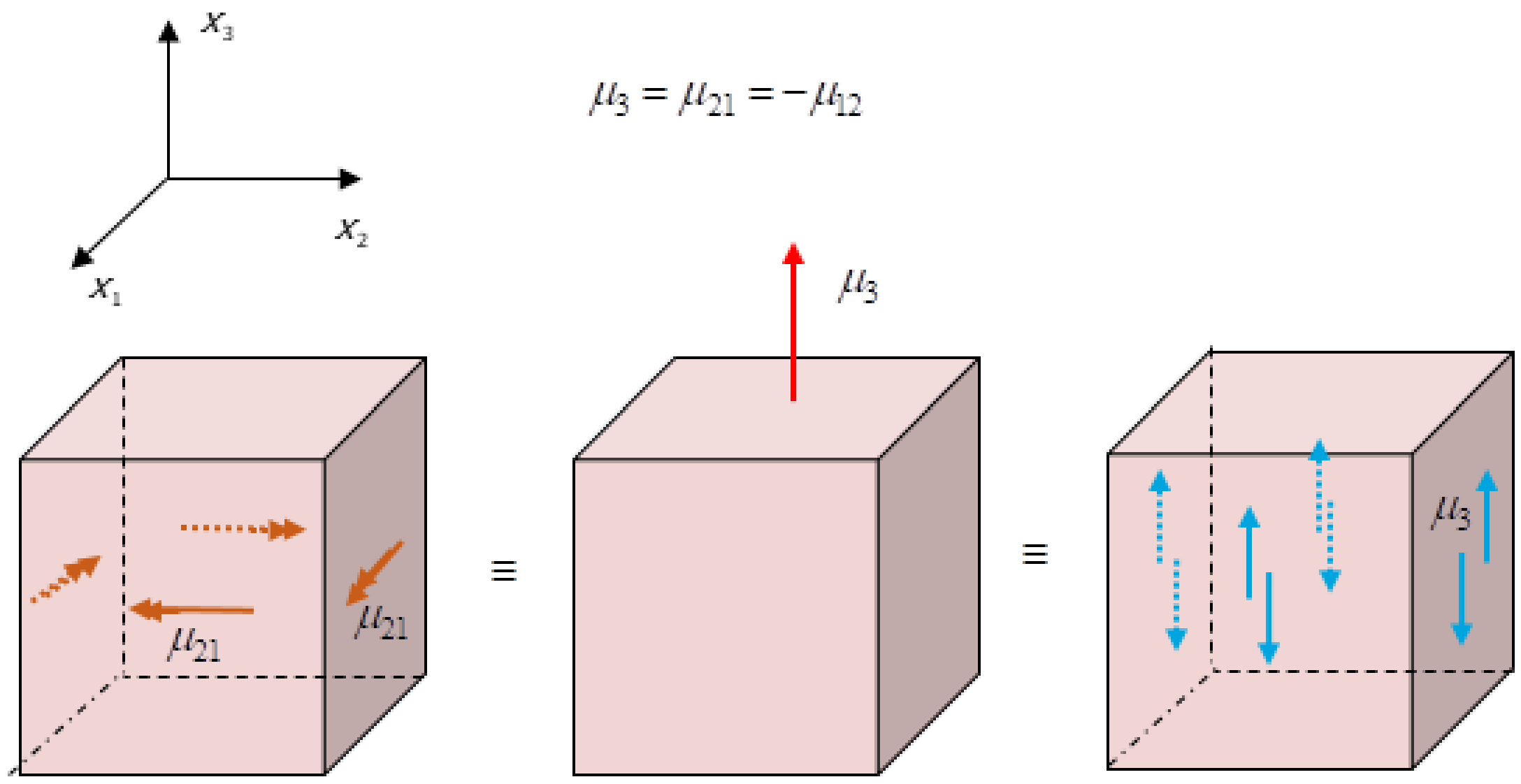

**Figure 35.** Couple stress component $\mu_3$ in the direction of its corresponding double layer shear force-stresses.

Figure 36 shows this fact from a two-dimensional perspective.

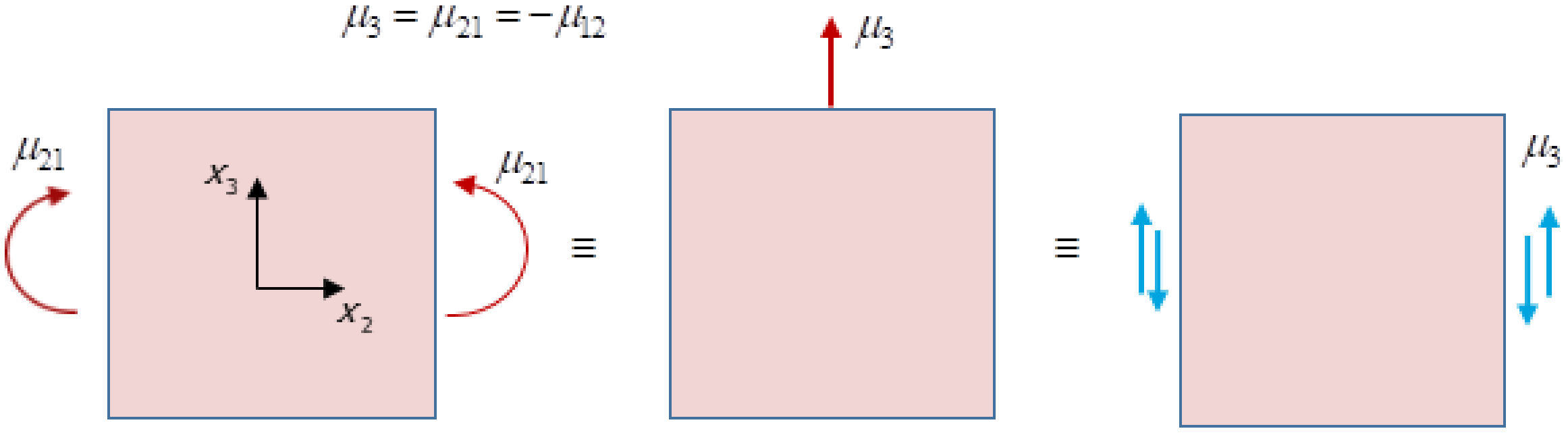

**Figure 36.** Couple stress component $\mu_3$ parallel to its corresponding double layer shear force-stresses in $x_1x_3$ plane.

Consequently, one can completely represent the effect of the couple-stresses by using double layer shear force-stresses. It is remarkable to note that in couple stress continuum mechanics, the bending can be created not only by couple of normal force-stresses (Figure 15), but also by couple of shear force-stresses in form of double layer shear force-tractions (Figure 36).

Interestingly, it is noted that the couple-stress vector $\vec{\boldsymbol{\mu}}$ is normal to the three couple-traction vectors $\vec{\mathbf{m}}^{(1)}$, $\vec{\mathbf{m}}^{(2)}$ and $\vec{\mathbf{m}}^{(3)}$, that is

$$\vec{\mathbf{m}}^{(1)} \bullet \vec{\boldsymbol{\mu}} = m_i^{(1)} \mu_i = 0 \tag{64a}$$

$$\vec{\mathbf{m}}^{(2)} \bullet \vec{\boldsymbol{\mu}} = m_i^{(2)} \mu_i = 0 \tag{64b}$$

$$\vec{\mathbf{m}}^{(3)} \bullet \vec{\boldsymbol{\mu}} = m_i^{(3)} \mu_i = 0 \tag{64c}$$

These relations indicate that the couple-traction vector $\vec{\boldsymbol{\mu}}$ is normal to the plane $\Gamma$. Remarkably, the couple-traction pseudo-vector $\vec{\mathbf{m}}^{(n)}$ can be expressed as

$$m_i^{(n)} = \mu_{ji} n_j = \varepsilon_{ijk} n_j \mu_k \tag{65}$$

or vectorial form

$$\vec{\mathbf{m}}^{(n)} = \mathbf{n} \times \vec{\boldsymbol{\mu}} \tag{66}$$

This obviously shows that the couple-traction pseudo-vector $\vec{\mathbf{m}}^{(n)}$ is in plane $\Gamma$ and perpendicular to the true couple-stress vector $\vec{\boldsymbol{\mu}}$ .

The singularity of the skew-symmetric couple-stress tensor $\mu_{ij}$ also shows that it has one and only one zero eigenvalue. Therefore, to comprehend the character of this zero eigenvalue and its corresponding eigenvector, the eigenvalue problem for the couple-stress tensor $\mu_{ij}$ is considered as follows.

Since $m_i^{(n)} n_i = 0$, there is no direction $n_i$ for which the couple-traction $m_i^{(n)}$ is parallel to the direction $n_i$. However, one can mathematically look for this direction, where

$$m_i^{(n)} = -\lambda n_i \,. \tag{67}$$

Therefore, the eigenvalue problem is obtained by using (67) in (46) as

$$\mu_{ij} n_j = \lambda n_i \,, \tag{68}$$

where $\lambda$ is the eigenvalue of the couple-stress tensor $\mu_{ij}$. This relation can be written as

$$\left(\mu_{ij} - \lambda\delta_{ij}\right)n_j = 0 \tag{69}$$

Note that the condition for (69) to have a non-trivial solution for $n_i$ is

$$\det\left(\mu_{ij} - \lambda\delta_{ij}\right) = 0 \tag{70}$$

This is the characteristic equation for the tensor $\mu_{ij}$, which can also be written as

$$\det\begin{bmatrix} -\lambda & -\mu_3 & \mu_2 \\ \mu_3 & -\lambda & -\mu_1 \\ -\mu_2 & \mu_1 & -\lambda \end{bmatrix} = 0 \tag{71}$$

As a result, the characteristic equation is the cubic equation

$$\lambda^3 + \left(\mu_1^2 + \mu_2^2 + \mu_3^2\right)\lambda = 0 \tag{72}$$

which can be written as

$$\lambda^3 + \mu^2\lambda = 0 \tag{73}$$

This equation shows that the tensor $\mu_{ij}$ has one zero eigenvalue, and two purely imaginary conjugate eigenvalues. This is consistent with our expectation that there is no non-zero real eigenvalue. Let us call the eigenvalues $\lambda_1$, $\lambda_2$ and $\lambda_3$, and arbitrarily choose the third eigenvalue to be the zero eigenvalue. As a result, for these eigenvalues the following relations hold:

$$\lambda_1 = i\mu\,, \qquad \lambda_2 = -i\mu\,, \qquad \lambda_3 = 0 \tag{74}$$

Note that only for $\lambda_3 = 0$, the associated unit eigenvector $n_i^{|3\rangle}$ is real, where

$$\left\{n_i\right\}^{|3\rangle} = \frac{1}{\mu}\begin{Bmatrix} \mu_1 \\ \mu_2 \\ \mu_3 \end{Bmatrix}. \tag{75}$$

This shows that the couple-stress vector $\vec{\boldsymbol{\mu}}$ is in the direction of the eigenvector $\mathbf{n}^{|3\rangle}$ of the tensor $\mu_{ij}$ corresponding to the zero eigenvalue $\lambda_3 = 0$, where

$$\vec{\boldsymbol{\mu}} = \mu\mathbf{n}^{|3\rangle} \qquad\qquad \mu_i = \mu n_i^{|3\rangle} \tag{76}$$

Now choose the orthogonal coordinate system $x_1'x_2'x_3'$ such that the axis $x_3'$ coincides with the direction of the real unit eigenvector $\mathbf{n}^{|3\rangle}$. Therefore, relative to coordinate system $x_1'x_2'x_3'$ the unit eigenvector $\mathbf{n}^{|3\rangle}$ is represented as

$$\{n'_i\}^{\langle 3\rangle} = \begin{Bmatrix} 0 \\ 0 \\ 1 \end{Bmatrix} \tag{77}$$

Interestingly, it is noticed that the plane $x'_1x'_2$ and the plane $\Gamma$ are the same. Therefore, one may choose the orthogonal axes $x'_1$ and $x'_2$ arbitrarily in the plane $\Gamma$. The representation of the couple-stress tensor and vector in this special coordinate system $x'_1x'_2x'_3$ become

$$\left[\mu'_{ij}\right] = \begin{bmatrix} 0 & \mu'_{12} & 0 \\ -\mu'_{12} & 0 & 0 \\ 0 & 0 & 0 \end{bmatrix}, \qquad \{\mu'_i\} = \begin{Bmatrix} 0 \\ 0 \\ \mu'_3 \end{Bmatrix} \tag{78}$$

where

$$-\mu'_{12} = \mu'_3 = \mu\,, \qquad \mu'_3 = |\vec{\boldsymbol{\mu}}| = \mu \tag{79}$$

Figure 37 shows the double layer of shear force-tractions for the total couple stress vector $\vec{\boldsymbol{\mu}}$ corresponding to $\mu'_{21}$.

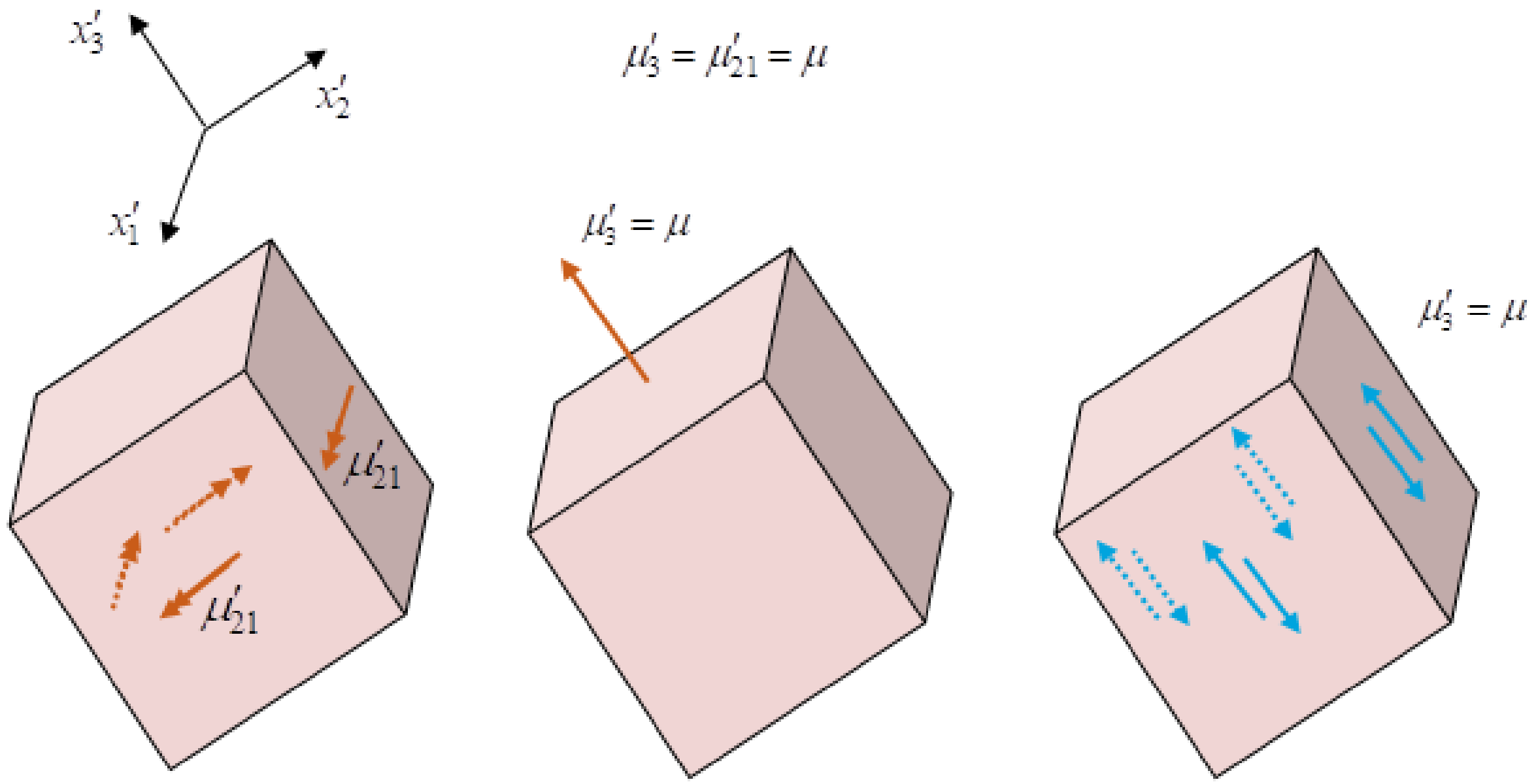

**Figure 37.** Couple stress vector in the direction of its double layer shear force-tractions.

The effect of this couple-stress on an infinitesimal cylindrical element along $x'_3$ axis has been represented in Figure 38, which causes pure bending specified with mean curvature vector along $x'_3$ axis (Hadjesfandiari and Dargush, 2011).

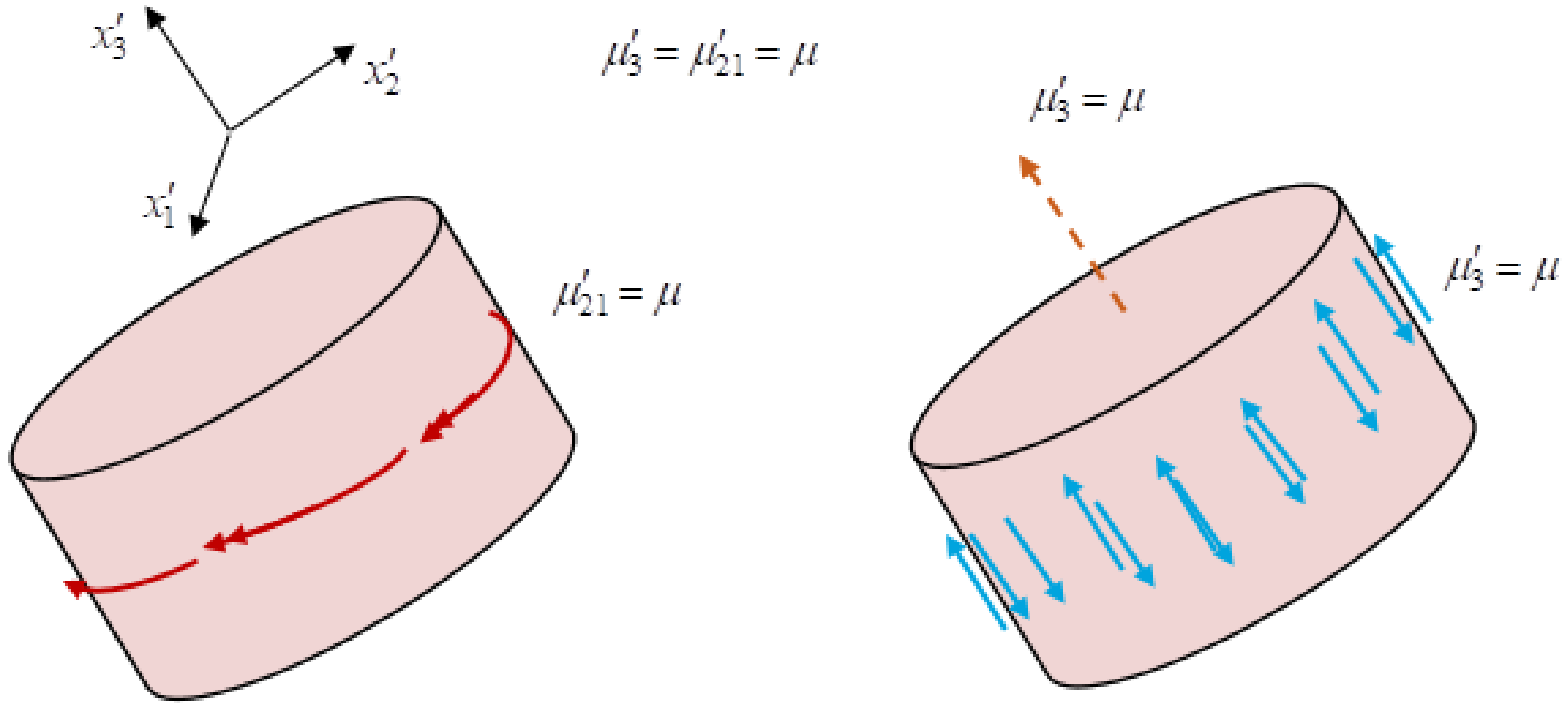

**Figure 38.** Couple stress vector in the direction of its double layer shear force-stresses creating pure bending along $x'_3$ axis.

### 6.4. Consistent fundamental governing equations of motion in differential form

The differential form of the governing equations of motion for an infinitesimal element of matter is finally obtained as

$$\sigma_{ji,j} + f_i = \rho a_i \tag{80}$$

$$\mu_{ji,j} + \varepsilon_{ijk}\sigma_{jk} = 0 \tag{81}$$

where the couple-stress moment tensor is skew-symmetric, that is $\mu_{ji} = -\mu_{ij}$.

It should be emphasized that the derivatives of stresses in the governing equations (80) and (81) are of first order. This is the character of the general fundamental laws of continuum mechanics that their basic form should have first derivatives of stresses, not second or higher orders.

The force-stress tensor is generally non-symmetric and can be decomposed as

$$\sigma_{ij} = \sigma_{(ij)} + \sigma_{[ij]} \tag{82}$$

where $\sigma_{(ij)}$ and $\sigma_{[ij]}$ are the symmetric and skew-symmetric parts, respectively. The relation (60) can be used to express the moment equation (81) as

$$\varepsilon_{ijk}\left(\mu_{k,j} + \sigma_{jk}\right) = 0 \tag{83}$$

which indicates that $\mu_{k,j} + \sigma_{jk}$ is symmetric. Therefore, its skew-symmetric part vanishes, so it follows that

$$\sigma_{[ji]} = \mu_{[j,i]} \tag{84}$$

Thus, the total force-stress tensor can be expressed as

$$\sigma_{ji} = \sigma_{(ji)} + \mu_{[j,i]} \tag{85}$$

Therefore, there are nine independent stress components in consistent couple stress theory or general size-dependent continuum mechanics. This includes six components of $\sigma_{(ji)}$ and three components of $\mu_i$.

Consequently, the force governing equation reduces to

$$\left[\sigma_{(ji)} + \mu_{[j,i]}\right]_{,j} + f_i = \rho a_i \tag{86}$$

which can be called the reduced force governing equation. Since this equation is a combination of the basic force and moment equations (80) and (81), it cannot be considered as a fundamental law by itself. This can be confirmed by noticing that the highest derivative in the governing equation (86) is of second order.

Interestingly, the relation (84) can be elaborated further by considering the pseudo (axial) vector $s_i$ dual to the skew-symmetric part of the force-stress tensor $\sigma_{[ij]}$, where

$$s_i = \frac{1}{2}\varepsilon_{ijk}\sigma_{[kj]} = \frac{1}{2}\varepsilon_{ijk}\sigma_{kj} \tag{87}$$

Then, by using (84) in (87), the following relation is obtained:

$$\vec{\mathbf{s}} = \frac{1}{2}\nabla \times \vec{\boldsymbol{\mu}} \qquad\qquad s_i = \frac{1}{2}\varepsilon_{ijk}\mu_{k,j} \tag{88}$$

It is amazing to notice that the apparently complicated moment equation (81) reduces to the simple curl relation (88). This is the result of the skew-symmetric character of the couple-stress tensor.

## 7. Conclusions

In this paper, it has been demonstrated that the confusion in the concept of couple and its moment vector has been the main reason for the troubles in the progress of couple stress continuum mechanics in the last century. Here, it has been shown that the representation of a couple by its pseudo moment vector in rigid body mechanics and to some extent in strength of materials and structural mechanics has been very misleading. This has given the false notion that the moment of a concentrated couple is a real concentrated vector and completely describes its effect. However, in continuum mechanics the effect of a couple cannot be completely represented by its moment vector when investigating the deformation and internal stresses. There are infinite equipollent couples with the same moment, which create different states of deformation and stresses. To represent the effect of a concentrated couple in continuum completely, the couple moment and the line of action of its opposite parallel forces must be specified. However, in the governing equations of motion only moments of body couple, couple-traction and couple-stresses appear. This requires that if body couple, couple-traction and couple-stresses exist, their effects are completely described by their moment densities $\vec{\mathbf{c}}$, $\vec{\mathbf{m}}^{(n)}$ and $\boldsymbol{\mu}$, respectively, without requiring the specification of the line of action of opposite parallel couple forces. This is the statement of uniqueness of interactions in continuum mechanics, which imposes some restrictions on the form of body couple, couple-traction and couple-stress distributions. Surprisingly, the indeterminacy of couple-stress tensor in Mindlin-Tiersten-Koiter couple-stress theory is the result of non-uniqueness of interactions within this theory. The uniqueness of interactions in the continuum has been used to establish:

1. A distribution of body couple with moment $\vec{\mathbf{c}}$ does not exist;

2. A distribution of surface twisting couple-traction with normal moment $\vec{\mathbf{m}}^{(nn)}$ does not exist on any arbitrary surface inside the body;
3. A distribution of surface bending couple-traction with tangential moment $\vec{\mathbf{m}}^{(nt)}$ can exist;
4. The surface bending couple-traction is a double layer of shear force-tractions ;
5. The pseudo couple-stress moment tensor $\boldsymbol{\mu}$ is skew-symmetric, and has a true vectorial character;
6. The effect of couple-stress is completely described by its skew-symmetric moment tensor $\boldsymbol{\mu}$, thus it can be called couple-stress tensor $\boldsymbol{\mu}$.

It is remarkable that the skew-symmetric character of couple-stress moment tensor has been systematically established by examining the concepts of moment and couple and the fundamental governing equations. Interestingly, by using elements of the work of Mindlin and Tiersten (1962) and Koiter (1964) in the indeterminate couple stress theory, Hadjesfandiari and Dargush (2011) realized that twisting couple-traction with moment $\vec{\mathbf{m}}^{(nn)}$ does not exist in continuum, and established the skew-symmetric character of couple-stress moment tensor $\boldsymbol{\mu}$. However, this original proof does not specify the mechanism of action of the bending couple-traction and couple-stresses. This clearly shows the superiority of the new more fundamental proof based on the uniqueness of interactions in continuum mechanics, which also reveals the structure of bending couple-traction as a double layer of shear force-tractions. Interestingly, this is the tensorial analogy of double layer in electrostatics, where the single layer of tangential shear traction-force distribution is analogous to a single layer of electric charge. As a result, couple-stresses act as double layer of tangential shear-force stresses on their associated planes.

Now it is realized that confusing continuum mechanics with structural mechanics (strength of material method) in many aspects has been the main reason for failure in developing a consistent couple stress theory for a long time. As examined, the confusion originates from the fact that the effect of a couple cannot be completely represented by its moment vector $\vec{\mathbf{M}}$ in continuum mechanics. The pioneers of indeterminate couple stress theory were incorrectly thinking that the moment-density vectors $\vec{\mathbf{c}}$ and $\vec{\mathbf{m}}^{(n)}$ are real vectors acting on the body similar to the force-density vectors $\mathbf{f}$ and $\mathbf{t}^{(n)}$. They did not realize that the moment tensor of a general couple-stress system is indeterminate, because the couple-stress system cannot be completely represented by its moment tensor $\boldsymbol{\mu}$. Only the effect of a couple-stress system with skew-symmetric moment tensor $\boldsymbol{\mu}$ can be completely represented by this skew-symmetric moment tensor $\boldsymbol{\mu}$.

The present systematic discovery of the skew-symmetric character of couple-stress tensor results in the complete form of size-dependent continuum mechanics consistently, which can give us more fundamental insights about the behavior of solids and fluids at the smallest scales for which continuum theory is valid. For example, it provides a fundamental basis for the development of size-dependent nonlinear elastic, elastoplastic and damage mechanics formulations that may govern the behavior of solid continua at the smallest scales.